%% file: draft_v2.tex
\documentclass[a4paper,11pt]{article}
\pdfoutput=1
\usepackage{./auxi/jheppub}
\usepackage[utf8]{inputenc}
\usepackage[most]{tcolorbox}
\usepackage{pifont}

\input{./auxi/packages.tex}
\input{./auxi/commands.tex}

\input{./auxi/topmatter}

\graphicspath{{./Pictures/}}

\definecolor{navyblue}{rgb}{0.0, 0.0, 0.5}

\usepackage[]{mdframed}
\newcommand{\f}{\phi}

\newcommand{\CH}[3]{\hat{\mathcal{C}}_{#1(#2,#3)}}
\newcommand{\C}[4]{\mathcal{C}_{#1,#2(#3,#4)}}

\newcommand{\ie}{\textit{i.e.}~}
\newcommand{\eg}{\textit{e.g.}}

\newcommand{\Mpld}{M_{{\rm Pl},d}}

\newcommand{\Mplfive}{M_{{\rm Pl},5}}
\newcommand{\RLd}{{\mathbb R}^{1,d-1}}

\newtcbox{\bluebox}{
 on line,
 colback=blue!15,
 colframe=blue!50,
 arc=4pt,
 boxrule=0pt,
 left=4pt,
 right=4pt,
 top=2pt,
 bottom=2pt
}
\newtcbox{\greenbox}{
 on line,
 colback=green!20,
 colframe=green!50,
 arc=4pt,
 boxrule=0pt,
 left=4pt,
 right=4pt,
 top=2pt,
 bottom=2pt
}
\newtcbox{\redbox}{
 on line,
 colback=red!20,
 colframe=red!50,
 arc=4pt,
 boxrule=0pt,
 left=4pt,
 right=4pt,
 top=2pt,
 bottom=2pt
}

\title{Multiplet Recombination and the CFT Distance Conjecture}

\author[a,b,c]{Fabio Mantegazza,}
\author[a]{Enrico Marchetto,}
\author[a,b,c]{Elli Pomoni,}
\author[a]{Torben Skrzypek,}
\author[b,c]{\\Timo Weigand}

\affiliation[a]{Deutsches Elektronen-Synchrotron DESY, Notkestr. 85, 22607 Hamburg, Germany}
\affiliation[b]{II. Institut f\"ur Theoretische Physik, Universit\"at Hamburg, Notkestr. 9,
22607 Hamburg, Germany}
\affiliation[c]{Zentrum f\"ur Mathematische Physik, Universit\"at Hamburg, Bundesstr. 55,
20146 Hamburg, Germany}

\emailAdd{fabio.mantegazza@desy.de}
\emailAdd{enrico.marchetto@desy.de}
\emailAdd{elli.pomoni@desy.de}
\emailAdd{torben.skrzypek@desy.de}
\emailAdd{timo.weigand@desy.de}

\abstract{ Motivated by quantum gravity and the CFT Distance Conjecture, we study infinite-distance limits in four-dimensional ${\cal N}=2$ superconformal field theories with higher-dimensional conformal manifolds and their AdS duals. We focus on partial decoupling limits where a gauge sector becomes weakly coupled while an interacting sector persists.

\noindent We analyse the structure of towers of states emerging in these limits. The weakly coupled sector contributes, among others, the massless higher-spin tower predicted by the CFT Distance Conjecture exhibiting polynomial degeneracy. The key novelty is the appearance of a protected BPS tower in the interacting sector, characterised by exponential degeneracy and masses at the AdS scale. This structure follows from multiplet recombination in the ${\cal N}=2$ superconformal algebra: As unprotected long multiplets hit the unitarity bound at weak coupling, they recombine into protected short multiplets. We verify this picture through an explicit one-loop computation in the simplest two-node quiver gauge theory with a two-dimensional conformal manifold.
}
\begin{document}
\begin{flushright}
 \small{DESY-26-026}\\
 \small{ZMP-HH/26-2}
\end{flushright}
 \maketitle
\flushbottom
\allowdisplaybreaks 

\title

\section{Introduction}

In the quest to identify universal properties of quantum gravity theories, asymptotic regions in moduli space are a natural starting point. Generically such regions are expected to exhibit some sort of weak coupling behaviour and are therefore not only candidate regimes of computational control, but are also particularly suitable as testing grounds for general ideas about the possible effective physics. In flat space, the Swampland Distance Conjecture \cite{Ooguri:2006in} characterises these asymptotic regions as regimes where an infinite tower of states becomes asymptotically massless with respect to the Planck scale, exponentially fast in the moduli space geodesic distance. By the Emergent String Conjecture \cite{Lee:2019wij} these towers are either Kaluza--Klein towers associated with a decompactifying dimension, in some dual sense, or excitations of a critical string that becomes asymptotically weakly coupled and massless.
 These claims are backed up by ample evidence in the context of string and M-theory, see \eg~
\cite{Grimm:2018ohb,Grimm:2018cpv,Lee:2019wij,Lee:2019xtm,Rudelius:2023odg,Baume:2019sry,Blumenhagen:2023yws,Xu:2020nlh,Basile:2022zee,Etheredge:2023odp,Aoufia:2024awo,Calderon-Infante:2024oed,Friedrich:2025gvs,Gkountoumis:2025btc} for an incomplete sample of works, 
 as well as by various bottom-up arguments \cite{Hamada:2021yxy,Basile:2023blg,Basile:2024dqq,Bedroya:2024ubj,Kaufmann:2024gqo}. 
 
 A priori, quantum gravity in spaces with a non-zero cosmological constant is qualitatively very different from flat space, 
 and this raises the question to what extent these general lessons apply in presence of a cosmological constant.
 For quantum gravity in AdS space, asymptotic regions in moduli space are mapped, by holography, to certain boundaries of the conformal manifold of a dual Conformal Field Theory (CFT). This has motivated the CFT Distance Conjecture \cite{Perlmutter:2020buo} (see also \cite{Baume:2020dqd}) as a holographically dual version of the Swampland Distance Conjecture: Infinite-distance points on the conformal manifold of a CFT in $d>2$ spacetime dimensions are conjectured to coincide with
 weak-coupling cusps of the conformal manifold, where one or more exactly marginal gauge couplings vanish, and to
 give rise to a higher-spin (HS) tower with asymptotically vanishing anomalous dimension.
 An HS tower is defined as a tower of spin-$J$ symmetric traceless tensor states with conformal dimension $\Delta = 2+J$, and
 by \cite{Maldacena:2011jn,Maldacena:2012sf} it represents a free sub-sector of theory. One direction of the conjecture, that HS towers arise only at infinite distance, has been proved for superconformal field theories (SCFTs) in $d>2$ in \cite{Perlmutter:2020buo} and more generally for unitary CFTs in $d>2$ with a stress energy tensor in \cite{Baume:2023msm}. For the infinite-distance behaviour in two-dimensional CFTs we refer to \cite{Kontsevich:2000yf,Roggenkamp:2003qp,Acharya:2006zw,Ooguri:2024ofs}.

In the context of ${\cal N}=4$ super Yang--Mills (SYM) theory, the conformal manifold is one-dimensional and parametrised by the gauge coupling.
The regime of vanishing gauge coupling corresponds to an infinite-distance locus. 
 In this limit, the anomalous conformal dimension for an HS tower of states vanishes at an exponential rate, according to distance measured by the Zamolodchikov metric of the conformal manifold 
\cite{Baume:2020dqd,Perlmutter:2020buo}. This is as required by the CFT Distance Conjecture.
 If we formally extrapolate the holographic dictionary, \eqref{eq:massgeneral}, for the mass in AdS \cite{Metsaev:2003cu} to this stringy regime,
 this tower can be interpreted as becoming massless with respect to the AdS scale. As we stress in Section \ref{sec: CFT DC 1d} and was also elaborated on in \cite{Calderon-Infante:2026rkj}, this tower has only polynomial, rather than exponential, degeneracy. 
 To recover exponential degeneracy, one has to take into account the vast tower of states with asymptotically vanishing anomalous dimension and $\Delta > 2 +J$. 
 This exponential degeneracy is responsible for the Hagedorn behaviour observed in \cite{Calderon-Infante:2024oed, Calderon-Infante:2026rkj}, but these states sit,
 according to the holographic dictionary, at the AdS mass scale.

 For four-dimensional SCFTs with less supersymmetry 
 the conformal manifold can be more than one-dimensional.
 For SCFTs with a one-dimensional conformal manifold, the CFT Distance Conjecture has been analysed in \cite{Calderon-Infante:2024oed}. 
In particular, three classes of such rank-one theories have been identified, distinguished by the exponential rate at which the anomalous dimension of the HS tower vanish.
 Moreover, limits in higher-dimensional conformal manifolds where all gauge couplings are sent to zero have been investigated in \cite{Calderon-Infante:2026rkj}.

In addition to such overall weak coupling limits, higher-rank four-dimensional SCFTs with ${\cal N}=2$ or ${\cal N}=1$ supersymmetry admit infinite-distance limits that do not necessarily trivialise the full dynamics: Sending a single coupling $\check{g} \to 0$ typically decouples the associated vector multiplet, which becomes free, while the remaining nodes form an interacting SCFT. This induces a reorganisation of the operator spectrum and results in operators associated to either of the decoupled theories. This paper is devoted to studying such partial weak coupling limits in the presence of a residual strongly coupled sector from the point of view of the Distance and the Emergent String Conjecture.

The decoupled vector multiplet produces an infinite tower of conserved HS currents, reflecting the weak coupling limit. The key result of this paper is that as this occurs, many operators in long multiplets have anomalous dimensions that vanish as $\check{g} \to 0$, approaching the unitarity bound. Through multiplet recombination, these long multiplets split into short multiplets, generating additional infinite towers of protected operators in the interacting sector. These towers emerge precisely at the cusp and their structure is fixed by representation theory.

\begin{figure}[t]
 \centering
 \includegraphics[width=\linewidth]{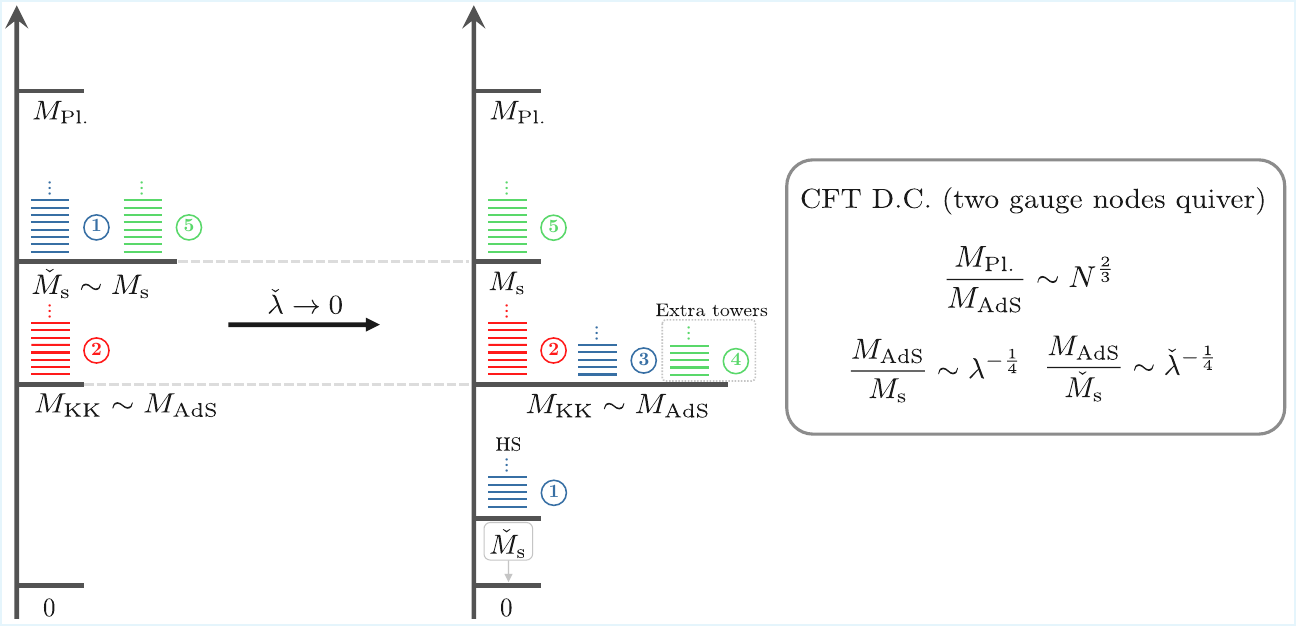}
 \caption{ The CFT Distance Conjecture in AdS space visualised in the case of a single decoupling gauge node. At large $N$, the initial scenario is completely analogous to the situation in $\mathcal{N}=4$ SYM, while the spectrum decomposes into five separate sectors in the limit $\check{\lambda} \to 0$. The precise mass of the HS tower in relation to $\check{M}_s$ is discussed in the main text.}
 \label{fig:N2DC}
\end{figure}

Consequently, weak-coupling infinite-distance limits generically produce five spectral sectors (see Figure \ref{fig:N2DC}): 
\begin{enumerate}
 \item[\textcircled{\footnotesize 1}] a free asymptotically massless higher-spin tower of polynomial degeneracy, which away from the weak-coupling limit becomes non-BPS,\footnote{In our nomenclature, a BPS state refers to a highest weight state in a superconformal multiplet that obeys a shortening condition. For the ${\cal N}=2$ superconformal algebra, the shortening conditions \cite{Dolan:2002zh} are listed in Table \ref{Tab:shortening}.}
 \item[\textcircled{\footnotesize 2}] the generic BPS tower of states present for every value of the couplings, including massless states like the graviton and their KK modes at the AdS scale, of polynomial degeneracy,
 \item[\textcircled{\footnotesize 3}] a tower of states (BPS and non-BPS) belonging to the free decoupled vector multiplet, asymptotically at the AdS scale and of exponential degeneracy, which
 become non-BPS away from the limit,
 \item[\textcircled{\footnotesize 4}] novel \emph{extra} BPS towers of the interacting sector, asymptotically at the AdS scale and of exponential degeneracy, due to multiplet recombination,
 \item[\textcircled{\footnotesize 5}] a non-BPS tower of massive string modes at the string mass scale of the interacting sector.
\end{enumerate}
 The key point is the appearance of the exponential tower of type \textcircled{\footnotesize 4}, whose mass is protected in the infinite-distance limit even though it lives in the interacting sector. This tower is a crucial observable to be matched by any dual description and thus poses an exciting challenge for holography.
\begin{figure}[t]
 \centering
 \includegraphics[width=0.6\linewidth]{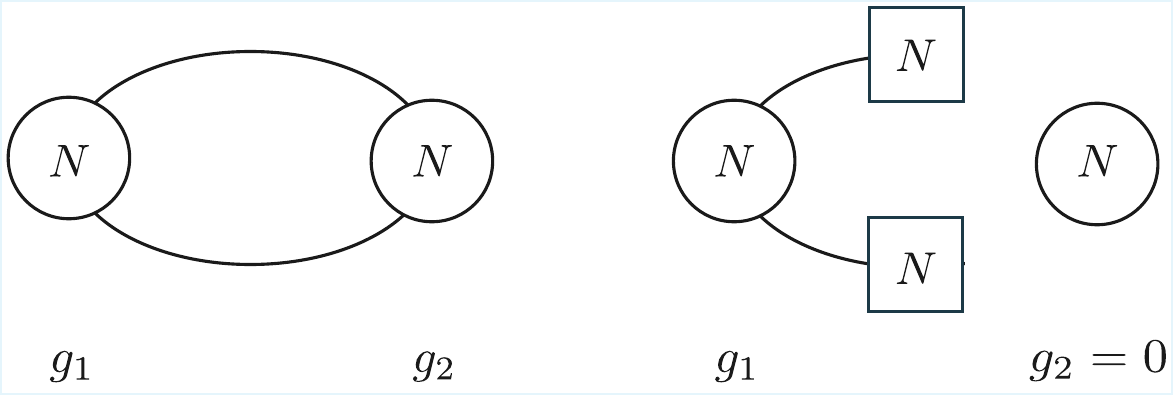}
 \caption{The $\mathcal{N}=2$ quivers of the two-node necklace theory and of SCQCD + decoupled vector multiplet. 
 \label{fig:Quivers}}
\end{figure}

The easiest example to explicitly illustrate the mechanism outlined above is a theory endowed with a two-dimensional conformal manifold. In the main part of this paper, we will consider marginal deformations of the ${\rm AdS}_5 \times S^5/\mathbb Z_2$ orbifold theory at large $N$ which preserves ${\cal N}=2$ supersymmetry and is dual to a two-node quiver gauge theory depicted on the left of Figure \ref{fig:Quivers}. In this case, the moduli space is two-dimensional and spanned by the gauge couplings $g_1$ and $g_2$. This setup is a special case of the more general family of ${\rm AdS}_5 \times S^5/\mathbb Z_k$ orbifolds with a $k$-dimensional conformal manifold.\footnote{In fact, quotienting by any discrete group $\Gamma\subset{SU(2)}$ results in a dual $\mathcal{N}=2$ SCFT. The $\mathbb{Z}_k$ family corresponds to the $A_{k-1}$ series in the usual ADE classification of subgroups $\Gamma$, its quiver diagram matching the associated affine Dynkin diagram. }
Sending one gauge coupling $\check{g} \coloneqq g_2\to0$ we arrive at superconformal QCD (SCQCD) plus a free vector multiplet as displayed on the right of Figure \ref{fig:Quivers}.  We provide an explicit analysis of the multiplet recombination mechanism leading to the extra tower \textcircled{\footnotesize 4} of BPS states appearing in the interacting SCQCD.

In our analysis of the BPS spectrum of SCQCD, we uniquely identify the extra tower of type \textcircled{\footnotesize 4} and separate it from tower \textcircled{\footnotesize 2} of BPS states inherited from the orbifold theory. This extra tower has previously been investigated in \cite{Gadde:2009dj}, where the superconformal index was used to argue for its existence, but a full characterisation of the spectrum of quantum numbers was obstructed by a cohomological ambiguity of the index. In this paper, we overcome this obstruction by a direct analysis via the one-loop dilatation operator in both the orbifold theory and SCQCD. This bottom-up approach enables us to compute the anomalous dimensions of all states, which allows us to both explicitly confirm the protection of the extra states in the weak-coupling limit and to identify their quantum numbers. This concludes the analysis initiated in \cite{Gadde:2009dj}.\footnote{In addition to giving rise to a cohomological ambiguity, the index might, a priori,  undercount the number of BPS states. This potential undercounting can always be addressed through the same one-loop computation by systematically scanning the operator spectrum. Indeed at all levels which we investigated, no such undercounting was observed.} Combined with the $\mathcal{N}=2$ recombination rules, our analysis demonstrates that these states arise from long multiplets splitting into shorter ones. We explicitly show this for the lightest extra multiplet; the extension to subsequent multiplets is straightforward. 

Thus, from the gauge-theory perspective, the infinite-distance limit generates not only free massless towers, but also \emph{interacting} BPS towers of states at the AdS scale. In our specific example, the degeneracy of these towers grows exponentially with the quantum numbers, providing a concrete gauge-theoretic realisation of the rich spectrum expected at infinite distance.

The structure of the paper is as follows. In \textbf{Section \ref{Sec:CFT_Distance}}, we review infinite-distance limits in flat space and AdS and reformulate them in terms of four-dimensional $\mathcal{N}=2$ superconformal representation theory. We explain how weak-coupling cusps on higher-dimensional conformal manifolds lead to additional protected towers in the interacting sector, beyond the higher-spin tower of the free vector multiplet. In \textbf{Section \ref{sec: twonode}}, we analyse this mechanism in the two-node $\mathcal{N}=2$ quiver obtained as a $\mathbb{Z}_2$ orbifold of $\mathcal{N}=4$ SYM. Using the superconformal index, we review that the inherited BPS spectrum (tower \textcircled{\footnotesize 2}) descending from the orbifold is incomplete and that extra protected multiplets appear at the cusp. We then identify these states via a one-loop analysis of the dilatation operator and determine their quantum numbers and degeneracies. In \textbf{Section \ref{Sec:Conclusion}}, we summarise our results and discuss future directions.

Some technical material is collected in the appendices. \textbf{Appendix \ref{App:Dolan-Osborn}} reviews the classification, shortening conditions, and recombination rules of four-dimensional $\mathcal{N}=2$ superconformal multiplets following \cite{Dolan:2002zh}. \textbf{Appendix \ref{app: Harmonic}} presents the oscillator representation and harmonic action used in the one-loop computation of anomalous dimensions and in the construction of the dilatation operator (for SCQCD this was worked out in \cite{Liendo:2011xb}).

\section{Recombination and CFT Distance Conjecture }
\label{Sec:CFT_Distance}

In this section we review some of the core statements of the Swampland and CFT Distance Conjectures. We then move on to describe the multiplet recombination that occurs in weak-coupling limits in 4d $\mathcal{N}=2$ SCFTs and make a case for the generic appearance of additional strongly coupled towers of states at the AdS scale.

\subsection{Distance Conjectures in flat space -- Review}

\begin{figure}
 \centering
 \includegraphics[width=0.75\linewidth]{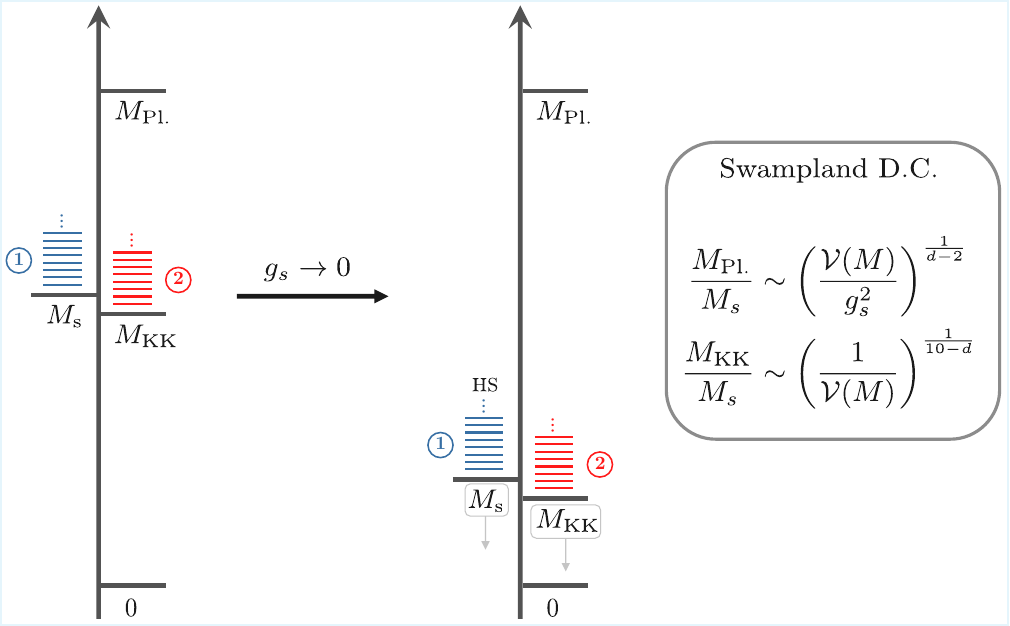}
 \caption{ 
 The Swampland Distance Conjecture in flat space visualised. Tower \emph{\textcircled{\footnotesize 1}} sits at the string scale, \emph{\textcircled{\footnotesize 2}} at the KK scale. The scale separation between them is determined by the volume of the internal manifold $\mathcal{V}(\mathcal{M})$, as prescribed by equation \eqref{eq: scales flat}. In this cartoon, we are already considering $\mathcal{V}(\mathcal{M}) \sim \mathcal{O}(1)$ and fixed, such that the string and the KK scales are not separated and are ``locked'' with each other. In the $g_s \to 0$ limit, both towers become asymptotically massless.}
 \label{fig:flatDC}
\end{figure}

To recall some of the main properties of infinite-distance limits in flat spacetime, 
 we consider a $d$-dimensional gravitational effective field theory (EFT) with action
\begin{equation}\label{eq:EFT}
 S = \frac{1}{2}(\Mpld)^{d-2} \int_{\RLd} {\sqrt{-g}}\left(  R + g_{ij}(\phi) \partial \f^i \partial \f^j + \ldots \right) \,.
\end{equation}
Here $g_{ij}$ denotes the metric on the moduli space spanned by the massless moduli fields $\f^i$.
 The Swampland Distance Conjecture \cite{Ooguri:2006in} constrains the behaviour of such a gravitational EFT near the boundary of the moduli space at infinite geodesic distance $ \delta$, measured in Planck units and with respect to $g_{ij}$: 
A tower of infinitely many, gravitationally weakly coupled states becomes light at an exponential rate, with scaling behaviour 
\begin{equation}
 \frac{m_{\rm tower}}{\Mpld} \sim {\rm exp}(-\alpha \, \delta ) \qquad {\rm as } \quad \delta \to \infty \, .
 \label{Eq:MassDC}
\end{equation}
The range for the exponential decay rate $\alpha$ as well as the nature of the towers are further constrained by the Emergent String Conjecture \cite{Lee:2019wij}: 
 The leading tower is either a Kaluza--Klein (KK) tower signalling one or several decompactifying spatial dimensions (decompactification limit) or it consists of the excitations of an asymptotically tensionless and weakly coupled critical string (emergent string limit). In this case, the tower of string excitations is accompanied by a Kaluza--Klein tower at the same parametric mass scale, unless $d$ corresponds to the critical dimension of the emergent string (in which case the EFT is defined in the maximal possible dimension of the string and hence, in absence of a compactification, no Kaluza--Klein tower can occur). 

 These two types of towers can be distinguished by the level dependence of the masses and degeneracy. For a KK-tower, the masses grow linearly with the level number $n$, while the degeneracy grows polynomially in $n$, at least for $n \gg 1$,
 \begin{equation}
 m_{{\rm KK},n}\sim n \,, \qquad {\rm deg}_{{\rm KK},n} \sim n^p \,.
 \label{Eq:MassKK}
 \end{equation}
 String towers, on the other hand, have a denser excitation spectrum and their degeneracy grows exponentially in the level number $\sqrt{n}$, for $n \gg 1$,
\begin{eqnarray} \label{mass-deg-string-flat}
m_{{\rm string},n}\sim \sqrt{n} \,, \qquad {\rm deg}_{{\rm string},n} \sim e^{c \sqrt{n}} \,, \qquad c\in \mathbb R \,.
\end{eqnarray}
This restriction on the types of infinite-distance limits is reflected in a lower bound on the exponential vanishing rate \cite{Etheredge:2022opl},
 \begin{equation}
 \alpha \geq \frac{1}{\sqrt{d-2}} \,,
\end{equation}
which is saturated precisely in emergent string limits. 
 For these, the infinite-distance direction can be interpreted as approaching the limit of vanishing string coupling for a dual critical string (even if the original quantum gravity was not formulated as a string theory).
Recall that for a compactification of ten-dimensional string theory on an internal space ${\cal M}$, the Planck mass $\Mpld$ and Kaluza--Klein scale are determined by 
\begin{equation}
\left(\frac{\Mpld}{M_s}\right)^{d-2} \sim \frac{{\cal V}({\cal M})}{g_s^2} \,, \qquad \left(\frac{M_{\rm KK}}{M_s}\right)^{10-d} \sim \frac{1}{{\cal V}({\cal M})} \,, \label{eq: scales flat}
\end{equation}
where $g_s$ denotes the ten-dimensional string coupling, 
${\cal V}({\cal M})$ measures the volume of ${\cal M}$ in units of the string length $\ell_s = M_s^{-1}$ and we are furthermore assuming for simplicity that ${\cal M}$ is isotropic, so that the Kaluza--Klein scale is set by the typical diameter of $\cal M$.
 It is then clear that in the limit $g_s \to 0$, with ${\cal V}({\cal M})$ fixed, both $M_{s}$ and $M_{\rm KK}$ jointly go to zero in Planck units,
 \begin{equation}\label{eq: flatlimit}
 \frac{M_s}{\Mpld} \sim \frac{M_{\rm KK}}{\Mpld} \to 0 \,.
 \end{equation}
This explains the characteristic locking between the string tower and the accompanying KK tower, visualised in Figure \ref{fig:flatDC}.

The validity of the Emergent String Conjecture in flat space has been confirmed in numerous analyses of compactifications of string or M-theory, for example \cite{Lee:2018urn,Lee:2019wij,Lee:2019xtm,Rudelius:2023odg,Lee:2021qkx,Lee:2021usk,Chen:2024cvc,Alvarez-Garcia:2023gdd,Alvarez-Garcia:2023qqj,Alvarez-Garcia:2021pxo,Marchesano:2019ifh,Baume:2019sry,Blumenhagen:2023yws,Xu:2020nlh,Lee:2019tst,Klaewer:2020lfg,Basile:2022zee,Etheredge:2023odp,Aoufia:2024awo,Calderon-Infante:2024oed,Friedrich:2025gvs,Gkountoumis:2025btc,Monnee:2025ynn,Monnee:2025msf}.
 It crucially rests on the intricate interplay of string dualities and non-trivial details of the compactification geometry. Bottom-up arguments for it have been developed in \cite{Basile:2023blg,Basile:2024dqq,Bedroya:2024ubj,Kaufmann:2024gqo}. 

Importantly, both the Swampland Distance Conjecture and the Emergent String Conjecture allow for the appearance of additional, non-gravitationally coupled towers to become massless in infinite-distance limits. Such strongly coupled 
field theory towers, decoupled from the weakly coupled gravitational KK or string towers, have been analysed in detail in \cite{Marchesano:2023thx,FierroCota:2023bsp,Marchesano:2024tod,Castellano:2024gwi,Blanco:2025qom, Monnee:2025ynn}. 

\subsection{Distance Conjecture in ${\rm AdS}_5$ -- One-dimensional conformal manifolds} \label{sec: CFT DC 1d}
\begin{figure}
 \centering
 \includegraphics[width=0.75\linewidth]{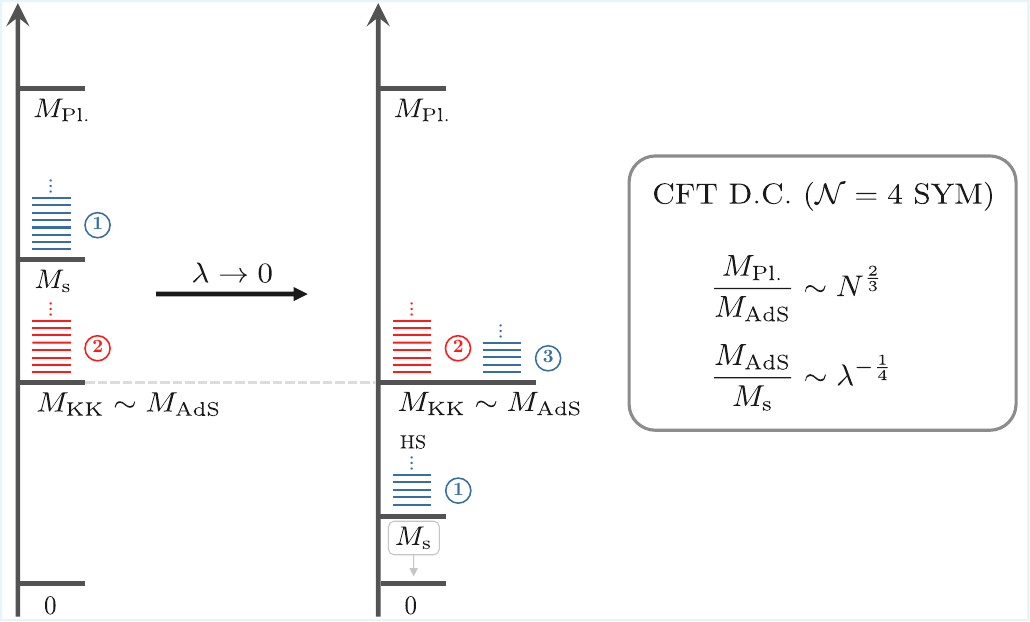}
 \caption{The CFT Distance Conjecture in AdS space visualised in the case of a one-dimensional conformal manifold on the boundary, with $\mathcal{N}=4$ SYM being employed as an illustrative example. At large $N$, the AdS and the Planck scales are separated as prescribed by equation \eqref{lpllAdS}. Tower \emph{\textcircled{\footnotesize 1}} becomes light with the rate given by \eqref{effective-mass}; tower \emph{\textcircled{\footnotesize 2}} is a tower of BPS states and sits at the KK scale. In this case, the separation between \emph{\textcircled{\footnotesize 1}} and \emph{\textcircled{\footnotesize 2}} is governed by $\lambda$ (see eq. \eqref{lslAds}), and both tower scales are not locked together. 
 In the infinite-distance limit, the AdS scale does not renormalise (as indicated by the dashed horizontal line), while the string tower goes asymptotically to zero, generating a tower of massless states (dual to HS currents on the gauge theory side). The tower \emph{\textcircled{\footnotesize 3}} represents all the new BPS states that belonged to the tower \emph{\textcircled{\footnotesize 1}} and got ``stuck'' at the AdS scale.}
 \label{fig:N4DC}
\end{figure}

Let us now turn to gravitational theories in AdS$_5$, related by holography to 4d CFTs.
The simplest example is, of course, Type IIB string theory on ${\rm AdS}_5 \times S^5$, which is dual to 4d $\mathcal{N}=4$ super Yang--Mills (SYM) theory.

The nature of infinite-distance limits in this context is discussed in detail in \cite{Baume:2020dqd, Perlmutter:2020buo}. 
 Recall first that it is important to distinguish between the five-dimensional Planck scale $\Mplfive = \ell_{\text{Pl}}^{-1}$, the 10d Type IIB string scale $M_s = \ell_s^{-1}$, and the AdS scale $M_{\rm AdS} = \ell^{-1}_{\rm AdS}$. 
The ratio of the latter two is set by the
 't Hooft coupling of the dual ${\cal N}=4$ SYM theory with gauge group $SU(N)$,
 \begin{equation} \label{lslAds}
 \lambda = g_{\text{YM}}^2N=4\pi g_s N \sim \frac{\ell^4_{\rm AdS}}{\ell_s^4} \,, \end{equation}
while the Planck scale follows from the relation
\begin{equation} \label{Planck-AdS}
\frac{M^3_{{\rm Pl},5}}{M_s^3} \sim \frac{{\cal V}(S^5)}{g_s^2} \sim \frac{\ell^5_{\rm AdS}}{\ell_s^5 g_s^2} \,.
\end{equation}
 In the second step we used 
 that the $S^5$ radius coincides with the AdS radius. Combining (\ref{lslAds}) and (\ref{Planck-AdS}) yields
\begin{equation} \label{lpllAdS}
\frac{M^3_{\rm AdS}}{M_{\rm Pl}^3} \sim \frac{1}{N^2} \,.
\end{equation}
 Recall that from the point of view of AdS/CFT, this identity is derived by matching the two-point correlator of the graviton, which is proportional to ${\ell^3_{\rm AdS}}/{\ell_{\rm Pl}^3}$, with the two-point correlator of the stress energy tensor in the dual CFT,
 \begin{equation}
 \langle T^{\mu\nu}(x)T^{\rho\sigma}(0)\rangle = \frac{C_T}{x^8}\mathcal{I}^{\mu\nu\rho\sigma}(x)\,,
\end{equation}
 where the prefactor is given by the ``central charge'' $C_T \sim N^2$. 

 The AdS/CFT correspondence is best understood in the planar large-$N$ limit, to which we will restrict our attention in this paper. In this regime the Planck scale $\Mplfive$ lies parametrically above the string scale $M_s$ and the AdS scale $M_{\rm AdS}$. 

To have access to an effective gravitational field theory on ${\rm AdS}_5$, 
 the string scale $M_s$ must lie parametrically above the AdS scale $M_{\rm AdS}$, as displayed in Figure \ref{fig:N4DC}. This is reminiscent of the ordering of scales in Minkowski space with compactifications in the parametric large volume regime 
 (see Figure \ref{fig:flatDC} left) 
 and occurs at strong coupling $\lambda \gg 1$.

 In this regime we can describe the low-energy modes in the bulk via an EFT akin to \eqref{eq:EFT}, supplemented by a cosmological constant to account for the AdS background. It is then possible to relate the mass parameters of bulk fields to their representation under the AdS symmetry group $SO(4,2)$, which coincides with the conformal group of the dual CFT. The necessary holographic dictionary was discussed in \cite{Metsaev:2003cu}. For bosonic and fermionic fields it takes the form
\begin{equation} \label{eq:massgeneral}
 \frac{m^2_{\text{bos}}}{M^2_{\rm AdS}}=(\Delta-j-\bar{\jmath}-2)(\Delta+j+\bar{\jmath}-2)\,,\qquad \frac{m^2_{\text{ferm}}}{M^2_{\rm AdS}}=(\Delta-j-\bar{\jmath}-2)^2\,,
\end{equation}
where $\Delta$ denotes the conformal dimension of the CFT dual operator, and $j\neq0$ and $\bar{\jmath}\neq0$ are the spin quantum numbers.\footnote{The special cases $j=0$ and/or $\bar{\jmath}=0$ can also be inferred from \cite{Metsaev:2003cu}:
For $j=0$ we need to use
\begin{equation} \label{eq:massjzero}
 m^2_{\text{bos}}=(\Delta-\bar{\jmath}-1)(\Delta+\bar{\jmath}-3)\,,\qquad m^2_{\text{ferm}}=(\Delta-\bar{\jmath}-1)^2\,,
\end{equation}
and similar for $\bar{\jmath}=0$. In three special boundary cases $(j,\bar{\jmath})=(0,0)$, $(0,\tfrac{1}{2})$ and $(\tfrac{1}{2},0)$ one is to apply the scalar and spinor mass relations
\begin{equation} \label{eq:massspecial1}
 m^2_{(0,0)}=\Delta(\Delta-4)\,,\qquad m^2_{(0,\tfrac{1}{2})}=m^2_{(\tfrac{1}{2},0)}=(\Delta-2)^2\,.
\end{equation}}

The $S^5$-factor of spacetime appears in the EFT through a KK tower of massive fields, which organises in spherical harmonics, \ie, in representations of $SO(6)$. Since the $S^5$ and $\mathrm{AdS}_5$ radii coincide, the KK mass scale is given by $M_{KK} \approx M_{\rm AdS}$, with the precise integer coefficients analysed in \cite{Kim:1985ez}. The dual operators transform accordingly in representations of the global $SU(4)_R$ $R$-symmetry of the CFT, with the $R$-charge contributing to $\Delta$.

 From a CFT point of view, conformal dimensions are the more natural observables to consider and they usually depend on the coupling $\lambda$ as
\begin{equation}
\Delta(\lambda)=\Delta_0+\gamma(\lambda)\,,
\end{equation}
in terms of a ``bare dimension'' $\Delta_0=\Delta(0)$ and an ``anomalous dimension'' $\gamma(\lambda)$. In $\mathcal{N}=4$ SYM, the integrability of the theory allows one to determine the spectrum of conformal dimensions at any coupling $\lambda$ (see \cite{Beisert:2010jr} for a review). 

So far we have only discussed effective fields in the strong coupling regime $\lambda \gg 1$. As we move to weak coupling the situation changes drastically. From \eqref{lslAds} we observe that at $\mathcal{\lambda} \approx \mathcal{O}(1)$ the string scale $M_s$ becomes comparable to the AdS scale $M_{\rm AdS}$, which suggests that excited string modes begin to play a role in the EFT. This results at first in higher-derivative corrections and ultimately in a breakdown of the EFT as a local gravitational field theory. As the string scale passes below the AdS scale (see Figure \ref{fig:N4DC}), a stringy regime emerges, which is poorly understood in the bulk, but has a natural dual description in terms of the weakly coupled CFT. In this regime the left-hand side of \eqref{eq:massgeneral} has lost its meaning, but the right-hand side is still perfectly meaningful in the dual CFT. With this understanding, we shall henceforth adopt the convention that \eqref{eq:massgeneral} defines an \emph{effective mass} $m_{\text{eff}.}$ at any coupling. This allows us to consider ``massless'' towers of states even in a regime where strictly no effective gravity description exists.\footnote{This convention is implicitly adopted in the CFT Distance Conjecture \cite{Baume:2020dqd,Perlmutter:2020buo}.}

Having reviewed these basic, but crucial points about the AdS/CFT correspondence, we can now step back and look at the theory through the lens of the Swampland Distance Conjecture \cite{Baume:2020dqd,Perlmutter:2020buo}. At fixed value of $N$, the moduli space of the theory is spanned by the 't Hooft coupling $\lambda$,
 \ie, the conformal manifold is one-dimensional.
 The weak coupling limit (Figure \ref{fig:N4DC}),
\begin{equation} \label{limitlambda}
\lambda \to 0 \,, 
\end{equation}
may be identified as the AdS version of an emergent string limit due to the appearance of a tower of string excitations \cite{Baume:2020dqd,Perlmutter:2020buo}.
 As we will see below, the mass of this tower is a non-trivial function of the 
 string scale defined as\footnote{The string scale governs the overall normalisation of the string sigma model on the exact background ${\rm AdS}_5\times S^5$ governed in turn by the AdS-scale. At strong curvature, \ie, $\lambda\to 0$, the target space interpretation is lost and we have to rely solely on the worldsheet sigma model \cite{Berkovits:2007rj, Tseytlin:2002gz}.  
}
\begin{equation} \label{mtowerAdSN4}
 \frac{M_{s}}{M_{\rm AdS}} \sim \lambda^{\frac{1}{4}} \to 0 \, .
\end{equation}
 The tower of string excitations in fact splits into two sub-towers: A tower of polynomial degeneracy becomes massless with respect to the AdS scale exponentially fast. This is the tower directly predicted by the CFT Distance Conjecture. In addition, a tower of exponential degeneracy remains at the AdS scale (tower {\textcircled{\footnotesize 3}} in Figure \ref{fig:N4DC}). The latter can be identified with the KK scale,
\begin{equation} \label{MKKvsMAdS}
 M_{\rm KK} \sim M_{\rm AdS} \gg M_{\rm s} \,.
 \end{equation}
This reflects the inherently ten-dimensional nature of the ${\rm AdS}_5 \times S^5$ background as opposed to a compactification to five dimensions with scale separation, where we would expect behaviour similar to \eqref{eq: flatlimit}.

More precisely, the states responsible for the asymptotically massless tower are string states with \begin{equation}\label{eq: HS quantum numbers}
\Delta_0=2+J \,, \qquad j=\bar{\jmath}=\tfrac{J}{2} \,, 
\end{equation}
 which are dual to HS currents of the schematic form (see \eg~\cite{Anselmi:1999bb} for the precise expression in $\mathcal{N}=4$ SYM)
\begin{equation} \label{eq:highspin-states}
 \tr[\phi^i \partial^J\phi_i] + \tr[\bar \lambda^I \partial^{J-1}\lambda_I]+ \tr[\bar \Fm \partial^{J-2} \Fm]. 
\end{equation}
Here, $\phi^i$ ($i=1,\dots,6$) denote the six real scalar fields, $\lambda^I$ ($I=1,\dots,4$) the Weyl fermions, and $\Fm$, $\bar{\Fm}$ the self-dual and anti-self-dual components of the field strength, respectively. All fields transform in the adjoint representation of the gauge group and together constitute the vector multiplet of $\mathcal{N}=4$ SYM theory. The derivatives $\partial^J$ are taken in the symmetric traceless representation, so that the operators carry definite spin $J$. From \eqref{eq:massgeneral} we can see that the effective mass of these states is proportional to the anomalous dimension $\gamma(\lambda)$,
\begin{equation} \label{effective-mass}
 \frac{m_{\text{eff.}}^2}{M^2_{\rm AdS}}=\gamma(\lambda)(2J+\gamma(\lambda))\sim 2J\lambda +{\cal O}({\lambda^{2}})\,.
\end{equation}
In the weak coupling limit $\lambda \to 0$, the anomalous conformal dimensions $\gamma(\lambda)$ vanish exponentially fast in the distance computed via the Zamolodchikov metric of the CFT \cite{Baume:2020dqd,Perlmutter:2020buo}, until they hit the unitarity bound at $\lambda=0$, as we will discuss below.
Indeed, the Zamolodchikov metric 
\begin{equation}
{\rm{d}}s^2  = \frac{1}{\alpha^2} \, \frac{{\rm{d}} \lambda^2}{\lambda^2} \,, \qquad \frac{1}{\alpha^2} = \frac{{\rm dim}G}{2c}\,,
\label{Eq:Zamolodchikov}
\end{equation}
 on the conformal 
manifold can be deduced, at weak coupling, from the gauge kinetic terms, and we follow the normalisation of \cite{Calderon-Infante:2024oed, Calderon-Infante:2026rkj}. The central charge $c$ of $\mathcal{N}=4$ SYM is given by $c=\tfrac{\text{dim}G}{4}$, which results in $\alpha=\tfrac{1}{\sqrt{2}}$. 
 The point $\lambda\ll 1$ on the conformal manifold hence lies at distance  
 \begin{equation}
 \delta = - \frac{1}{\alpha}  \,{\rm log} \lambda + \ldots \,,
\end{equation}
and the HS tower thus obeys the mass relation\footnote{We keeping $N$ large and fixed so that the ratio $M_{\rm Pl}$ and $M_{\rm AdS}$ does not change as we vary $\lambda$.}
\begin{equation} \label{eq:massformula}
 \frac{m_{\text{eff.}}^2}{M^2_{\rm AdS}} = 2 J \, {\rm exp} \left(- \alpha \,  \delta\right) \,, \qquad (\alpha = \frac{1}{\sqrt{2}} \quad {\rm in } \quad {\cal N} =4 \, \, {\rm SYM}) \,. 
\end{equation}
 As a result, these higher spin current states form exponentially massless towers.
 
 Note that the scaling of the string tower mass with $\lambda$ as deduced in (\ref{effective-mass}) indicates a substantial mass renormalisation compared to the naive string scale (\ref{mtowerAdSN4}). This is no surprise given that the theory is in a deeply stringy regime.\footnote{At strong coupling on the other hand, the anomalous dimension of generic unprotected operators scales as $\gamma(\lambda)\sim \lambda^{\frac{1}{4}}$ \cite{Gubser:1998bc}, which inserted in \eqref{eq:massgeneral} results in $\tfrac{m_{\rm eff.}}{M_{\rm AdS}}\sim\gamma \sim\lambda^{\frac{1}{4}}$, reproducing the expected string scale hierarchy \eqref{mtowerAdSN4}.}

The CFT Distance Conjecture \cite{Baume:2020dqd,Perlmutter:2020buo}    generalises this pattern by claiming an equivalence between limits at infinite distance in the Zamolodchikov limit of a CFT in $d > 2$ dimensions and limits in which the anomalous dimension of a tower of HS currents vanishes and becomes free. One direction of this conjecture, the fact that a tower of free HS currents appears only at infinite distance, has been proven, for unitary CFTs in $d>2$ with a stress energy tensor, in \cite{Baume:2023msm}. Infinite-distance limits in CFTs of dimension $d=2$ are discussed in \cite{Ooguri:2024ofs}.

From the mass formula (\ref{eq:massformula}) we observe that the HS current exhibits the characteristic string-like mass spacing $m_{\text{eff.},J}\sim \sqrt{J}$. However, the degeneracy of states only grows polynomially, \begin{equation} \label{polynomial-deg}
\text{deg}(m_{\text{eff.}})\sim J^2 \,,
\end{equation}
since these states furnish symmetric traceless representations of $SO(1,3)$ whose dimensions grow quadratically. This point has also been stressed in the present context in \cite{Calderon-Infante:2026rkj}. The polynomial degeneracy of the HS tower is to be contrasted with the exponential degeneracy (\ref{mass-deg-string-flat}) of the asymptotically massless string tower in flat space.

This difference is due to the full string tower splitting in two sub-towers (\textcircled{\footnotesize 1} and \textcircled{\footnotesize 3} in Fig. \ref{fig:N4DC}), as anticipated already.
 This becomes evident by analysing in more detail the spectrum of excited string states in AdS \cite{Beisert:2003te}, which organise into superconformal multiplets. The HS currents (\ref{eq:highspin-states}) correspond to the leading Regge trajectory, picking out the maximum spin states at each level which satisfy $\Delta_0=2+J$ \eqref{eq: HS quantum numbers}. All other states are generated by turning on string excitations that do not contribute to the spin of the state, such as KK-momentum on the $S^5$ or massive string modes.\footnote{In case of KK-mode excitations the $R$-charge of the dual operator is raised. This does not apply for massive string mode excitations, which only raise the bare dimension $\Delta_0$.} 
 It is clear that in both cases, the states receive a contribution to their mass of the order of the AdS scale, where in the case of KK momentum this is due to the identification (\ref{MKKvsMAdS}).
 Indeed, such excitations raise the bare dimension by integer numbers but leave the spin unchanged. A generic string state thus has a bare mass contribution to the effective mass \cite{Berenstein:2002jq} 
 \begin{equation}\label{eq: intmasses}
 \frac{m_{\text{eff.}}^2}{M^2_{\rm AdS}}=m_1+m_2\gamma(\lambda)+\dots\,,\qquad m_1,m_2 \in \mathbb{N}\,,
\end{equation}
 where $m_1=0$ generically requires vanishing $R$-charge.\footnote{A sparse set of massless modes with non-vanishing $R$-charge may occur at low spins $j=0$ and/or $\bar{\jmath}=0$, which follow special mass relations as indicated in \cite{Metsaev:2003cu} and in the footnote to \eqref{eq:massgeneral}.} As the string scale $M_s$ passes below the AdS scale $M_{\rm AdS}$, an \emph{exponentially degenerate} tower of string modes thus ``gets stuck'' at the AdS scale $M_{\rm AdS}$. This applies in particular to the KK-tower associated to the $S^5$ subspace. Only the leading Regge trajectory satisfying \eqref{eq: HS quantum numbers} can generate a tower of exactly massless states, but this part of the spectrum exhibits only polynomial degeneracy (\ref{polynomial-deg}).
By contrast, as already observed in \cite{Calderon-Infante:2024oed, Calderon-Infante:2026rkj}, the exponential degeneracy of states responsible for the Hagedorn temperature and the divergence of the partition function is due to states at the AdS scale.

To summarise, compared to emergent string limits in flat space, we observe a qualitative difference in AdS. While it is true that a tower of string excitations becomes weakly coupled, this tower splits in two sub-towers, a tower of asymptotically massless higher-spin (HS) currents with polynomial growth in its degeneracy, and the remaining exponentially degenerate tower at the AdS scale, which is parametrically above the HS tower and sits at the KK scale.
We will encounter similar and in fact even richer behaviour in theories with a higher-dimensional conformal manifold.

\subsection{Higher-dimensional conformal manifolds}
\label{ssec: recombination}

Even though presented in the specific context of ${\cal N}=4$ SYM,
 most arguments of the previous section are only dependent on representation theory of $SO(4,2)$ and the existence of a weak-coupling limit in moduli space. The same discussion may thus be applied to holographic SCFTs preserving less supersymmetry \cite{Calderon-Infante:2024oed}. Once we break supersymmetry to $\mathcal{N} \leq 2$, a generic SCFT has a higher-dimensional conformal manifold, in which a plethora of infinite-distance limits become accessible.

The focus of this paper is to investigate the additional effects of  infinite-distance limits on $\mathcal{N}=2$ SCFTs endowed with a \emph{higher-dimensional} conformal manifolds.\footnote{In order to make contact with holography, we require a large central charge such that the theories allow for a dual description in terms of an effective gravitational theory in ${\rm AdS}_5$, with a parametric scale separation between Planck and AdS scale $\Mplfive\gg M_{\rm AdS}$. In large-rank gauge theories we may equally parameterise the conformal manifold by 't Hooft couplings as  we did in $\mathcal{N}=4$ SYM.} The paradigmatic example are superconformal 4d $\mathcal{N}=2$ gauge theories with multiple exactly marginal gauge couplings $ (g_1, g_2, \dots, g_{k} )$, spanning the associated conformal manifold. 
An important class of infinite-distance limits in this moduli space are weak-coupling limits of gauge multiplets with components ($\phi$, $\lambda$, $\Fm$), which may couple to hypermultiplets ($Q$, $\psi$) in certain specific combinations, as classified by \cite{Bhardwaj:2013qia}. 
 Overall weak-coupling limits in such theories, sending all couplings to zero, have recently been studied in \cite{Calderon-Infante:2026rkj}.
 
 More generally, it is possible to send only a subset of the gauge couplings to zero, in the simplest case only one coupling, 
 \begin{equation}
 \check{g} \coloneqq g_k \to 0 \,.
 \end{equation}
 As predicted by the CFT Distance Conjecture, this limit is associated with the appearance of an infinite tower of massless HS currents always of the schematic form (see also  \eqref{eq: Cvv} below)
\begin{equation}
\label{eq: HScurrent}
 \tr[\check{\bar \phi} \partial^J\check\phi] + \tr[\check{\bar \lambda}^I \partial^{J-1}\check\lambda_I]+ \tr[\check{\bar \Fm} \partial^{J-2} \check \Fm].
\end{equation}
All fields belong to the decoupling vector multiplet of a 4d $\mathcal{N}=2$ theory, which throughout this paper we will denote with a check. In particular, $\check\phi$ and $\check{\bar{\phi}}$ denote the complex scalar fields, $\check\lambda^I$ and $\check{\bar{\lambda}}^I$ ($I=1,2$) are Weyl fermions, and $\check \Fm$ and $\check{\bar{\Fm}}$ represent the self-dual and anti-self-dual components of the field strength, respectively. The derivatives $\partial^J$ are projected onto the symmetric traceless representation, ensuring that the operators have definite spin $J$. From the gauge theory point of view, the vector multiplet associated with the coupling $\check{g}=0$ becomes free and decouples from the theory, in which case it gains a higher-spin symmetry \cite{Maldacena:2011jn} responsible for the tower of massless currents. As in the discussion of Section \ref{sec: CFT DC 1d}, 
 the decoupled theory therefore contributes 
 the usual asymptotically massless tower of HS currents with polynomial degeneracy (\textcircled{\footnotesize 1} in Fig. \ref{fig:N2DC}) as well as a free
 massive string tower at the AdS scale (\textcircled{\footnotesize 3} in Fig. \ref{fig:N2DC} with exponential degeneracy). 
 The HS tower continues to become massless at an exponential rate in the moduli space distance, which can be determined from the Zamolodchikov metric, see \cite{Calderon-Infante:2026rkj}.

 The element of novelty is that we are now left with a new interacting 4d $\mathcal{N}=2$ SCFT, whose conformal manifold is still parametrised by the remaining couplings $( g_1, \dots, g_{k-1} )$.
 As it turns out, there 
will appear \emph{extra towers of BPS operators} in the strongly-coupled remainder theory whose mass is protected and sits at the AdS scale (\textcircled{\footnotesize 4} in Fig. \ref{fig:N2DC}). 

\paragraph{Extra towers: general mechanism.}  
We now sketch the generating mechanism for these towers. Our notation for $\mathcal{N}=2$ superconformal multiplets follows \cite{Dolan:2002zh} and is reviewed in Appendix \ref{App:Dolan-Osborn}. Consider a long superconformal multiplet $\mathcal{A}_{R,r(j,\bar{\jmath})}^{\Delta}$ identified by its conformal dimension $\Delta$, the $SU(2)_R\times U(1)_r$ $R$-charges $(R,r)$ and the spin quantum numbers $(j,\bar{\jmath})$. Let us assume that in the infinite-distance limit $\check{g} \to 0$ this multiplet hits at least one of the two (conjugate) unitarity bounds, \ie, for
\begin{equation}
 \check{g} \to 0: \quad \Delta \to 2+2j+2R +r \ \quad {\rm and/or} \qquad \Delta \to 2+2\bar{\jmath} +2R -r\,. \label{eq: unit bounds}
\end{equation}
In this case, the long multiplet splits into BPS multiplets according to the following recombination rules: If the multiplet hits both unitarity bounds simultaneously, it splits into $\tfrac{1}{4}$-BPS multiplets as \cite{Dolan:2002zh}
\begin{equation}\label{eq: recomb1}
 \mathcal{A}^{2R + j + \bar{\jmath} + 2}_{R,\;\bar{\jmath}-j\,(j,\bar{\jmath})}
\simeq
\hat{\mathcal{C}}_{R(j,\bar{\jmath})}
\;\oplus\;
\hat{\mathcal{C}}_{R+\frac{\scriptscriptstyle 1}{\scriptscriptstyle 2}
\,(j-\frac{\scriptscriptstyle 1}{\scriptscriptstyle 2},\bar{\jmath})}
\;\oplus\;
\hat{\mathcal{C}}_{R+\frac{\scriptscriptstyle 1}{\scriptscriptstyle 2}
\,(j,\bar{\jmath}-\frac{\scriptscriptstyle 1}{\scriptscriptstyle 2})}
\;\oplus\;
\hat{\mathcal{C}}_{R+1\,(j-\frac{\scriptscriptstyle 1}{\scriptscriptstyle 2},
\bar{\jmath}-\frac{\scriptscriptstyle 1}{\scriptscriptstyle 2})}\,;
\end{equation}
if it hits only one unitarity bound (\eg~the first one in (\ref{eq: unit bounds})) it splits into $\tfrac{1}{8}$-BPS multiplets as \cite{Dolan:2002zh}
\begin{equation}\label{eq: recomb2}
 \mathcal{A}^{2R + r + 2j + 2}_{R,r\,(j,\bar{\jmath})}
\simeq
\mathcal{C}_{R,r\,(j,\bar{\jmath})}
\;\oplus\;
\mathcal{C}_{R+\frac{\scriptscriptstyle 1}{\scriptscriptstyle 2},
\,r+\frac{\scriptscriptstyle 1}{\scriptscriptstyle 2}
\,(j-\frac{\scriptscriptstyle 1}{\scriptscriptstyle 2},\bar{\jmath})} \ .
\end{equation}
Given that we have two decoupled theories, an interacting one and a free vector multiplet, the Hilbert space of single-trace operators decomposes, such that the operators constituting the superconformal multiplets fall into one of three separate classes:
\begin{itemize}
 \item \textbf{Class I} corresponds to the operators of the decoupled vector multiplet. In a free field realisation they may be constructed exclusively from fields $(\check{\phi},\check{\lambda},\check{\Fm})$ of the decoupled vector multiplet.
 \item \textbf{Class II} corresponds to the operators of the interacting remainder theory. In a free field realisation at sufficiently weak coupling, they may be constructed exclusively from fields $(\phi_i,\lambda_i,\Fm_i,Q_j,\psi_j)$ of the remainder theory.
 \item \textbf{Class III} corresponds to mixed operators. In a free field realisation at sufficiently weak coupling, they may be constructed from a mixture of fields from both theories.
\end{itemize}
The massless HS tower arising in the infinite-distance limit necessarily falls into Class I and we will show in the following that it will always be accompanied by a tower of Class II, which contributes to the BPS spectrum of the interacting remainder theory. We will also refer to these additional Class II BPS states as ``extra states'' in the following. Operators of Class III decompose into contractions of Class I and Class II operators with open indices, which may be interpreted as open strings in the respective dual theories. Since their correlation functions factorise and the open string coupling is suppressed at large-$N$, we can safely ignore operators of Class III.

Of course, operators from different classes can enter the multiplets in linear combinations, but since the superconformal algebra decomposes into two independent representations acting within the two decoupled theories, we can equally split the superconformal multiplets into their respective decoupled subsets, \eg
\begin{equation}
 \hat{\mathcal{C}}_{R(j,\bar{\jmath})}= \hat{\mathcal{C}}^I_{R(j,\bar{\jmath})}\oplus \hat{\mathcal{C}}^{II}_{R(j,\bar{\jmath})}\oplus \hat{\mathcal{C}}^{III}_{R(j,\bar{\jmath})}\,,
\end{equation}
where the superscript denotes the respective class.

\begin{figure}
 \centering
 \includegraphics[width=\linewidth]{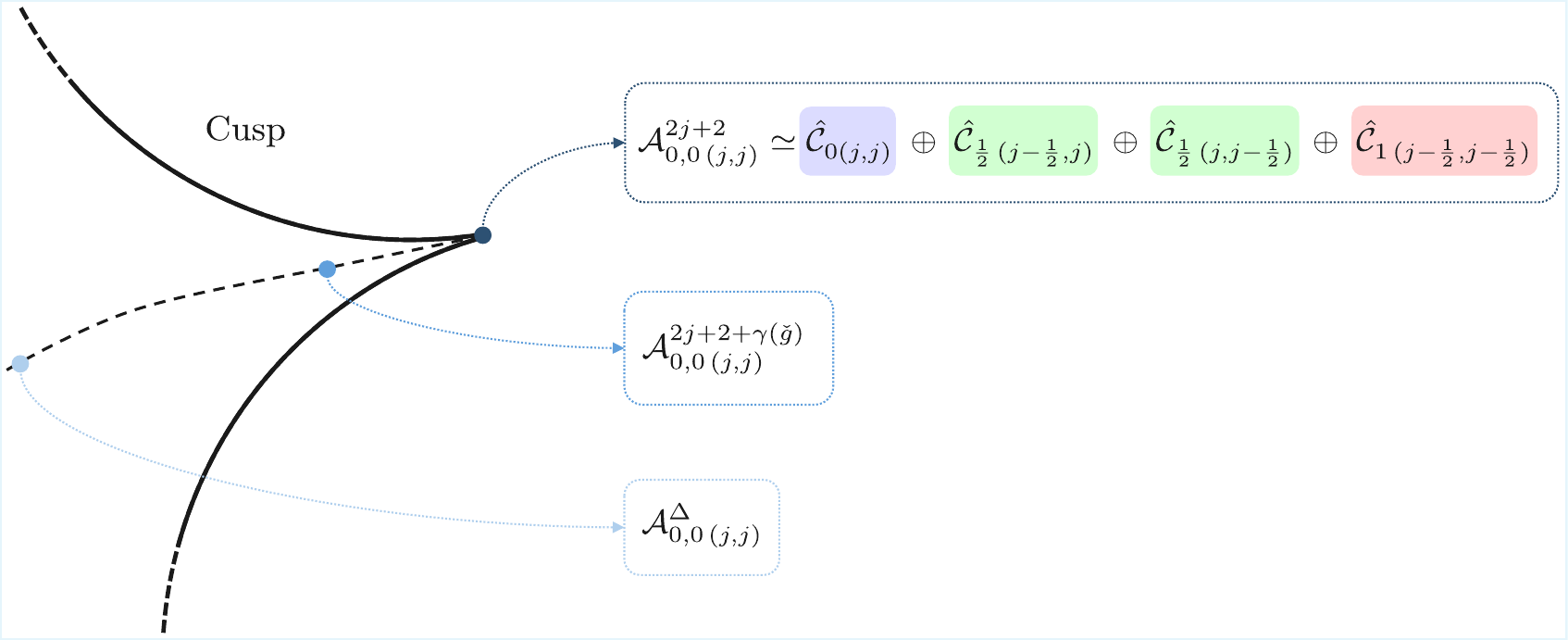}
 \caption{We illustrate the generation of extra BPS states from the long multiplet $\mathcal{A}^{\Delta}_{0,0(j,\bar{\jmath})}$. We start by considering the long multiplet at a finite distance point on the conformal manifold, \ie away from the cusp. When we move towards the cusp, $\check{g} \ll 1$, and the conformal dimension becomes perturbative in the coupling $\check{g}$. The infinite-distance limit coincides with reaching the unitarity bound, and the long multiplet breaks down as shown in equation \eqref{eq: longHS}. The same reasoning applies to other long multiplets reaching the unitarity bound in the same limit.}
 \label{fig:genmech}
\end{figure}

Let us now demonstrate how the extra towers arise in this multiplet recombination. Of particular interest is the long multiplet $\mathcal{A}^{2j+ 2}_{0,0\,(j,j)}$ with $R=0$ and $j=\bar{\jmath}$, which descends from the HS current primary \eqref{eq: HScurrent} associated with the decoupling gauge node. As described above, the anomalous dimension of this operator vanishes in the limit $\check{g} \to 0$, which causes the following recombination: 
 \begin{equation}
 \mathcal{A}^{2j+ 2}_{0,0\,(j,j)}
\simeq
\mathop{\bluebox{$\hat{\mathcal{C}}_{\vphantom{\frac{\scriptscriptstyle 1}{\scriptscriptstyle 2}}0(j,j)}$}}
\limits_{\overset{\rotatebox{90}{$\in$}}{\text{HS currents}}}
\;\oplus\;
\greenbox{$\hat{\mathcal{C}}_{\frac{\scriptscriptstyle 1}{\scriptscriptstyle 2}
\,(j-\frac{\scriptscriptstyle 1}{\scriptscriptstyle 2},j)}$}
\;\oplus\;
\greenbox{$\hat{\mathcal{C}}_{\frac{\scriptscriptstyle 1}{\scriptscriptstyle 2}
\,(j,j-\frac{\scriptscriptstyle 1}{\scriptscriptstyle 2})}$}
\;\oplus\;
\mathop{\redbox{$\hat{\mathcal{C}}_{1\,(j-\frac{\scriptscriptstyle 1}{\scriptscriptstyle 2},
j-\frac{\scriptscriptstyle 1}{\scriptscriptstyle 2})}$}}\limits_{\overset{\rotatebox{90}{$\in$}}{\text{Extra states}}} \ . \label{eq: longHS}
\end{equation}
The recombination is represented in Fig. \ref{fig:genmech} in tandem with the infinite-distance limit.
Since the primary operator \eqref{eq: HScurrent} consists of fields from the decoupled vector multiplet, the whole leading multiplet (highlighted in \bluebox{blue} in the equation \eqref{eq: longHS}) will be of Class I. This is due to the supersymmetry generators $\mathcal{Q}_0$ at $\check{g}=0$ acting strictly within the two decoupled theories and thus not mixing operators of different classes. However, in order to determine the content of the other multiplets we have to consider the long multiplet just before decoupling takes place, \ie at $0<\check{g}\ll 1$, where we may expand the supersymmetry generators as\footnote{This expansion of the supersymmetry generators can be performed more explicitly \cite{Beisert:2004ry,Liendo:2011xb}, but here we only highlight its mixing properties.}
\begin{equation}
 \mathcal{Q}(\check{g})=\mathcal{Q}_0+\check{g}\mathcal{Q}_1+\dots\,,
\end{equation}
with $\mathcal{Q}_1$ mixing the two theories. Since $\mathcal{Q}$ satisfies the Leibniz rule, and since the HS current \eqref{eq: HScurrent} is quadratic in the fields, we see that repeated action with $\mathcal{Q}_1$ generically generates all classes of operators. For example the schematic action on operators involving two gauginos results in the following descendants: 
\begin{equation}
\check{\bar{\lambda}}\check{\lambda}\quad\underset{\mathcal{Q}_1}{\longrightarrow} \quad \check{\bar{\lambda}}\check{\bar{\lambda}}\check{\lambda} + \check{\bar{\lambda}}\bar{Q}Q+\dots \quad\underset{\mathcal{Q}_1}{\longrightarrow} \quad \check{\bar{\lambda}}\check{\bar{\lambda}}\check{\bar{\lambda}}\check{\lambda} + \check{\bar{\lambda}}\check{\bar{\lambda}}\bar{Q}Q+\bar{Q}Q\bar{Q}Q+\dots\,.
\end{equation}
This suggests that the multiplets $\hat{\mathcal{C}}_{\frac{\scriptscriptstyle 1}{\scriptscriptstyle 2}
\,(j-\frac{\scriptscriptstyle 1}{\scriptscriptstyle 2},j)}$ and $\hat{\mathcal{C}}_{\frac{\scriptscriptstyle 1}{\scriptscriptstyle 2}
\,(j,j-\frac{\scriptscriptstyle 1}{\scriptscriptstyle 2})}$ generically consist of components in Class I and III, while Class II operators first appear after two $\Qm_1$-actions in 
$\hat{\mathcal{C}}_{1\,(j-\frac{\scriptscriptstyle 1}{\scriptscriptstyle 2},
j-\frac{\scriptscriptstyle 1}{\scriptscriptstyle 2})}$. These form an ``extra'' BPS multiplet of the interacting theory that only arises in the strict decoupling limit. In conclusion, in theories endowed with a higher dimensional conformal manifold, we expect weak coupling limits, which sit at infinite distance, to produce not only towers of HS currents (along with exponentially degenerate towers from the free sector at the AdS scale), but also extra interacting BPS towers in the remainder theory.

The higher-spin multiplet contains, in particular, a tower with vanishing $R$-charge 
whose effective mass
goes to zero in AdS units, in exactly the same manner as for the unique HS tower in ${\cal N}=4$ SYM (see \textcircled{\footnotesize 1} in Fig. \ref{fig:N2DC} and Fig. \ref{fig:N4DC}). 
 Again, the asymptotically massless sector has only polynomial degeneracy.
The extra BPS towers hitting the unitarity bound, by contrast, have a non-vanishing $R$-charge (see Section \ref{ssec: scqcdhamiltonian}) and their mass sits exactly at the AdS scale (\textcircled{\footnotesize 4} in Fig. \ref{fig:N2DC}). Supersymmetry protects these multiplets from mass renormalisation, so even if the remainder theory is taken to its strongly-coupling regime, the mass of these extra towers stays fixed. As the BPS spectrum is one of the most crucial quantities to match in holography, these towers have to be taken into account when exploring possible holographic dual descriptions of the strongly-coupled remainder theory.

\begin{figure}
 \centering
 \includegraphics[width=0.75\linewidth]{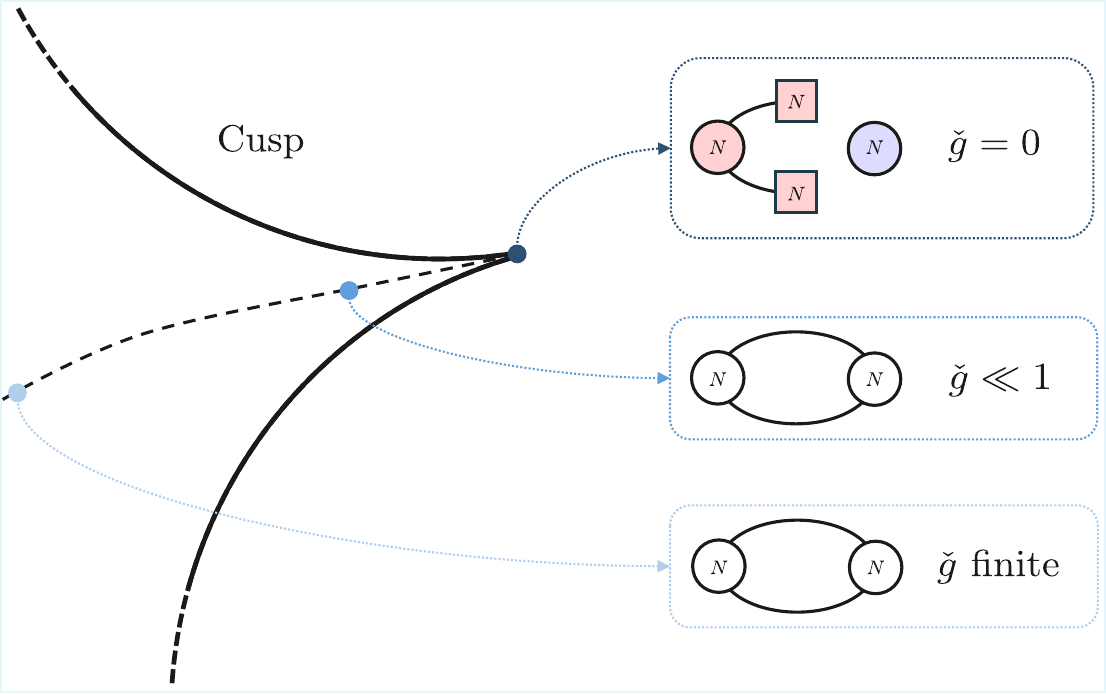}
 \caption{We illustrate the decoupling of the vector multiplet from the two-node quiver theory. In the infinite-distance limit, $\check{g} \to 0$, the decoupled vector multiplet, highlighted in \bluebox{blue}, generates the well-known tower of HS currents; SCQCD, highlighted in \redbox{red}, features the extra towers of BPS states investigated in this paper. }
 \label{fig:sqcdmanifold}
\end{figure}

It is worth noticing that the long multiplet \eqref{eq: longHS} is only one of a higher number of long multiplets hitting the unitarity bound in the limit $\check{g}\to0$. While these do not give rise to extra massless HS towers because of non-vanishing $R$-charge, they will generically contribute with additional protected BPS towers at the AdS scale. We therefore expect the spectrum of all such protected BPS towers at the AdS scale to exhibit exponential degeneracy.
We confirm this behaviour for a simple example in Section \ref{sec: twonode}. 
\paragraph{Two-dimensional conformal manifolds.} 
The easiest example to explicitly illustrate the mechanism outlined above is a theory endowed with a two-dimensional conformal manifold. In Section \ref{sec: twonode}, we will consider marginal deformations of the ${\rm AdS}_5 \times S^5/\mathbb Z_2$ orbifold theory at large $N$ which preserves ${\cal N}=2$ supersymmetry. In this case, the moduli space is two-dimensional and spanned by the gauge couplings $g_1$ and $g_2$. We can then consider two inequivalent types of infinite-distance limits:
 \begin{eqnarray}
 & i) \qquad & g_2 \to 0 \,, \qquad g_1 \to 0 \,. \\
 & ii) \qquad & g_2 \to 0 \,, \qquad g_1 \quad {\rm fixed}\,. \label{inflimit-1}
 \end{eqnarray}
 The first type of limit is the ${\cal N}=2$ version of the infinite-distance limit $g_{\rm YM} \to 0$ of the ${\cal N}=4$ SYM theory, and it shows the same qualitative behaviour (\eg~an HS tower satisfying \eqref{eq:massformula} with $\alpha=\tfrac{1}{\sqrt{2}}$). The key novelty is the second possibility of an infinite-distance limit where $\check{g} \coloneqq g_2 \to 0$, with $g_1$ fixed and finite.\footnote{The central charge of this theory is given by $c\sim\tfrac{N^2}{2}$, which by \eqref{Eq:Zamolodchikov} implies a  decay rate $\alpha=1$, governing the mass of the resulting HS tower \eqref{eq:massformula}\cite{Calderon-Infante:2026rkj}.} From the gauge theory point of view, we witness a gauge node decoupling, which can be visualised in terms of the associated quiver diagrams in Figure \ref{fig:Quivers}, and on the conformal manifold in Figure \ref{fig:sqcdmanifold}. While the decoupled vector multiplet represents a free theory, the finite coupling remainder theory features an $SU(N)$ gauge group and $2N$ hypermultiplets and can be identified with SCQCD; its large-$N$ limit goes under the name of Veneziano limit.
The example of SCQCD is well suited to our purposes, as it is the simplest theory in which one can completely identify and list the additional BPS towers.

\section{An example: The two-node quiver and SCQCD} \label{sec: twonode}

Even though the observations in Section \ref{ssec: recombination} hold for generic supersymmetric theories with a conformal manifold, we would like to present a concrete calculation to demonstrate how this mechanism works out in detail. We therefore restrict ourselves to the simplest possible $\mathcal{N}=2$ example, namely the two-node quiver SCFT whose quiver diagram is depicted in Figure \ref{fig:Quivers}. As already discussed at the end of the previous section, this theory enjoys a two-dimensional conformal manifold spanned by the couplings $g_1$ and $g_2$ and allows us to take the infinite distance limit 
\eqref{inflimit-1}, thereby providing a clean setting in which we see the recombination rules in action.

In this section we review a few basic facts about this theory, familiar from \cite{Gadde:2009dj, Gadde:2010zi}, which are indispensable for our arguments.

\paragraph{The interpolating theory.}
We consider a superconformal $\mathcal{N}=2$ quiver gauge theory with two exactly marginal gauge couplings $g_1$ and $g_2$, gauge groups $SU(N_c)_1$ and $SU(N_c)_2$, and $N_f = 2N_c$ hypermultiplets. The $\mathbb{Z}_2$ orbifold of $\mathcal{N}=4$ SYM can then be identified with 
the 
\begin{equation}
{\rm orbifold \, \, locus} \qquad g_1 = g_2
\end{equation}
on the two-dimensional conformal manifold parametrised by the couplings $g_1$ and $g_2$. 
 Another special locus is provided by the decoupled regime $g_2=0$ (or equivalently $g_1=0$), in which one of the gauge nodes completely decouples from the remaining fields, which in turn constitute $\mathcal{N}=2$ SCQCD. By varying $g_2$ from the orbifold point $g_2 = g_1$ down to the decoupling limit $g_2 = 0$, while keeping $g_1$ fixed, we obtain a continuous family of theories \emph{interpolating} between the orbifold theory and $\mathcal{N}=2$ SCQCD, supplemented by an additional decoupled vector multiplet (see Figure \ref{fig:Quivers}). 

At the orbifold point, the gauge group is $SU(N_c)_1 \times SU(N_c)_2 \times U(1)$. The presence of the additional $U(1)$ factor breaks conformal invariance and therefore it must be removed by hand \cite{Dymarsky_2005}. The $SU(4)_R$ $R$-symmetry of the parent $\mathcal{N}=4$ SYM theory is broken by the orbifold projection to the $\mathcal{N}=2$ superconformal $R$-symmetry group $SU(2)_R \times U(1)_r$, together with an extra global $SU(2)_L$ symmetry under which the preserved supercharges are neutral.\footnote{Since $SU(2)_L$ acts only on integer-spin representations, it may equivalently be viewed as an $SO(3)$ symmetry. }

The field content of the theory consists of two vector multiplets, $\Vm$ and $\check{\Vm}$, transforming in the adjoint representations of the gauge groups $SU(N_c)_1$ and $SU(N_c)_2$, respectively, as well as two bifundamental hypermultiplets, $\Hm_{12}$ and $\Hm_{21}$.\footnote{To write the harmonic action we will adopt a slightly different splitting of the multiplets; we refer the reader to Appendix \ref{app: Harmonic} for a detailed discussion of this choice.} The vector multiplet $\Vm$ contains the complex spacetime scalar $\phi$, the gaugino $\lambda_{\alpha}^{\mathcal{I}}$, and the gauge field $A_{\alpha\dot{\alpha}}$, while the second gauge node $\check{\Vm}$ contains the analogous fields $\check{\phi}$, $\check{\lambda}_{\alpha}^{\mathcal{I}}$, and $\check{A}_{\alpha\dot{\alpha}}$. Here $\mathcal{I} = 1,2$ denotes the $SU(2)_R$ index. 

In addition to $SU(2)_R$ charge, the two hypermultiplets are labeled by an $SU(2)_L$ index $\hat{\mathcal{I}} = \hat{1}, \hat{2}$ and consist of the complex spacetime scalars $Q_{\mathcal{I}\hat{\mathcal{I}}}$ together with two Weyl fermions $\psi_{\alpha\hat{\mathcal{I}}}$ and $\tilde{\psi}_{\alpha\hat{\mathcal{I}}}$. Here the spacetime spinor indices $\alpha$ and $\dot{\alpha}$ take values $\pm$ and $\dot{\pm}$, respectively. Gauge indices have been suppressed throughout; for completeness, they range over $a, \check{a} = 1, \ldots, N_c$. The global symmetries and the quantum numbers of all fields in the interpolating theory are summarised in Table~\ref{tab:Interpolating}.

\begin{table}[]
 \centering
 \begin{tabular}{c||c|c|c|c|c}
 & $SU(N_{c})_{1}$ & $SU(N_{c})_{2}$ & $SU(2)_{R}$ & $SU(2)_{L}$ & $U(1)_{r}$\tabularnewline
\hline
\hline 
$A_{\alpha\dot\alpha}$ & Adj & \textbf{${\bf 1}$} & \textbf{${\bf 1}$} & \textbf{${\bf 1}$} & 0\tabularnewline
$\check{A}_{\alpha\dot\alpha}$ & \textbf{${\bf 1}$} & Adj & \textbf{${\bf 1}$} & \textbf{${\bf 1}$} & 0\tabularnewline
$\f$ & Adj & \textbf{${\bf 1}$} & \textbf{${\bf 1}$} & \textbf{${\bf 1}$} & --1\tabularnewline
$\check{\f}$ & \textbf{${\bf 1}$} & Adj & \textbf{${\bf 1}$} & \textbf{${\bf 1}$} & --1\tabularnewline 
$\lambda^{\mathcal{I}}$ & Adj & \textbf{${\bf 1}$} & \textbf{${\bf 2}$} & \textbf{${\bf 1}$} & --1/2\tabularnewline
$\check{\lambda}^{\mathcal{I}}$ & \textbf{${\bf 1}$} & Adj & \textbf{${\bf 2}$} & \textbf{${\bf 1}$} & --1/2\tabularnewline
$Q_{\mathcal{I}\hat{\mathcal{I}}}$ & $\Box$ & $\overline{\Box}$ & \textbf{${\bf 2}$} & \textbf{${\bf 2}$} & 0\tabularnewline
$\psi_{\hat{\mathcal{I}}}$ & $\Box$ & $\overline{\Box}$ & \textbf{${\bf 1}$} & \textbf{${\bf 2}$} & +1/2\tabularnewline
$\tilde{\psi}_{\hat{\mathcal{I}}}$ & $\overline{\Box}$ & $\Box$ & \textbf{${\bf 1}$} & \textbf{${\bf 2}$} & +1/2
\end{tabular}
\caption{Symmetries and representations of the $\mathbb{Z}_2$ orbifold of ${\cal N} = 4$ SYM and of the interpolating family of ${\cal N} = 2$ SCFTs. We show the quantum numbers of the elementary component fields. Conjugate objects (such as $\bar{\phi}$) are not written explicitly. 
\label{tab:Interpolating}}
\end{table}

\paragraph{${\cal N}=2$ SCQCD.} 
We obtain $\mathcal{N}=2$ SCQCD by moving to the infinite-distance locus (\ref{inflimit-1}) on the conformal manifold of the interpolating theory, namely by decoupling the second vector multiplet $(\check{\phi}, \check{\lambda}^{\mathcal{I}}, \check{A}_{\alpha\dot{\alpha}})$ by setting $g_2 = 0$. In this limit, the global $SU(2)_L$ symmetry combines with the decoupled $SU(N_c)_2$ gauge symmetry and enhances to a $U(N_f = 2N_c)$ flavour symmetry. The resulting theory is therefore an $\mathcal{N}=2$ SYM theory with gauge group $SU(N_c)$ and $N_f = 2N_c$ fundamental hypermultiplets, which lies in the conformal window.

The full global symmetry group is thus $U(N_f) \times SU(2)_R \times U(1)_r$, where $SU(2)_R \times U(1)_r$ is the superconformal $R$-symmetry. Since the original $SU(2)_L$ global symmetry has now enhanced to a flavour symmetry, it is convenient to reorganise the indices in a more uniform manner. In particular, the $SU(2)_L$ index $\hat{I} = \hat{1}, \hat{2}$ and the gauge index $\check{a} = 1, \ldots, N_c$ combine into a single flavour index $i = 1, \ldots, 2N_c$. In what follows, we will suppress both gauge and flavour indices, reinstating them only when an explicit notation is required. The global symmetries and the quantum numbers of the fields of $\mathcal{N}=2$ SCQCD are summarised in Table~\ref{Tab:SCQCD}.

\begin{table}
\begin{centering}
\begin{tabular}{c||c|c|c|c}

 & $SU(N_{c})$ & $U(N_{f})$ & $SU(2)_{R}$ & $U(1)_{r}$\tabularnewline
\hline
\hline 
$A_{\alpha\dot\alpha}$ & Adj & \textbf{$\mathbf{1}$} & \textbf{$\mathbf{1}$} & $0$\tabularnewline
$\f$ & Adj & \textbf{$\mathbf{1}$} & \textbf{$\mathbf{1}$} & $-1$\tabularnewline
$\lambda_{\alpha}^{\mathcal{I}}$ & Adj & \textbf{$\mathbf{1}$} & \textbf{$\mathbf{2}$} & $-1/2$\tabularnewline
$Q_{\mathcal{I}}$ & $\Box$ & $\Box$ & \textbf{$\mathbf{2}$} & $0$\tabularnewline
$\psi_{\alpha}$ & $\Box$ & $\Box$ & \textbf{$\mathbf{1}$} & $+1/2$\tabularnewline
$\tilde{\psi}_{\alpha}$ & $\overline{\Box}$ & $\overline{\Box}$ & \textbf{$\mathbf{1}$} & $+1/2$\\
\end{tabular}
\par\end{centering}
\caption{\label{Tab:SCQCD} Symmetries and representations of $\mathcal{N}=2$ SCQCD. We show the quantum numbers of the elementary components fields. Conjugate objects (such as $\bar{\phi}$) are not written explicitly.}
\end{table}

\subsection{Inherited BPS spectrum}
In this subsection, we present the BPS spectrum of the interpolating theory at the orbifold point, as well as the \textit{inherited} SCQCD spectrum from the orbifold construction. These states correspond to tower \textcircled{\footnotesize 2} in Figure \ref{fig:N2DC}. The complete spectrum at the orbifold point is obtained by considering the untwisted states inherited from $\Nm =4$ SYM together with the twisted operators, exhausted by the generators of the chiral ring of the theory. We refer the reader to Appendix B of \cite{Gadde:2009dj} for a detailed analysis of this object for both the interpolating theory and for SCQCD. In principle, these multiplets could recombine into long ones. However, the recombination rules in which they participate involve short multiplets with different $SU(2)_L$ quantum numbers. Since the supercharges are neutral under this residual symmetry, such short multiplets are forbidden from recombining and therefore remain short for all values of the coupling.
 The evolution of these states away from the orbifold point toward the SCQCD decoupling limit can be followed through a one-loop calculation \cite{Gadde:2010zi},
 and gives the complete spectrum of the interpolating theory until one reaches the SCQCD point. At this point, the spectrum inherited from the orbifold theory in this manner will turn out to be incomplete and is therefore referred to as the inherited part of the spectrum at the SCQCD point.

\paragraph{Orbifold point.}

As already mentioned, at the orbifold point we split the $\mathcal{N}=2$ spectrum into twisted and untwisted sectors. The untwisted sector is obtained directly by decomposing the $\mathcal{N}=4$ multiplets into $\mathcal{N}=2$ multiplets. The protected single-trace operators of $\mathcal{N}=4$ SYM consist only of the half-BPS multiplets $\Bm_{[0,p,0]}^{\frac{1}{2},\frac{1}{2}}$,\footnote{Here $\Bm_{[0,p,0]}^{\frac{1}{2}\frac{1}{2}}$ indicates that the highest weight of the multiplet is annihilated by half of the $\Qm$ and $\bar \Qm$ supercharges and sits in the $p$-index symmetric representation $[0,p,0]$ of $SU(4)_R$. The multiplet notation used here and in the rest of the paper follows the conventions of \cite{Dolan:2002zh} and is reviewed in Appendix \ref{App:Dolan-Osborn}.} which on the gravity side map to the KK-reduction of type IIB supergravity on ${\rm AdS}_5 \times S^5$ \cite{Kim:1985ez, Kinney:2005ej}. To obtain the untwisted spectrum in the $\mathcal{N}=2$ theory, we simply need to project out states that are odd under the $\mathbb{Z}_2$ symmetry. On the gauge theory side, this can be achieved by decomposing the $\mathcal{N}=4$ half-BPS multiplets into $\mathcal{N}=2$ multiplets as \cite{Dolan:2002zh}
\begin{eqnarray}
\Bm_{[0,p,0]}^{\frac{1}{2},\frac{1}{2}} & \simeq & (p+1)\hat{\Bm}_{\frac{1}{2}p}\oplus\Em_{p(0,0)}\oplus\bar{\Em}_{-p(0,0)}\nonumber \\
 & & \oplus(p-1)\hat{\Cm}_{\frac{1}{2}p-1(0,0)}\oplus p(\mathcal{D}_{\frac{1}{2}(p-1)(0,0)}\oplus\bar{\mathcal{D}}_{\frac{1}{2}(p-1)(0,0)})\nonumber \\
 & & \oplus\bigoplus_{k=1}^{p-2}(k+1)(\Bm_{\frac{1}{2}k,p-k(0,0)}\oplus\bar{\Bm}_{\frac{1}{2}k,k-p(0,0)})\nonumber \\
 & & \oplus\bigoplus_{k=0}^{p-3}(k+1)(\Cm_{\frac{1}{2}k,p-k-2(0,0)}\oplus\bar{\Cm}_{\frac{1}{2}k,k-p+2(0,0)})\nonumber \\
 & & \oplus\bigoplus_{k=0}^{p-4}\bigoplus_{l=0}^{p-k-4}(k+1)\Am_{\frac{1}{2}k,p-k-4-2l(0,0)}^{p}\label{recombination} \, .
\end{eqnarray}
Retaining orbifold invariant operators boils down to keeping only the states with integer $SU(2)_R$ quantum numbers (see Sections 3.2 and 4.1.1 of \cite{Gadde:2009dj} for a detailed analysis). We thus obtain the following  spectrum of untwisted states at the orbifold point: 
\begin{equation}
\begin{split}
&\hat{\Bm}_{R+1}\,, \quad 
\bar \Em_{-(\ell+2)(0,0)}\,, \quad 
\hat{\Cm}_{R(0,0)}\,, \quad 
\bar{\mathcal{D}}_{R+1(0,0)}\,, \quad 
\bar \Bm_{R+1,-(\ell+2)(0,0)}\,, \\
& \bar \Cm_{R,-(\ell+1)(0,0)}\,, \quad 
\Am_{R,-\ell(0,0)}^{\Delta=2R+\ell+2n}\,, \quad R\,,\ell \geq 0\,, \quad n\geq 2 \,,
\end{split}
\label{Eq:Orbifold_Untwisted}
\end{equation}
plus the conjugate multiplets. The long multiplets $\Am$ appearing in this decomposition are not strictly BPS states in the sense of shortening but their conformal dimension is nevertheless protected. The completeness of this spectrum is then ensured by an index analysis, as confirmed in the next subsection. By viewing $S^5 / \mathbb{Z}_2$ as an $S^1 \times S^3 / \mathbb{Z}_2$ fibration over an interval $I$, with degenerate fibers at the boundaries \cite{Aharony:1998xz}, we can match the multiplets above with the KK-reduction of type IIB supergravity on ${\rm AdS}_5 \times (S^1 \times S^3/ \mathbb{Z}_2) \ltimes I$, as illustrated in Section 6 of \cite{Gadde:2009dj}. In particular, the momentum along $S^1$ corresponds to the $U(1)_r$ charge, the $SU(2)_R$ subgroup of the $SO(3)_L \times SU(2)_R$ isometry group of $S^3 / \mathbb{Z}_2$ is interpreted as the $SU(2)_R$ $R$-charge, and the integer $n$ labelling the harmonics on the interval is dual to the power of the $R$-charge neutral combination of scalar fields, which in \cite{Gadde:2009dj} was denoted by $\Tm$. An explicit KK-reduction was carried out in \cite{Aharony:1998xz}, where it was found that the KK-modes have scaling dimension $\Delta = |r| + 2R + 2n$, in perfect agreement with the list \eqref{Eq:Orbifold_Untwisted}. More precisely the multiplets have the following interpretation:
\begin{itemize}
 \item The $\bar \Em_{-(\ell+2)(0,0)}$ multiplets, carrying arbitrary $U(1)_r$ charge and all other quantum numbers vanishing, correspond to KK-modes with increasing momentum along $S^1$;
 \item the $\hat{\Bm}_{R+1}$ multiplets, carrying arbitrary $SU(2)_R$ charge and all other quantum numbers vanishing, correspond to KK-modes with increasing angular momentum on $S^3$;
 \item the $\Am_{0,0(0,0)}^{\Delta=2n}$ multiplets, carrying a conformal dimension scaling with $n$ and all other quantum numbers vanishing, correspond to the KK-tower of harmonics on the interval $I$;
 \item all other multiplets arise from mixed KK-towers carrying momentum along $S^1$, angular momentum on $S^3$, and arbitrary $n$.
\end{itemize}

The twisted sector is obtained by considering the generators of the chiral ring of the orbifold theory, as performed in Appendix B of \cite{Gadde:2009dj}. This leads to the following states:
\begin{equation}
 \hat{\Bm}_1\,, \quad \bar \Em_{-(\ell+2)(0,0)}\,, \quad \ell \geq 0\,,
 \label{Eq:Orbifold_Twisted}
\end{equation}
where each multiplet is the twisted copy of the corresponding one in the untwisted sector. Again, the completeness of this list is confirmed by the index analysis. At the level of the holographic mapping, this twisted sector should map on the dual side to twisted closed strings localised at the fixed locus of the orbifold, namely ${\rm AdS}_5 \times S^1$. The massless twisted states of type IIB supergravity on this singularity fit into a massless six-dimensional tensor multiplet. As analysed in \cite{Gukov:1998kk} and in Appendix D of \cite{Gadde:2009dj}, the KK-reduction of these states on ${\rm AdS}_5 \times S^5$ reproduces precisely the list in \eqref{Eq:Orbifold_Twisted}.

\paragraph{SCQCD.}

We are now ready to present the spectrum for SCQCD. The spectrum at the orbifold point splits into $SU(2)_L$ singlets and non-singlets. In the Veneziano limit (in which one sends $N_f, N_c \rightarrow \infty$, with the ratio fixed $\tfrac{N_f}{N_c}=2$), and restricting to flavour singlets, all single-trace protected operators of SCQCD arise from the $SU(2)_L$ singlet subset, since the $SU(2)_L$ non-singlet states would be multi-trace operators dual to open strings, \ie, Class III operators as discussed below \eqref{eq: recomb2}. By tracking the evolution of these singlets through a one-loop calculation \cite{Gadde:2010zi}, one finds that the Class II part of the BPS spectrum \emph{inherited} from the orbifold consists of the multiplets 
\begin{equation}
 \hat{\Bm}_1\,, \quad \bar \Em_{-(\ell+2)(0,0)}\,, \quad \hat{\Cm}_{0(0,0)}\,, \quad 
 \bar \Cm_{0,-(\ell+1)(0,0)}\,, \quad \ell \geq 0\,,
 \label{Eq:Naive_SCQCD_Spectrum}
\end{equation}
built from fields living in the interacting SCQCD sector. Already this inherited spectrum highlights an interesting feature of these states: they do not contain multiplets with increasing $SU(2)_R$ charge. This fact, together with their being singlets under $SU(2)_L$, implies that geometrically they do not descend from states carrying angular momentum on $S^3/\mathbb{Z}_2$, signaling the ``loss'' of three spatial dimensions in the dual description. More precisely, we can draw the following conclusions:
\begin{itemize}
 \item The $\bar \Em_{-(\ell+2)(0,0)}$ multiplets, carrying arbitrary $U(1)_r$ charge and all other quantum numbers vanishing, correspond to KK-modes with increasing momentum along $S^1$.
 \item The single appearance of one $\hat{\Bm}_{1}$ multiplet, carrying finite $SU(2)_R$ charge and with all other quantum numbers vanishing, hints at the absence of a large $S^3$ direction in the gravity dual.
 \item The $\hat{\Cm}_{0(0,0)}$ multiplet is the stress-energy tensor multiplet, mapped to the dual graviton multiplet, and the $\Cm_{0,r(0,0)}$ multiplets arise from the KK-reduction of these states carrying momentum along $S^1$.
\end{itemize}
As explained in \cite{Gadde:2009dj}, and reviewed in the next section, this inherited spectrum is not complete, and a thorough analysis via the superconformal index is necessary. In fact we will find additional BPS states with non-vanishing $SU(2)_R$-charges, which however do not derive from geometric KK modes but instead from string excitations. This has a clear interpretation in the proposed non-critical string dual \cite{Gadde:2009dj}. We will comment on this proposition in the conclusion.

\subsection{The index and extra states}
To check the BPS spectra of the $\mathbb{Z}_2$ orbifold of $\mathcal{N}=4$ SYM and the class II part contributing to the SCQCD spectrum, we employ the superconformal index. The strategy is the following: the index can be computed autonomously for the two theories and be compared with the index built using the inherited BPS spectrum.

We will observe a mismatch for SCQCD, which highlights the appearance of additional towers of protected multiplets of extra states (labelled \textcircled{\footnotesize 4} in Figure \ref{fig:N2DC})  in the limit $g_2\to0$. The precise identification of such states will be presented in Section \ref{sec: identify}, employing the one-loop dilatation operator of the interpolating quiver theory. 

For our purposes, the (superconformal) index has two inequivalent definitions, labelled as $\mathcal{I}^{L}$ (left index) and $\mathcal{I}^{R}$ (right index):
\begin{align}
 \mathcal{I}^{L}(t,y,v)&=\text{Tr} (-1)^{F} t^{2(\Delta+j)}y^{2 \bar{\jmath}}v^{r-R} \ , \\
 \mathcal{I}^{R}(t,y,v)&=\text{Tr} (-1)^{F} t^{2(\Delta+\bar{\jmath})}y^{2 j}v^{-r-R} \ ,
 \label{Eq:Indices}
\end{align}
where the fugacities $t,y,v$ were introduced; the left index should be regarded as the Witten index with respect to the supercharge $\mathcal{Q}^{1}_{-}$, counting states with $\delta^{L}=0$ , while the right one reflects $\Bar{\mathcal{Q}}_{2,+}$, counting states with $\delta^R=0$, with
\begin{equation}
 \begin{split}
 & \delta^L = \Delta -2 j - 2R -r\,, \\
 & \delta^R = \Delta - 2 \bar{\jmath} - 2 R +r\,.
 \end{split}
\end{equation}
In the following, we will mainly focus on the left index, identifying \begin{equation}
\mathcal{I}\coloneqq \mathcal{I}^{L} \,.
\end{equation}
\paragraph{The index of the interpolating theory.} 
The procedure to explicitly compute the index for an $\mathcal{N}=2$ quiver theory is well established. Since the index is invariant under marginal deformations, it is sufficient to compute it at the orbifold point, where the theory descends from $\mathcal{N}=4$ SYM; moreover, it is sufficient to compute it in the free theory limit. The result is:
\begin{multline}
 \mathcal{I}_{\text{orb.}}
= 2\!\left[
\frac{t^{2} v}{1 - t^{2} v}
- \frac{t^{3} y}{1 - t^{3} y}
- \frac{t^{3} y^{-1}}{1 - t^{3} y^{-1}}
\right]
+ \frac{t^{4} w^{2}/v}{1 - \frac{t^{4} w^{2}}{v}}
+ \frac{t^{4}/(v w^{2})}{1 - \frac{t^{4}}{v w^{2}}}
- 2 f_V(t,y,v) \ , \label{eq: indexorb}
\end{multline}
where we added the fugacity $w$ to the definition \eqref{Eq:Indices} to keep track of the $SU(2)_L$ quantum numbers and we defined the single letter index of the vector and hyper multiplet as
\begin{equation}
 \begin{split}
 f_{V}(t,y,v)=&\frac{t^{2}v-t^{3}\left(y+y^{-1}\right)-t^{4}v^{-1}+2t^{6}}{\left(1-t^{3}y\right)\left(1-t^{3}y^{-1}\right)}, \\
f_{H}(t,y,v)=&\frac{t^{2}}{v^{1/2}}\frac{(1-t^{2}v)}{\left(1-t^{3}y\right)\left(1-t^{3}y^{-1}\right)} \, .
 \end{split}
\end{equation}
A standard calculation \cite{Kinney:2005ej} gives the result for the single-trace index of $\Nm=4$ SYM:
\begin{equation}
\begin{split}
\Im_{{\cal N} = 4} =& \frac{t^2v}{1-t^2v}+\frac{\frac{t^2w}{\sqrt v}}{1-\frac{t^2w}{\sqrt v}}+\frac{\frac{t^2}{w\sqrt v}}{1-\frac{t^2}{w\sqrt v}}-\frac{t^{3}y}{1-t^{3}y}-\frac{t^{3}y^{-1}}{1-t^{3}y^{-1}} +\\
&-f_V(t,y,v)-(w+\frac{1}{w})f_H(t,y,v)\,.
\end{split}
\label{Eq:N4index}
\end{equation}
The orbifold acts on the index through $w \to -w$, and allows us to identify the index counting untwisted and twisted BPS multiplets. Extracting the untwisted index from the part of \eqref{Eq:N4index} invariant under the action of the orbifold and subtracting it from the total index given by \eqref{eq: indexorb}, we obtain 
\begin{align}
 \mathcal{I}_{\text{orb.}}^{\text{untwisted}}
&= \frac{t^{2} v}{1 - t^{2} v}
- \frac{t^{3} y}{1 - t^{3} y}
- \frac{t^{3} y^{-1}}{1 - t^{3} y^{-1}}
+ \frac{t^{4} w^{2} v^{-1}}{1 - \frac{t^{4} w^{2}}{v}}
+ \frac{t^{4} v^{-1} w^{-2}}{1 - \frac{t^{4}}{v w^{2}}}
- f_V(t,y,v) \ , \label{eq:untwist} \\ 
\mathcal{I}_{\text{orb.}}^{\text{twisted}}
&= \frac{t^{2} v}{1 - t^{2} v}
- \frac{t^{3} y}{1 - t^{3} y}
- \frac{t^{3} y^{-1}}{1 - t^{3} y^{-1}}
- f_V(t,y,v) \label{eq:twist}\ .
\end{align}
The two indices \eqref{eq:untwist} and \eqref{eq:twist} exactly reproduce the inherited BPS spectrum listed in \eqref{Eq:Orbifold_Untwisted} and \eqref{Eq:Orbifold_Twisted}.
\paragraph{The index of SCQCD.} As shown explicitly in \cite{Gadde:2009dj}, the index counting the number of single-trace BPS multiplets in SCQCD is given by
\begin{equation}
 \mathcal{I}_{\text{SCQCD}}
= - \sum_{n=1}^{\infty} \frac{\varphi(n)}{n}
\log\!\Big[(1 - f_V(t^n, y^n, v^n)) - f_H^{\,2}(t^n, y^n, v^n)\Big]
- f_V(t, y, v) \ , \label{eq: indexqcd}
\end{equation}
where the sum cannot be explicitly computed, in contrast to the interpolating theory index. As discussed previously, the $SU(2)_L$ symmetry of the interpolating theory forbids the recombination of short multiplets, so the naive expectation would have been for the SCQCD index to enumerate the $SU(2)_L$ singlets listed in \eqref{Eq:Naive_SCQCD_Spectrum}. Hence, the expected inherited index would have been 
\begin{equation}
\begin{split}
 & \mathcal{I}^{\text{inh.}}_{\text{SCQCD}}
= \\
&\frac{1}{(1 - t^{3} y)\left(1 - \frac{t^{3}}{y}\right)}
\Bigg[
- t^{6}\!\left(1 - \frac{t}{v}\left(y + \frac{1}{y}\right)\right)
- \frac{t^{10}}{v}
+ \frac{t^{4} v^{2}\!\left(1 - \frac{t}{v y}\right)\!\left(1 - \frac{t y}{v}\right)}{1 - t^{2} v}
+ \frac{t^{4}}{v}(1 - t^{2} v)
\Bigg] \ . 
\end{split}
\label{eq: naiveindex}
\end{equation}
The discrepancy between the inherited index \eqref{eq: naiveindex} and the SCQCD index \eqref{eq: indexqcd}, 
\begin{equation} \label{eq:DeltaIdef}
 \Delta {\cal I} = {\cal I}_{\rm SCQCD} - {\cal I}^{\rm inh.}_{\rm SCQCD} \,,
 \end{equation}
has a physical explanation, spelled out in \cite{Gadde:2009dj}. If we keep the coupling $g_1$ fixed, but we send $g_2 \to 0$, we are moving away from the orbifold point on the conformal manifold of the interpolating theory and we are reaching one of the two ``cusps'', corresponding to SCQCD plus an additional, free vector multiplet (see Figure \ref{fig:sqcdmanifold}). As explained in Section \ref{Sec:CFT_Distance}, an infinite number of long multiplets reach their unitarity bound and split into new BPS multiplets according to their recombination rules \eqref{eq: recomb1}, \eqref{eq: recomb2}. This splitting generates ``extra'' BPS states (\textcircled{\footnotesize 4} in fig.~\ref{fig:N2DC}) in the SCQCD theory, which are counted by the the difference $\Delta {\cal I}$.

\begin{figure}
 \centering
 \includegraphics[width=0.75\linewidth]{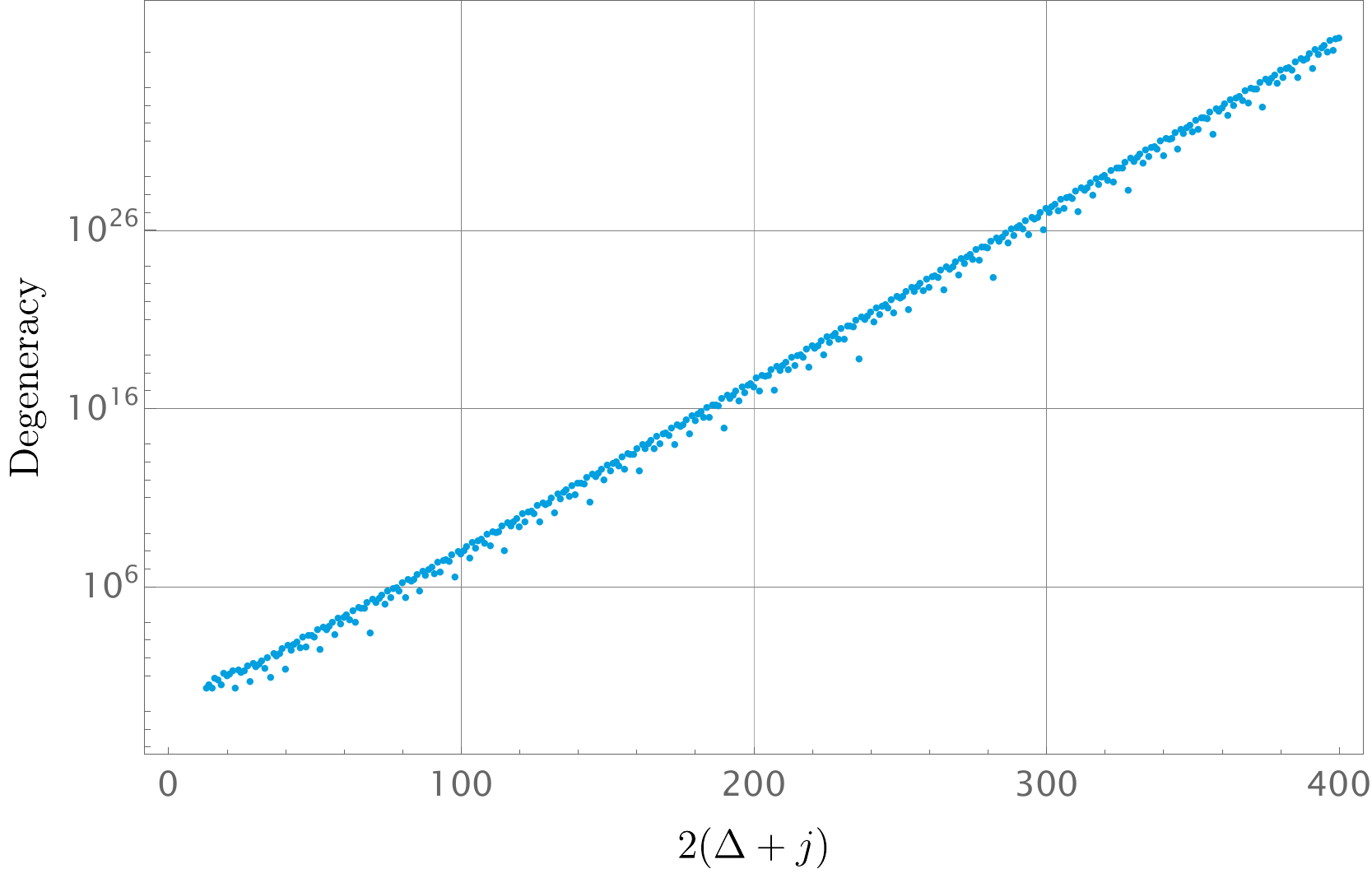}
 \caption{This figure shows the coefficient of $t^{2(\Delta+j)}$
 in $\Delta {\cal I}$ defined in (\ref{eq:DeltaIdef}), corresponding to the degeneracy of the extra states with a given value of the charge associated with $t$. The scale on the vertical axis is logarithmic, and the growth is therefore exponential.}
 \label{fig:indexfull}
\end{figure}

\begin{figure}
 \centering
 \includegraphics[width=0.75\linewidth]{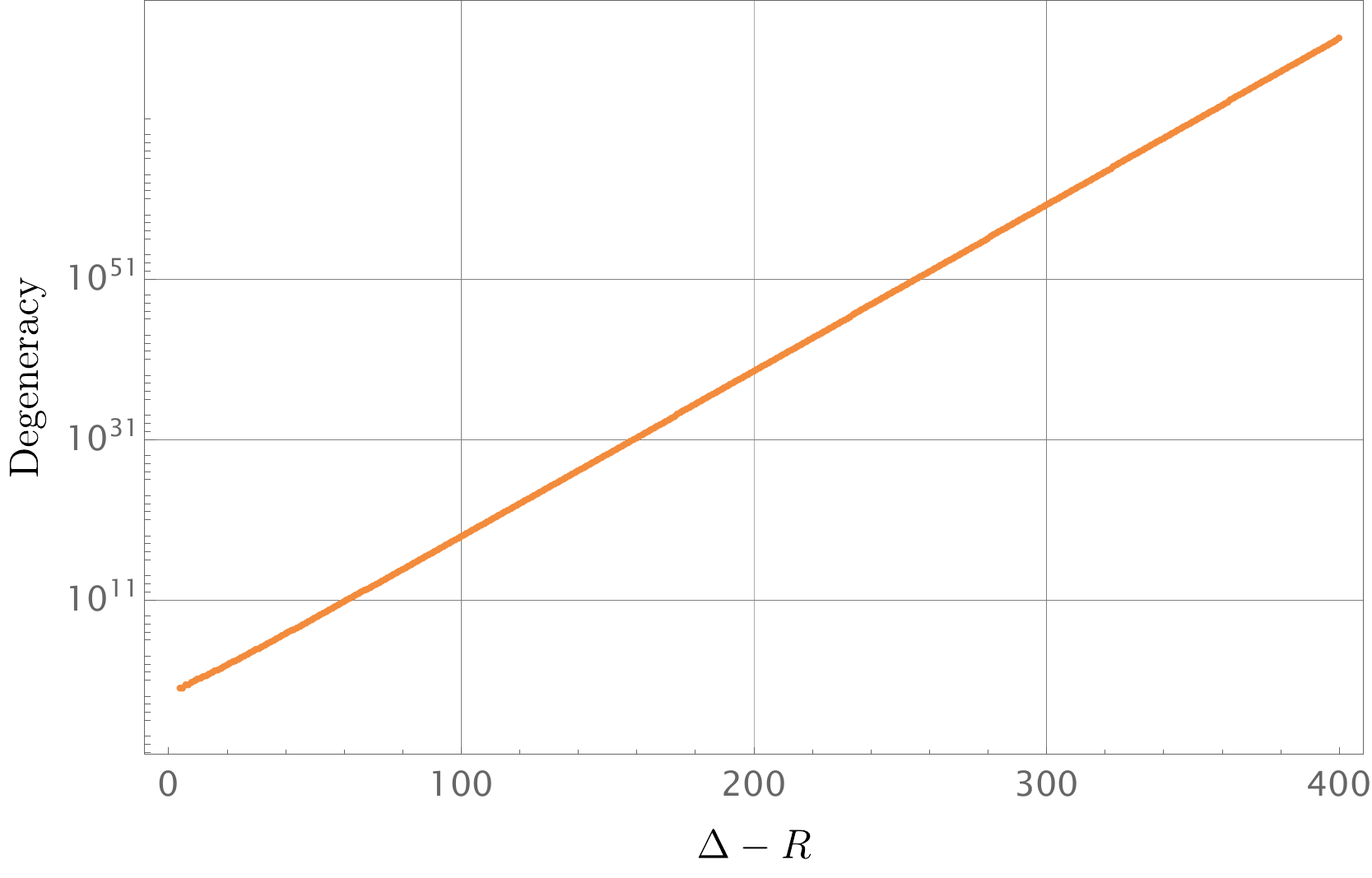}
 \caption{In this figure, we plot the coefficient of $q^{\Delta-R}$ when taking the difference between the Schur limits of $\mathcal{I}_{\text{SCQCD}}$ and $\mathcal{I}^{\text{inh.}}_{\text{SCQCD}}$. This choice makes sure that only the contributions of extra $\Hat{\mathcal{C}}$ multiplets are enumerated. Also in this case, the growth of the number of extra multiplets is exponential in the chosen fugacity.}
 \label{fig:indexschur}
\end{figure}

To establish a physical interpretation of this phenomenon, seen through the lens of the AdS/CFT correspondence (hence, from the point of view of the CFT Distance Conjecture), it is important to verify that the density of BPS states contributing to $\Delta {\cal I}$ grows exponentially, \ie, that it displays the Hagedorn behaviour \eqref{mass-deg-string-flat} of an emergent string limit. For this purpose, we plot the degeneracy of such states in Figure \ref{fig:indexfull}, clearly exhibiting exponential growth. Note in particular that this exponential growth is not due to the growing quantum numbers of the multiplets (\eg~$R,j,\bar{\jmath}$ in $\CH{R}{j}{\bar{\jmath}}$), which come with polynomially growing representation size, but instead due to the exponential growth of the number of multiplets themselves. An even sharper exponential growth can be identified by considering the Schur limit of the index, \ie, reducing the fugacities according to $y=\frac{v}{t}$: In this case, the index turns out to depend just on one fugacity $q = \frac{t^4}{v}$. The Schur index enumerates only the number of extra $\tfrac{1}{4}$-BPS $\hat{\mathcal{C}}$ multiplets (see Table \ref{Tab:shortening} for the list of shortening conditions), and their degeneracies are plotted in Figure \ref{fig:indexschur}. This matches our interpretation of the extra BPS towers as resulting from an emergent string limit.


\subsection{Identifying the extra states}
\label{sec: identify}

Having observed the appearance of additional towers of protected multiplets in SCQCD, one would like to identify the precise particle (or rather multiplet) content of the extra states. As explained in Section \ref{ssec: recombination} we expect these extra states to arise from the recombination of long multiplets, but in this section we shall stay oblivious to this reasoning and instead consider the concrete evidence at our disposal.

The first approach which was outlined in \cite{Gadde:2009dj} is to assess the index mismatch $\Delta \mathcal{I}$ order by order in $t$. Since we can compute the index for every possible short multiplet, we can attempt to decompose $\Delta \mathcal{I}$ into contributions from candidate multiplets. Unfortunately, this procedure, dubbed ``sieve algorithm'', does not uniquely fix the multiplets but provides us with a finite number of candidates. We need another tool to distinguish which multiplet is the correct one. 

We resolve this degeneracy by applying the one-loop Hamiltonian to the set of candidate states, which provides us with the spectrum of anomalous dimensions. Protected states are characterised by a vanishing anomalous dimension $\gamma=0$, so we have a direct criterion to distinguish the true BPS multiplets. This process requires the construction and diagonalisation of rather large mixing matrices, but already proved successful for some of the lightest extra states, which we list in Table \ref{tab: results} at the end of section \ref{ssec: scqcdhamiltonian}. An extension to heavier operators is simply a matter of computational power. 

To present the main technical steps we will work through the example of the lightest extra multiplet, which turns out to be a $\CH{1}{\frac{1}{2}}{\frac{1}{2}} $ multiplet. The other multiplets in Table \ref{tab: results} have been determined using the same procedure.

\subsubsection{The sieve algorithm}

Let us first review the sieve algorithm employed in \cite{Gadde:2009dj} to determine suitable candidate multiplets to account for the index mismatch 
\begin{equation}\label{eq: difffirst}
 \Delta \mathcal{I}=-\frac{t^{13}}{v}\left(y+\frac{1}{y}\right)+\dots\,.
\end{equation}
We can compute the (left) index of short multiplets which are given by 
\begin{align}\label{eq: indivindices}
 & \Im(\C{R}{r}{j}{\bar{\jmath}})= (-1)^{2j+2\bar{\jmath}+1}t^{6+4 R+2 r+6j}v^{-2-R+r} \nonumber \\
 &\phantom{\Im_{[\hat R, \bar{\jmath}]_\pm}=\qquad\qquad\qquad }\times\frac{(1-t^2 v)(t-\frac{v}{y})(t-vy)}{(1-t^3 y)(1-t^3/y)}(y^{2\bar{\jmath}}+\ldots +y^{-2\bar{\jmath}})\,,\\
 &\Im(\CH{R}{j}{\bar{\jmath}})=(-1)^{ 2j+2\bar{\jmath}}\,\,\frac{t^{6+4R+4 j+2\bar{\jmath}}v^{-1-R-j+\bar{\jmath}}(1-t^2 v)}{(1-t^3 y)(1-t^3 /y)}\nonumber \\
 & \phantom{\Im_{[\hat R, \bar{\jmath}]_\pm}=\qquad\qquad\qquad\qquad }\times(t(y^{2\bar{\jmath}+1}+\ldots+y^{-(2\bar{\jmath}+1)})-v(y^{2\bar{\jmath}}+\ldots+y^{-2\bar{\jmath}}))\,.
\end{align}
The expression for $\Im(\C{R}{r}{j}{\bar{\jmath}})$ holds unless a further shortening occurs for $r=\bar{\jmath}-j$, in which case $\C{R}{\bar{\jmath}-j}{j}{\bar{\jmath}}$ should be identified with $\CH{R}{j}{\bar{\jmath}}$. Expanding these expressions in $t$ we find the contributions
\begin{equation}\begin{split}
 \Im(\C{R}{r}{j}{\bar{\jmath}})\sim (-1)^{2j+2\bar{\jmath}+1}t^{6+4 R+2 r+6j}v^{-R+r}(y^{2\bar{\jmath}}+\ldots +y^{-2\bar{\jmath}})+\dots\,,\\
 \Im(\CH{R}{j}{\bar{\jmath}})\sim(-1)^{ 2j+2\bar{\jmath}+1}t^{6+4R+4 j+2\bar{\jmath}}v^{-R-j+\bar{\jmath}}(y^{2\bar{\jmath}}+\ldots+y^{-2\bar{\jmath}})+\dots\,.
\end{split}\end{equation}
We can then compare to the leading order term in \eqref{eq: difffirst}. In order to match the fugacities we have to impose the constraints $\bar{\jmath}=\tfrac{1}{2}$, $r=\tfrac{1}{2}-j$ and $R=\tfrac{3}{2}-j$ on the quantum numbers of the highest weight state, which leaves $j$ unfixed. We note that indeed the shortening condition $r=\bar{\jmath}-j$ is satisfied and thus the leading contribution is necessarily supplied by a $\hat{\mathcal{C}}$ multiplet. We end up with the following choices 
\begin{equation}\label{eq: candidatemults}
 \CH{0}{\frac{3}{2}}{\frac{1}{2}}\,,\quad \CH{1}{\frac{1}{2}}{\frac{1}{2}}\,,\quad \CH{2}{-\frac{1}{2}}{\frac{1}{2}}=\mathcal{D}_{\frac{5}{2}(0,\frac{1}{2})}\,.
\end{equation}
All these multiplets have the same index and are thus indistinguishable at this stage. Nevertheless, we know that one of these multiplets is responsible for the leading part of the mismatch $\Delta\mathcal{I}$ \eqref{eq: difffirst}. Without loss of generality we may therefore subtract
\begin{equation}
 \Delta \mathcal{I}-\Im(\CH{R}{\frac{3}{2}-R}{\frac{1}{2}})=\order{t^{18}}\,.
\end{equation}
We can then repeat the same procedure and order by order ``sieve'' through the index mismatch $\Delta\mathcal{I}$. We list the first few mismatch levels and the resulting constraints in Table \ref{tab: constraints}. At every iteration we find a finite number of candidate multiplets which can be grouped into equivalence classes, as explained in \cite{Gadde:2009dj}. Within each equivalence class the quantum numbers are partially constrained such that each representative provides an identical index contribution. In order to break the ambiguity among the representatives of each equivalence class, we need to employ slightly more sensitive technology. 

\begin{table}[]
 \centering
 \begin{tabular}{c||c|c|c}
 Order & $\Delta+j$ &$\bar{\jmath}$ & $r-R$\\\hline\hline
 $t^{13}$ & $\tfrac{13}{2}$ & $\tfrac{1}{2}$ & $-1$\\
 $t^{18}$ & $9$ & $0$ & $0$\\
 $t^{19}$ & $\tfrac{19}{2}$ & $\tfrac{3}{2}$ & $-1$\\
 $t^{20}$ & $10$ & $0$ & $1$\\
 $t^{20}$ & $10$ & $1$ & $-2$
 \end{tabular}
 \caption{Contributions to the index mismatch $\Delta\mathcal{I}$ up to order $\mathcal{O}(t^{20})$ and the resulting constraints on the quantum numbers of the extra states. For each order, there are multiple candidate multiplets that could generate this contribution. This ambiguity will be lifted in the following.}
 \label{tab: constraints}
\end{table}

\subsubsection{Zero anomalous dimensions at one-loop for SCQCD}\label{ssec: scqcdhamiltonian}

Let us return to the index mismatch \eqref{eq: difffirst} and remember that the index counts states satisfying the restriction 
\begin{equation}\label{eq: BPS}
 \Delta=2R+r+2j\,.
\end{equation}
The leading term in \eqref{eq: difffirst} thus signals the presence of a fermionic BPS operator (that in general will not be the highest weight, see Figures \ref{fig:CH_Index} and \ref{fig:C_Index}) with quantum numbers
\begin{equation}\label{eq: ambiguity}
 \Delta= \frac{13}{2}-j \,,\quad R=\frac{5}{2}-j\,,\quad r=\frac{3}{2}-j\,,\quad j\in\{0,1,2\}\,,\quad\bar{\jmath}=\frac{1}{2}\,,
\end{equation}
where the restriction on $j$ follows from the positivity of $R$ and from the fermionic nature of the state.
We recognise a similar ambiguity as before, when we tried to identify multiplets. This comes at no surprise given that these BPS operators would be part of corresponding short multiplets. In order to resolve this ambiguity we now directly test whether a BPS operator with such quantum numbers exists in the theory. To this end we expand the dilatation operator of SCQCD, 
\begin{equation}\label{eq:DilatationSCQCD}
 D(g_1)=\Delta_0+g_1^2 \mathcal{H}+\order{g_1^4} \,,
\end{equation}
and consider the one-loop Hamiltonian $\mathcal{H}$, which has been explicitly derived in \cite{Liendo:2011xb}. The BPS operator $\mathcal{O}_{\text{BPS}}$ in question should satisfy \eqref{eq: BPS} at any coupling $g_1$, which in particular means that
\begin{equation}
[\mathcal{H},\mathcal{O}_{\text{BPS}}]=0
\end{equation}
and thus no anomalous dimension is generated at one loop. This will be a strong enough condition to resolve the ambiguity in \eqref{eq: ambiguity} and determine the precise operator $\mathcal{O}_{\text{BPS}}$ picked up by the index. The appearance of $\mathcal{O}_{\text{BPS}}$ in the index conversely guarantees a vanishing of the anomalous dimension at higher-loop order.

To isolate the BPS operator we first need to generate a basis of operators with quantum numbers according to \eqref{eq: ambiguity}. Since the spin $j$ is a conserved quantity we may consider each choice of $j$ separately. We then need to combine letters of the SCQCD theory, which we summarise in Table \ref{tab: SCQCD}, into single-trace operators such as to match the desired quantum numbers. In order to build single traces, we have to make sure to always contract hypermultiplet fields into $SU(2N)_F$ flavour-singlets such that they form mesonic letters in the adjoint representation of the $SU(N)$ gauge group, \eg~$Q\bar{Q}$. It will furthermore be useful to characterise single-trace operators by their ``length'' $L$, which corresponds to the number of fields inserted and is bounded by $L\leq \Delta-j-\bar{\jmath}$. This length is preserved under the action of the one-loop Hamiltonian, so we may use it as additional super-selection rule. 

\begin{table}[]
 \centering
 \begin{tabular}{c||c|c|c|c|c}
 Letter & $\Delta$ & $R$ & $r$ & $j$ & $\bar{\jmath}$\\\hline\hline
 $\phi$ & $1$ & $0$ & $-1$ & $0$ & $0$\\
 $\bar\phi$ & $1$ & $0$ & $1$ & $0$ & $0$\\
 $\lambda$ & $\tfrac{3}{2}$ & $\tfrac{1}{2}$ & $-\tfrac{1}{2}$ & $\tfrac{1}{2}$ & $0$\\
 $\bar\lambda$ & $\tfrac{3}{2}$ & $\tfrac{1}{2}$ & $\tfrac{1}{2}$ & $0$ & $\tfrac{1}{2}$\\
 $\Fm$ & $2$ & $0$ & $0$ & $1$ & $0$\\
 $\bar \Fm$ & $2$ & $0$ & $0$ & $0$ & $1$ \\\hline
 $Q$, $\bar{Q}$ & $1$ & $\frac{1}{2}$ & $0$ & $0$ & $0$\\
 $\psi$, $\tilde{\psi}$ & $\tfrac{3}{2}$ & $0$ & $\tfrac{1}{2}$ & $\tfrac{1}{2}$ & $0$ \\
 $\bar{\tilde{\psi}}$, $\bar{\psi}$ & $\tfrac{3}{2}$ & $0$ & $-\tfrac{1}{2}$ & $0$ & $\tfrac{1}{2}$ \\\hline
 $D$ & 1 & 0 & 0& $\tfrac{1}{2}$ & $\tfrac{1}{2}$\\
 \end{tabular}
 \caption{Letters for building operators in $\mathcal{N}=2$ SCQCD. $D$ denotes a derivative while all other letters are fields. We list their contributions to the various quantum numbers. Note that the $R,j$ and $\bar{\jmath}$ quantum numbers are not strictly additive but obey the familiar $SU(2)$ tensor-product recombination rules. The hypermultiplet fields with equal quantum numbers are distinguished by their gauge representation. 
 \label{tab: SCQCD}}
\end{table}

Let us first consider the case of $j=2$ in \eqref{eq: ambiguity}, where we observe the bound $L\leq 2$ and thus only find a single operator with the correct quantum numbers, namely
\begin{equation}
 \tr(\Fm D\lambda)\,.
\end{equation}
Acting with the one-loop Hamiltonian \cite{Liendo:2011xb} on this operator yields a non-vanishing one-loop anomalous dimension $\gamma=g_1^2$. We thus rule out this state and the candidate $j=2$. 

Moving on to $j=1$ we find $L\leq4$, but in order to build up sufficient $R$-charge we also require $L\geq 3$, resulting in a plethora of operators such as \eg
\begin{equation}\label{eq: seeds1}
\tr(\bar{\phi}\lambda\lambda\bar\lambda)\,,\quad \tr(\lambda \bar{\lambda}D\bar{\lambda})\,,\dots \,.
\end{equation}
Finally, at $j=0$ we find $L\leq 6$, but $R$-charge again constrains $L\geq 5$, leading to states such as
\begin{equation}\label{eq: seeds2}
 \tr(Q\bar QQ\bar Q\bar{\phi}\bar{\lambda})\,,\quad \tr(Q\bar Q\bar{\lambda}\bar{\lambda}\bar{\lambda})\,,\dots \,.
\end{equation}

We thus find four sub-sectors of states at lengths $L\in\{3,4,5,6\}$ on which we need to diagonalise the one-loop Hamiltonian and search for Eigenstates to the Eigenvalue 0. An efficient way to implement the one-loop Hamiltonian is to represent the various letters in terms of an oscillator representation \cite{Beisert:2003jj,Liendo:2011xb} which we discuss in Appendix \ref{app: Harmonic} and extend to the interpolating theory. We provide a corresponding Wolfram Mathematica notebook as ancillary file to the ar$\chi$iv-submission of this paper. 

Once the Hamiltonian has been implemented it provides a short-cut to constructing the basis of operators of given quantum numbers. We may simply act with the Hamiltonian on seed states such as the ones provided in \eqref{eq: seeds1} and \eqref{eq: seeds2}, which will result in a superposition of various other single-trace operators with the same quantum numbers. Iteratively adding these to the basis of known operators and repeating the process until no additional operators arise creates a closed subset of operators which we can then check for completeness, for example by acting with the quadratic Casimir operators of the various $SU(2)$-symmetries. As a by-product of this procedure we can also extract the coefficient matrix which has to be diagonalised in order to extract the anomalous dimensions. 

Running the numbers, one finds two states with vanishing anomalous dimension, one at $L=3$ and one at $L=4$ (and thus both at spin $j=1$). The state at length $L=3$ is a descendant of $\tr (\mathcal{T}\bar{\phi})$, which belongs to the inherited spectrum of BPS states \eqref{Eq:Naive_SCQCD_Spectrum} and is therefore not one of the genuinely new BPS multiplets. The BPS state at length $L=4$ may be determined explicitly and is given by the fairly involved combination 
\begin{align}\label{eq:extrastate}
\mathcal{O}&_{\text{BPS}}=\nonumber\\&-\frac{7}{2}\tr\left(\bar{\phi} \, D_{+\dot{+}}\left(Q_1\right) \, \bar{Q}_1 \, \lambda_{+1}\right)
-\frac{1}{2}\tr\left(\bar{\phi} \, D_{+\dot{+}}\left(\lambda_{+1}\right) \, Q_1 \, \bar{Q}_1\right)
+\frac{7}{2}\tr\left(\bar{\phi} \, Q_1 \, D_{+\dot{+}}\left(\bar{Q}_1\right) \, \lambda_{+1}\right)
\nonumber\\ &
+\frac{1}{2}\tr\left(\bar{\phi} \, Q_1 \, \bar{Q}_1 \, D_{+\dot{+}}\left(\lambda_{+1}\right)\right)
-\frac{7}{2}\tr\left(\bar{\phi} \, \lambda_{+1} \, D_{+\dot{+}}\left(Q_1\right) \, \bar{Q}_1\right)
+\frac{7}{2}\tr\left(\bar{\phi} \, \lambda_{+1} \, Q_1 \, D_{+\dot{+}}\left(\bar{Q}_1\right)\right)
\nonumber\\ &
+2\tr\left(\bar{\phi} \, \lambda_{+1} \, \lambda_{+1} \, \bar{\lambda}_{\dot{+}1}\right)
-2\tr\left(\bar{\phi} \, \bar{\lambda}_{\dot{+}1} \, \lambda_{+1} \, \lambda_{+1}\right)
+\frac{13}{2}\tr\left(Q_1 \, \tilde{\psi}_+ \, Q_1 \, D_{+\dot{+}}\left(\bar{Q}_1\right)\right)
\nonumber\\ &
+\frac{7}{2}\tr\left(Q_1 \, \tilde{\psi}_+ \, \lambda_{+1} \, \bar{\lambda}_{\dot{+}1}\right)
-\frac{7}{2}\tr\left(Q_1 \, \tilde{\psi}_+ \, \bar{\lambda}_{\dot{+}1} \, \lambda_{+1}\right)
+\tr\left(Q_1 \, \bar{Q}_1 \, D_{+\dot{+}}\left(\bar{\phi}\right) \, \lambda_{+1}\right)
\nonumber\\ &
+\frac{7}{2}\tr\left(Q_1 \, \bar{Q}_1 \, D_{+\dot{+}}\left(Q_1\right) \, \tilde{\psi}_+\right)
-\frac{1}{2}\tr\left(Q_1 \, \bar{Q}_1 \, \Fm_{++} \, \bar{\lambda}_{\dot{+}1}\right)
-2\tr\left(Q_1 \, \bar{Q}_1 \, Q_1 \, D_{+\dot{+}}\left(\tilde{\psi}_+\right)\right)
\nonumber\\ &
-\tr\left(Q_1 \, \bar{Q}_1 \, \lambda_{+1} \, D_{+\dot{+}}\left(\bar{\phi}\right)\right)
-\frac{7}{2}\tr\left(Q_1 \, \bar{Q}_1 \, \psi_+ \, D_{+\dot{+}}\left(\bar{Q}_1\right)\right)
+\frac{1}{2}\tr\left(Q_1 \, \bar{Q}_1 \, \bar{\lambda}_{\dot{+}1} \, \Fm_{++}\right)
\nonumber\\ &
-6\tr\left(\bar{Q}_1 \, Q_1 \, D_{+\dot{+}}\left(\bar{Q}_1\right) \, \psi_+\right)
+6\tr\left(\bar{Q}_1 \, Q_1 \, \tilde{\psi}_+ \, D_{+\dot{+}}\left(Q_1\right)\right)
-2\tr\left(\bar{Q}_1 \, Q_1 \, \bar{Q}_1 \, D_{+\dot{+}}\left(\psi_+\right)\right)
\nonumber\\ &
-\frac{7}{2}\tr\left(\bar{Q}_1 \, \lambda_{+1} \, \bar{\lambda}_{\dot{+}1} \, \psi_+\right)
+\frac{13}{2}\tr\left(\bar{Q}_1 \, \psi_+ \, \bar{Q}_1 \, D_{+\dot{+}}\left(Q_1\right)\right)
+\frac{7}{2}\tr\left(\bar{Q}_1 \, \bar{\lambda}_{\dot{+}1} \, \lambda_{+1} \, \psi_+\right) \,.
\end{align}
Here we fixed a specific polarisation for simplicity. The $+,\dot +,1$ subscripts denote positive polarisations in the Lorentz $SU(2)_j\cross SU(2)_{\bar{\jmath}}$ and internal $SU(2)_R$ symmetries, respectively. Once the explicit operator is known, we would also like to identify which multiplet it belongs to. Since multiplets are defined by their highest weight states, we simply act with all possible ``raising'' supercharges $\mathcal{S}$ until the state is annihilated by all of them. In this case, there is only one supercharge that does not vanish so we find the highest-weight state
\pagebreak
\begin{align}
 \mathcal{O}_{\text{HW}}&=\mathcal{S}^{+}_{ 2}\mathcal{O}_{\text{BPS}}=\nonumber\\
&\frac{7}{2}\tr\left(\bar{\phi}\,\phi\,D_{+\dot{+}}\left(Q_{1}\right)\,\bar{Q}_{1}\right)
-\frac{7}{2}\tr\left(\bar{\phi}\,\phi\,Q_{1}\,D_{+\dot{+}}\left(\bar{Q}_{1}\right)\right)
-2\tr\left(\bar{\phi}\,\phi\,\lambda_{+1}\,\bar{\lambda}_{\dot{+}1}\right)
\nonumber\\ &
+\tr\left(\bar{\phi}\,D_{+\dot{+}}\left(\phi\right)\,Q_{1}\,\bar{Q}_{1}\right)
+\frac{7}{2}\tr\left(\bar{\phi}\,D_{+\dot{+}}\left(Q_{1}\right)\,\bar{Q}_{1}\,\phi\right)
-\frac{7}{2}\tr\left(\bar{\phi}\,Q_{1}\,D_{+\dot{+}}\left(\bar{Q}_{1}\right)\,\phi\right)
\nonumber\\ &
-\tr\left(\bar{\phi}\,Q_{1}\,\bar{Q}_{1}\,D_{+\dot{+}}\left(\phi\right)\right)
+\frac{7}{2}\tr\left(\bar{\phi}\,Q_{1}\,\bar{\psi}_{+}\,\lambda_{+1}\right)
+2\tr\left(\bar{\phi}\,\lambda_{+1}\,\phi\,\bar{\lambda}_{\dot{+}1}\right)
\nonumber\\ &
-\frac{7}{2}\tr\left(\bar{\phi}\,\lambda_{+1}\,Q_{1}\,\bar{\psi}_{+}\right)
-\frac{7}{2}\tr\left(\bar{\phi}\,\lambda_{+1}\,\bar{\tilde{\psi}}_{\dot{+}}\,\bar{Q}_{1}\right)
-2\tr\left(\bar{\phi}\,\bar{\lambda}_{\dot{+}1}\,\phi\,\lambda_{+1}\right)
\nonumber\\ &
+2\tr\left(\bar{\phi}\,\bar{\lambda}_{\dot{+}1}\,\lambda_{+1}\,\phi\right)
+\frac{7}{2}\tr\left(\bar{\phi}\,\bar{\tilde{\psi}}_{\dot{+}}\,\bar{Q}_{1}\,\lambda_{+1}\right)
+\frac{7}{2}\tr\left(Q_{1}\,\bar{\tilde{\psi}}_{\dot{+}}\,\phi\,\bar{\lambda}_{\dot{+}1}\right)
\nonumber\\ &
-\frac{13}{2}\tr\left(Q_{1}\,\bar{\tilde{\psi}}_{\dot{+}}\,Q_{1}\,\bar{\psi}_{+}\right)
+\frac{7}{2}\tr\left(Q_{1}\,\bar{\tilde{\psi}}_{\dot{+}}\,\bar{\lambda}_{\dot{+}1}\,\phi\right)
+\tr\left(Q_{1}\,\bar{Q}_{1}\,\phi\,D_{+\dot{+}}\left(\bar\phi\right)\right)
\nonumber\\ &
-\tr\left(Q_{1}\,\bar{Q}_{1}\,D_{+\dot{+}}\left(\bar\phi\right)\,\phi\right)
-\frac{7}{2}\tr\left(Q_{1}\,\bar{Q}_{1}\,D_{+\dot{+}}\left(Q_{1}\right)\,\bar{Q}_{2}\right)
+4\tr\left(Q_{1}\,\bar{Q}_{1}\,Q_{1}\,D_{+\dot{+}}\left(\bar{Q}_{2}\right)\right)
\nonumber\\ &
-\frac{7}{2}\tr\left(Q_{1}\,\bar{Q}_{1}\,Q_{2}\,D_{+\dot{+}}\left(\bar{Q}_{1}\right)\right)
+\tr\left(Q_{1}\,\bar{Q}_{1}\,\lambda_{+1}\,\bar{\lambda}_{\dot{+}2}\right)
-\tr\left(Q_{1}\,\bar{Q}_{1}\,\lambda_{+2}\,\bar{\lambda}_{\dot{+}1}\right)
\nonumber\\ &
+\frac{7}{2}\tr\left(Q_{1}\,\bar{Q}_{1}\,\psi_{+}\,\bar{\psi}_{+}\right)
-\tr\left(Q_{1}\,\bar{Q}_{1}\,\bar{\lambda}_{\dot{+}1}\,\lambda_{+2}\right)
+\tr\left(Q_{1}\,\bar{Q}_{1}\,\bar{\lambda}_{\dot{+}2}\,\lambda_{+1}\right)
\nonumber\\ &
-\frac{7}{2}\tr\left(Q_{1}\,\bar{Q}_{1}\,\bar{\tilde{\psi}}_{\dot{+}}\,\tilde{\psi}_{+}\right)
-\frac{13}{2}\tr\left(Q_{1}\,\bar{Q}_{2}\,Q_{1}\,D_{+\dot{+}}\left(\bar{Q}_{1}\right)\right)
-\frac{7}{2}\tr\left(Q_{1}\,\bar{Q}_{2}\,\lambda_{+1}\,\bar{\lambda}_{\dot{+}1}\right)
\nonumber\\ &
+\frac{7}{2}\tr\left(Q_{1}\,\bar{Q}_{2}\,\bar{\lambda}_{\dot{+}1}\,\lambda_{+1}\right)
-\frac{7}{2}\tr\left(\bar{Q}_{1}\,\phi\,\bar{\lambda}_{\dot{+}1}\,\psi_{+}\right)
+6\tr\left(\bar{Q}_{1}\,Q_{1}\,D_{+\dot{+}}\left(\bar{Q}_{1}\right)\,Q_{2}\right)
\nonumber\\ &
-6\tr\left(\bar{Q}_{1}\,Q_{1}\,\tilde{\psi}_{+}\,\bar{\tilde{\psi}}_{\dot{+}}\right)
+4\tr\left(\bar{Q}_{1}\,Q_{1}\,\bar{Q}_{1}\,D_{+\dot{+}}\left(Q_{2}\right)\right)
+6\tr\left(\bar{Q}_{1}\,Q_{1}\,\bar{Q}_{2}\,D_{+\dot{+}}\left(Q_{1}\right)\right)
\nonumber\\ &
+6\tr\left(\bar{Q}_{1}\,Q_{1}\,\bar{\psi}_{+}\,\psi_{+}\right)
-\frac{13}{2}\tr\left(\bar{Q}_{1}\,Q_{2}\,\bar{Q}_{1}\,D_{+\dot{+}}\left(Q_{1}\right)\right)
+\frac{7}{2}\tr\left(\bar{Q}_{1}\,\lambda_{+1}\,\bar{\lambda}_{\dot{+}1}\,Q_{2}\right)
\nonumber\\ &
-\frac{13}{2}\tr\left(\bar{Q}_{1}\,\psi_{+}\,\bar{Q}_{1}\,\bar{\tilde{\psi}}_{\dot{+}}\right)
-\frac{7}{2}\tr\left(\bar{Q}_{1}\,\bar{\lambda}_{\dot{+}1}\,\phi\,\psi_{+}\right)
-\frac{7}{2}\tr\left(\bar{Q}_{1}\,\bar{\lambda}_{\dot{+}1}\,\lambda_{+1}\,Q_{2}\right)\,.
\end{align}
We can then read off the charges and determine the BPS multiplet to be $\CH{1}{\frac{1}{2}}{\frac{1}{2}}$. 

We have thus demonstrated how to resolve the ambiguity intrinsic to the sieve algorithm by explicit diagonalisation of the one-loop Hamiltonian. The leading contribution to the index is produced by the short multiplet $\CH{1}{\frac{1}{2}}{\frac{1}{2}}$. Admittedly, this conclusion could have been reached by observing that the leading contribution to both the left and the right index $\Delta\mathcal{I}^L$ and $\Delta\mathcal{I}^R$ are of the form \eqref{eq: difffirst} and therefore $j=\bar{\jmath}=\tfrac{1}{2}$. However, such arguments do not apply to the subleading contributions where a treatment with the one-loop Hamiltonian becomes necessary. Repeating this procedure, we were able to identify the short multiplets contributing to $\Delta\mathcal{I}^L$ up to $\order{t^{20}}$. We list the length, quantum numbers of the extra state and associated multiplet in Table \ref{tab: results}. At higher order in $t$ some sporadic data points have been considered but a full survey is still ongoing.

\begin{table}[]
 \centering
 \renewcommand{\arraystretch}{1.2}
 \begin{tabular}{c||c|c|c|c|c|c||c}
 Order & L &$\Delta$ & $R$ & $r$ & $j$ & $\bar{\jmath}$ & multiplet\\\hline\hline
 $t^{13}$ & $4$ & $\tfrac{11}{2}$ & $\tfrac{3}{2}$ & $\tfrac{1}{2}$ & $1$ & $\tfrac{1}{2}$ & $\CH{1}{\frac{1}{2}}{\frac{1}{2}}$\\
 $t^{18}$ & $6$ & $\tfrac{15}{2}$ & $\tfrac{3}{2}$ & $\tfrac{3}{2}$ & $\tfrac{3}{2}$ & $0$ & $\C{1}{1}{1}{0}$\\
 $t^{19}$ & $4$ & $\tfrac{15}{2}$ & $\tfrac{3}{2}$ & $\tfrac{1}{2}$ & $2$ & $\tfrac{3}{2}$ & $\CH{1}{\frac{3}{2}}{\frac{3}{2}}$\\
 $t^{20}$ & $7$ & $\tfrac{17}{2}$ & $\tfrac{3}{2}$ & $\tfrac{5}{2}$ & $\tfrac{3}{2}$ & $0$ &$\C{1}{2}{1}{0}$\\
 $t^{20}$ & $6$ & $\tfrac{17}{2}$ & $\tfrac{5}{2}$ & $\tfrac{1}{2}$ & $\tfrac{3}{2}$ & $1$ & $\CH{2}{1}{1}$\\
 \end{tabular}
 \caption{The precise extra states contributing to $\Delta \mathcal{I}^L$ up to order $\mathcal{O}(t^{20})$ (compare to Table \ref{tab: constraints}). We list the length and quantum numbers of the extra state and the short multiplet to which it belongs. Note, that a similar analysis of $\Delta \mathcal{I}^R$ would result in a parallel spectrum with $\Cm\rightarrow\bar{\Cm}$ multiplets. }
 \label{tab: results}
\end{table}

\subsubsection{One-loop anomalous dimensions in the interpolating theory}

Having identified the extra BPS multiplets in SCQCD, we now investigate their origin in the interpolating theory. As outlined in Section \ref{ssec: recombination}, we expect that certain long multiplets in the interpolating theory saturate the BPS bound precisely in the limit $\kappa=\frac{g_2}{g_1}\to0$, where they break down into short multiplets \eqref{eq: longHS}. Conversely, if we know the short multiplets in the limit, we should be able to observe the long multiplet recombination away from $\kappa=0$. In particular, all states contributing to the same long multiplets should develop the same anomalous dimension $\gamma(\kappa)$. 

In this section we will confirm this behaviour for the first extra multiplet $\CH{1}{\frac{1}{2}}{\frac{1}{2}}$, which we discussed in detail before. It is supposed to arise in the recombination \eqref{eq: longHS}
\begin{equation}\label{eq:exprecombination}
 \mathcal{A}^{4}_{0,0(1,1)} \simeq \hat{\mathcal{C}}_{0(1,1)} \oplus \hat{\mathcal{C}}_{\frac{1}{2}(\frac{1}{2},1)} \oplus \hat{\mathcal{C}}_{\frac{1}{2}(1,\frac{1}{2})} \oplus \hat{\mathcal{C}}_{1(\frac{1}{2},\frac{1}{2})}\,,
\end{equation}
where the highest weight state of $\mathcal{A}^{4}_{0,0(1,1)}$ contributes to the HS current multiplet $\CH{0}{1}{1}$. At $\kappa=0$, we expect $\CH{0}{1}{1}$ to consist entirely of fields from the decoupling gauge multiplet, \ie, to belong to Class I in the classification of Section \ref{ssec: recombination}. Meanwhile our observed Class II extra multiplet should contribute to the right-most component in \eqref{eq:exprecombination}.

In order to analyse anomalous dimensions at arbitrary $\kappa$ we need to extend the dilatation operator \eqref{eq:DilatationSCQCD} to allow for a second coupling constant
\begin{equation}\label{eq:Dilatation}
 D(g_1,g_2)=\Delta_0+g_1^2 (\mathcal{H}_0+\kappa\mathcal{H}_1+\kappa^2\mathcal{H}_2)+\order{g_1^{4-l}g_2^{l}}\,.
\end{equation}
The one-loop Hamiltonian thus receives additional contributions, which have been discussed in \cite{Liendo:2011xb}. We review its action on a suitable oscillator representation in Appendix \ref{app: Harmonic}. This Hamiltonian now couples the fields in SCQCD to the vector multiplet ($\check{\phi},\check{\lambda},\check{\Fm}$) of the second gauge node, breaking the global $SU(2N)_F$ flavour symmetry of the hypermultiplets into the gauged $SU(N)$ symmetry and a residual global $SU(2)_L$ symmetry. We have to extend our Hilbert space to account for novel combinations of fields, such as \eg~$Q\check{\phi}\bar{Q}$, which were not present in SCQCD. Similarly the $SU(2)_L$ quantum numbers of hypermultiplets have to be kept track of: In SCQCD combinations like $Q\bar{Q}$ formed singlets under $SU(2N)_F$, and were by extension singlets of $SU(2)_L$. In the interpolating theory also triplet combinations emerge. With these subtleties in mind, we were able to implement the one-loop Hamiltonian for any value of $\kappa$ and could use it to generate the desired closed subsectors of the Hilbert space. The anomalous dimensions can be determined numerically. This enables us to confirm the matching of anomalous dimensions among states in the same long multiplet.

\begin{figure}
 \centering
 \includegraphics[width=0.5\linewidth]{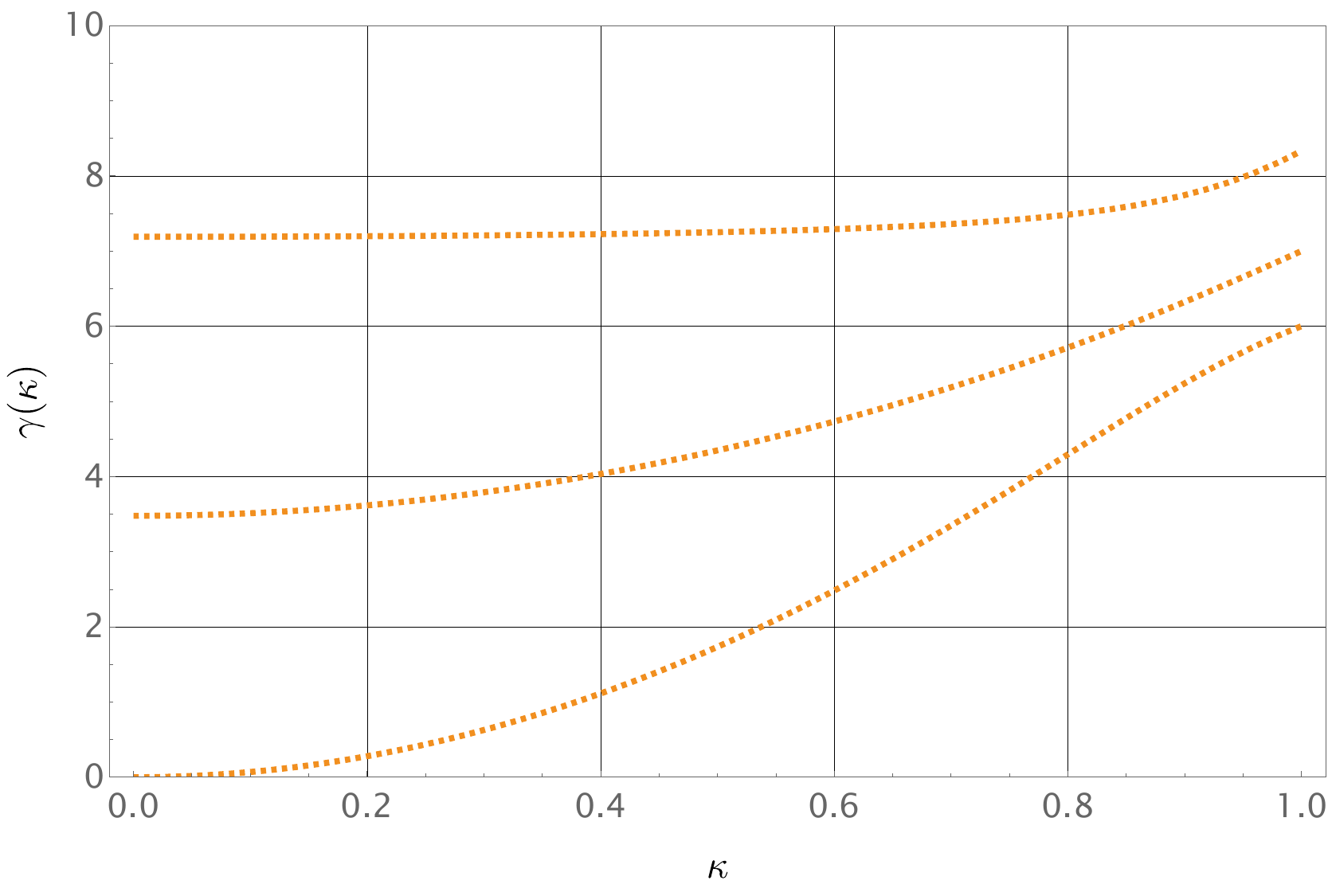}\includegraphics[width=0.5\linewidth]{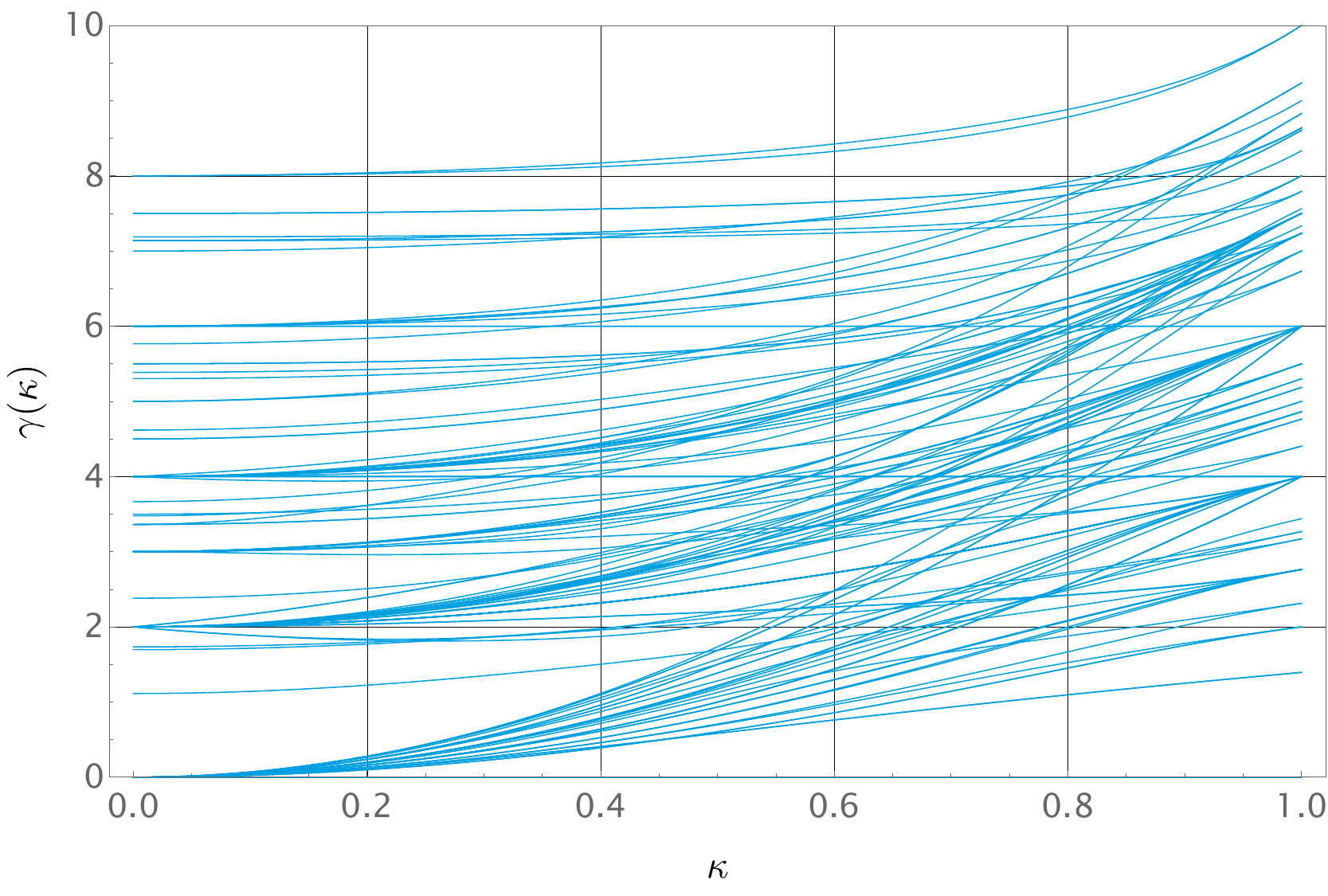}
 \caption{We plot the anomalous dimensions $\gamma(\kappa)$ for the highest weight states (left) and the closed sector including the extra state \eqref{eq:extrastate} (right).}
 \label{fig:anomalousdimensions}
\end{figure}

Let us start by investigating the highest-weight operator of the HS multiplet $\CH{0}{1}{1}$, which will also become the highest-weight operator of the long-multiplet $\mathcal{A}^{4}_{0,0(1,1)}$ after recombination. This is the first level of the HS tower predicted by the CFT Distance Conjecture and denoted by \textcircled{\footnotesize 1} in Figure \ref{fig:N2DC}. It has to be of length $L=2$. In order to find an explicit representation of this highest-weight state in terms of fields, we may first combine the various fundamental fields at strictly vanishing coupling $g_1=g_2=0$ and build free-field realisations of $\hat{\mathcal{C}}_{0,(\frac{q}{2},\frac{q}{2})}$ multiplets in the sectors $\Vm \times \Vmb$, $\Vmc \times \bar{\Vmc}$, and $\Hm \times \Hmb$ (in the $SU(2)_L$ singlet representation). The highest weight states are constructed by writing down all possible operators with the correct charges and requiring annihilation under the conformal supercharges $\mathcal{S}$. The appropriate combinations are \cite{Liendo:2011xb}\footnote{In a slight abuse of notation we will denote the highest-weight state of a multiplet $\mathcal{M}$ as $\vert \mathcal{M}\rangle$.}
\begin{align}
\vert\hat{\mathcal{C}}_{0,(\frac{q}{2},\frac{q}{2})}\rangle_{\Vm\times \Vmb}=& \sum_{k=0}^{q}(-1)^k\binom{q}{k}^2D^{q-k}\phi D^k \bar{\phi}+
q\sum_{k=0}^{q-2}\frac{(-1)^k}{k+2}\binom{q}{k}\binom{q-1}{k+1} D^{q-k-2}\mathcal{F}_{++} D^k \bar{\mathcal{F}}_{\dot{+}\dot{+}}+\notag\\
 &-q\sum_{k=0}^{q-1}\frac{(-1)^k}{k+1}\binom{q-1}{k}\binom{q}{k} \left(D^{q-k-1} \lambda_{+1}D^k \bar{\lambda}_{\dot{+}2}-D^{q-k-1}\lambda_{+2}D^k \bar{\lambda}_{\dot{+}2}\right)\label{eq: Cvv}\\
\vert\hat{\mathcal{C}}_{0,(\frac{q}{2},\frac{q}{2})}\rangle_{\Vmc\times \bar{\Vmc}}=&\sum_{k=0}^{q}(-1)^k\binom{q}{k}^2D^{q-k}\check{\phi} D^k \bar{\check{\phi}}+
q\sum_{k=0}^{q-2}\frac{(-1)^k}{k+2}\binom{q}{k}\binom{q-1}{k+1}D^{q-k-2}\mathcal{\check{F}}_{++} D^k \bar{\mathcal{\check{F}}}_{\dot{+}\dot{+}}+\notag\\
&-q\sum_{k=0}^{q-1}\frac{(-1)^k}{k+1}\binom{q-1}{k}\binom{q}{k} \left( D^{q-k-1} \check{\lambda}_{+1} D^k \bar{\check{\lambda}}_{\dot{+}2}-D^{q-k-1}\check{\lambda}_{+2} D^k \bar{\check{\lambda}}_{\dot{+}1}\right)\label{eq: hs-hw}\\
\vert\hat{\mathcal{C}}_{0,(\frac{q}{2},\frac{q}{2})}\rangle_{\Hm\times \Hmb}=&\sum_{k=0}^{q}(-1)^k\binom{q}{k}^2\left(D^{q-k}Q_1 D^k \bar{Q}_2-D^{q-k}Q_2 D^k \bar{Q}_1\right)\notag\\
&+q\sum_{k=0}^{q-1}\frac{(-1)^k}{k+1}\binom{q-1}{k}\binom{q}{k} \left( D^{q-k-1} \psi_{+} D^k \bar{\psi}_{\dot{+}}-D^{q-k-1}\tilde{\psi}_{+} D^k \bar{\tilde{\psi}}_{\dot{+}}\right)\,.\label{eq: Chh}
\end{align}
Turning on the one-loop term in \eqref{eq:Dilatation} these operators mix but crucially only among themselves, as we can confirm with our Hamiltonian implementation. Combining \eqref{eq: Cvv}, \eqref{eq: hs-hw}, and \eqref{eq: Chh} into a vector $\vec{\mathcal{C}}_q$, the mixing matrix takes the relatively simple form 
\begin{align}\label{eq:matrix}
\mathcal{H}\,\,\vec{\mathcal{C}}_q=
\begin{pmatrix}
\frac{4}{q+2}+4h(q) & 0 & \frac{2}{q+2}\\
0 & \kappa^2\left(\frac{4}{q+2}+4h(q)\right) & -\frac{2\kappa^2}{q+2}\\
\frac{4}{q+1} & -\frac{4\kappa^2}{q+1} & 2 h(q+1)(1+\kappa^2)
\end{pmatrix}
\vec{\mathcal{C}}_q\,,
\end{align}
where $h(q)=\sum_{k=1}^q\tfrac{1}{k}$ is the $q$\textsuperscript{th}-harmonic number. A protected state appears whenever $\Hm(q)$ has a zero Eigenvalue. For the case at hand ($q=2$) we can compute the determinant 
\begin{align}
 &\det\mathcal{H} =\frac{1}{12} \left(2100 \kappa ^4+2100 \kappa ^2\right),
\end{align}
which only vanishes at $\kappa=0$, where $\vert\hat{\mathcal{C}}_{0,(1,1)}\rangle_{\Vmc\times \bar{\Vmc}}$ becomes protected. This is precisely the decoupled HS current we expect to find. Now, however, we can also determine its anomalous dimension for any value of $\kappa$, as it mixes with the other currents. The anomalous dimensions are depicted on the left of Figure \ref{fig:anomalousdimensions}.

\begin{figure}
 \centering
 \includegraphics[width=0.80\linewidth]{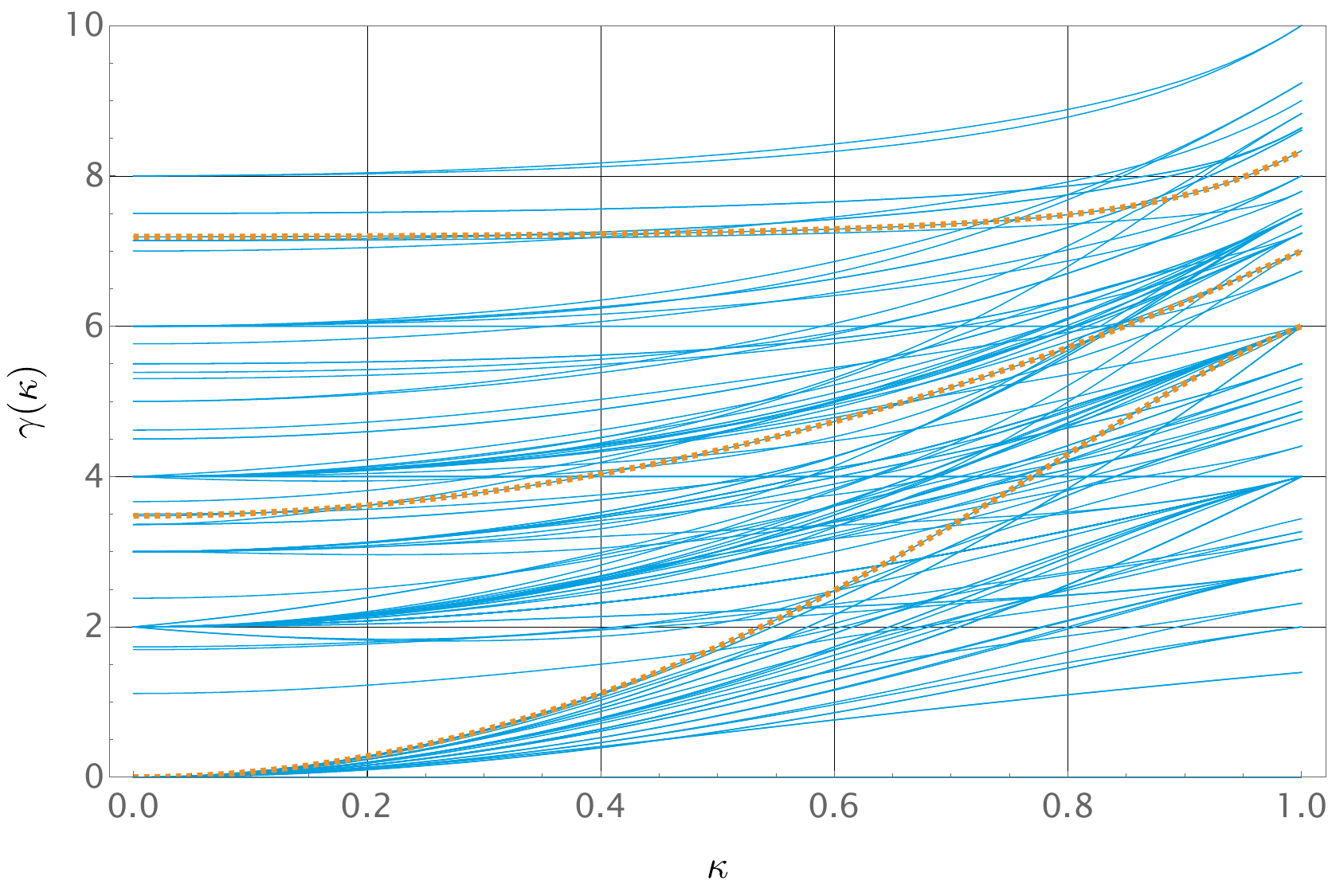}
 \caption{Overlap of the plots in Figure \ref{fig:anomalousdimensions}. We observe perfect overlap for three trajectories, demonstrating the recombination of highest weight and extra states into the same long multiplet. }
 \label{fig:anomalousoverlap}
\end{figure}

For the extra-state determined in \eqref{eq:extrastate} applying the Hamiltonian at $\kappa\neq0$ generates an abundance of states. Numerically diagonalising the Hamiltonian (see right-hand-side of Figure \ref{fig:anomalousdimensions}) on this subset of states we find three states which precisely match the anomalous dimensions from \eqref{eq:matrix}. This can be demonstrated most clearly by overlaying the plots of $\gamma(\kappa)$ for the highest-weight and extra-state sectors and observing perfect overlap of three trajectories, as is shown in Figure \ref{fig:anomalousoverlap}.

We take this perfect match of anomalous dimensions as evidence that the HS multiplet $\hat{\mathcal{C}}_{0(1,1)}$ and the extra multiplet $\hat{\mathcal{C}}_{1(\frac{1}{2},\frac{1}{2})}$ combine to the same long multiplet $\mathcal{A}^{4}_{0,0(1,1)}$ as predicted in \eqref{eq:exprecombination}. Of course we expect additional Class I and Class III multiplets to arise in this recombination, for example to generate the multiplets $\hat{\mathcal{C}}_{\frac{1}{2}(\frac{1}{2},1)} \oplus \hat{\mathcal{C}}_{\frac{1}{2}(1,\frac{1}{2})}$. In fact a direct computation in the decoupled gauge theory allows us to find exactly those additional Class I BPS states, which further underlines the validity of this recombination picture. Class III operators such as \eg
\begin{equation}
 \tr\left(\check\phi\,D_{+\dot{+}}\left(\bar{Q}_{1}\right)\,Q_{1}\right)= (\check\phi)_{\check{a}}^{\ph{\check{a}}\check{b}} \times \left(D_{+\dot{+}}\left(\bar{Q}_{1}\right)\,Q_{1}\right)_{\check{b}}^{\ph{\check{b}}\check{a}} \,,
\end{equation}
can only be found at finite coupling. We see that these operators decompose into operators with free flavour symmetry indices, which should correspond to open string states in the dual theory. The interaction of these operators with single trace operators within the two decoupled theories is suppressed at large-$N$ and can therefore be ignored \cite{Gadde:2009dj}.

A full proof of the indicated multiplet structure would require the analysis of the one-loop corrected supercharges $\mathcal{Q}$, which, acting on the highest weight state at $\kappa\neq 0$, would generate the entire long multiplet. Although the one-loop correction to $\mathcal{Q}$ has been identified for a subsector of fields in \cite{Liendo:2011xb}, its full expression remains to be determined. This would allow us to directly identify the short multiplets that arise in the decoupling limit, without requiring an indirect argument via matching of anomalous dimensions. We would like to return to this challenge in future work.

\section{Conclusions and Outlook}
\label{Sec:Conclusion}

In this work we have investigated the CFT Distance Conjecture \cite{Perlmutter:2020buo} in the context of four-dimensional $\mathcal{N}=2$ quiver SCFTs with higher-dimensional conformal manifolds. 
Our main interest has been in a partial weak coupling limit where one vector multiplet decouples as $\check{g} \to 0$ while the remainder theory stays at finite coupling. As predicted by the CFT Distance Conjecture a tower of massless states is found to emerge \cite{Baume:2020dqd,Perlmutter:2020buo}, where the effective mass is computed via the holographic dictionary \eqref{eq:massgeneral}. The associated operators are HS currents \cite{Maldacena:2011jn} in the free vector multiplet that at $\check{g} =0$ decouples completely from the interacting remainder SCFT. 
 As our main result, we have established the existence of an extra BPS tower at the AdS scale in the interacting sector of the theory whose existence is linked to this massless HS tower, via representation theory and multiplet recombination.

Specifically, in $\mathcal{N}=2$ theories the HS currents sit in long superconformal multiplets which as $\check{g} \to 0$ hit the unitarity bound \eqref{eq: unit bounds} and recombine into short multiplets. This recombination generically results in a plethora of protected BPS states in the interacting remainder theory, which would not have been present at generic points of the conformal manifold with $\check{g}\neq0$. 

In order to reconcile these towers with the usual intuition of infinite-distance limits, we have shown that the massless tower of HS currents exhibits polynomial growth in degeneracy, while the extra towers, whose mass is determined by the AdS scale $M_{\rm{AdS}}$, grow exponentially. This is in contrast to the flat-space intuition that emergent string limits should come with a massless tower of exponential degeneracy. As discussed below equation \eqref{eq: intmasses}, we thus interpret the weak-coupling limits in this paper as emergent string limits in which the background geometry nevertheless contributes an AdS-scale mass to all but the leading Regge trajectories.
The leading Regge trajectory, which demonstrates polynomial degeneracy growth, is thus free to become massless and generate the higher-spin tower, while all other Regge trajectories ``get stuck'' at the AdS scale $M_{\rm{AdS}}$ and contribute an exponentially degenerate tower of new BPS states to the remainder theory.

Although the phenomena we find are generic, we have provided explicit computations for the $\mathcal{N}=2$ two-node quiver SCFT depicted in Figure \ref{fig:Quivers}, which has a two-dimensional conformal manifold and allows us to decouple a free gauge node from the remainder, known as SCQCD. The extra BPS multiplets in this theory were previously observed in \cite{Gadde:2009dj}, but there a cohomological ambiguity of the superconformal index was obstructing the full determination of the short multiplet spectrum. Here, in tandem with the index, we have diagonalised the one-loop dilatation operator and fixed the BPS spectrum completely. Following the Eigenvalues of the dilatation operator to finite coupling we furthermore found direct evidence for the multiplet recombination process: We observed that the anomalous dimensions of the lightest HS current and the lightest extra state coincide, confirming their common origin in the same long multiplet. The exponential degeneracy of the extra states has been established explicitly (see \eg~Figures \ref{fig:indexfull} and \ref{fig:indexschur}), providing evidence for our interpretation as an emergent string tower.

As explained above, in this paper we have focused on the partial weak coupling limits, where $g_2 \to 0$ with $g_1$ finite. If in addition, one takes the limit $g_1 \to 0$, the anomalous dimensions of the operators in \eqref{eq: Cvv} and \eqref{eq: Chh} go to zero, and the operators hit the BPS bound $\Delta = 2 +j + \bar{\jmath}$.
While for non-zero values of $g_1$ these operators are heavy (they are part of the tower \textcircled{\footnotesize 5} in Fig. \ref{fig:N2DC}), as $g_1\to 0$ they contribute additional higher-spin towers of vanishing mass. More precisely, the tower \eqref{eq: Cvv} is a gauge invariant HS tower of flavour singlets,  while the tower \eqref{eq: Chh} is an $N_f \times N_f$ open string HS current.  
Furthermore, from \eqref{eq:extrastate} it is clear that the destiny of the extra tower for $g_1 \to 0$ is to break apart in open string towers (analogous to Class III states) whose mass remains at the AdS scale because they carry non-trivial $R$-charge. \\

This work opens several directions for future research:
\begin{itemize}
\item Having determined the BPS spectrum of SCQCD with a one-loop computation (from the gauge theory side), a natural question is how this tower arises in the holographic dual theory. At the orbifold point $\kappa=\tfrac{g_2}{g_1}=1$ of the quiver gauge theory, the holographic dual theory is given by type IIB string theory on $AdS_5\times S^5/\mathbb{Z}_2$ background. This orbifold features a non-trivial two-cycle within the resolution of the orbifold singularity, which can carry a Kalb--Ramond field $B_2= b \omega_2$, where $\omega_2$ is the two-form wrapping the resolution cycle. Excitations of this field generate twisted sector scalars \cite{Douglas:1996sw,Gukov:1998kk,Klebanov:1999rd,Skrzypek:2023fkr,Martinez:2025jjq}, while the constant zero-mode of $b$ is related to the difference in couplings \cite{Aspinwall:1994ev}
\begin{equation}
 b=\frac{g_2}{g_1+g_2}=\frac{\kappa}{1+\kappa}\,.
\end{equation}
Given that this background field $b$ does not generate a field-strength, it is rather difficult to determine its effect on the bulk theory, requiring an analysis of worldsheet instantons and D-branes wrapping the resolution cycle. This analysis could be matched to localisation data that is available even at strong coupling \cite{Zarembo:2020tpf}. 
\item The BPS data we computed here does not run with the remaining coupling $g_1$ and thus provides us with information about the dual theory at the extremal limit $b\to0$. In fact, we can do even better. The weak coupling gauge theory calculations presented here allow us to follow these multiplets all the way to the orbifold locus $b=\tfrac{1}{2}$, at which point the theory becomes integrable \cite{Beisert:2005he}. Integrability allows us to determine the spectrum at finite coupling \cite{Skrzypek:2022cgg} and to extrapolate the results to strong coupling \cite{Gromov:2013pga, Gromov:2015wca,Gromov:2023hzc}. This would provide us with spectral data both at $b=0$ and $b=\tfrac{1}{2}$, which the string dual should interpolate between (see Fig. \ref{fig:idea}). This raises the question whether we can identify the appropriate string states and determine their evolution. The novel hidden symmetries discovered in \cite{Bertle:2024djm} (and further elaborated on in \cite{Pomoni:2021pbj,Pomoni:2019oib,Bozkurt:2024tpz,lePlat:2025eod,Bozkurt:2025exl,Klabbers, Abedin}) should be instrumental to this task. 

\begin{figure}[h]
\centering
\includegraphics[scale=0.3]{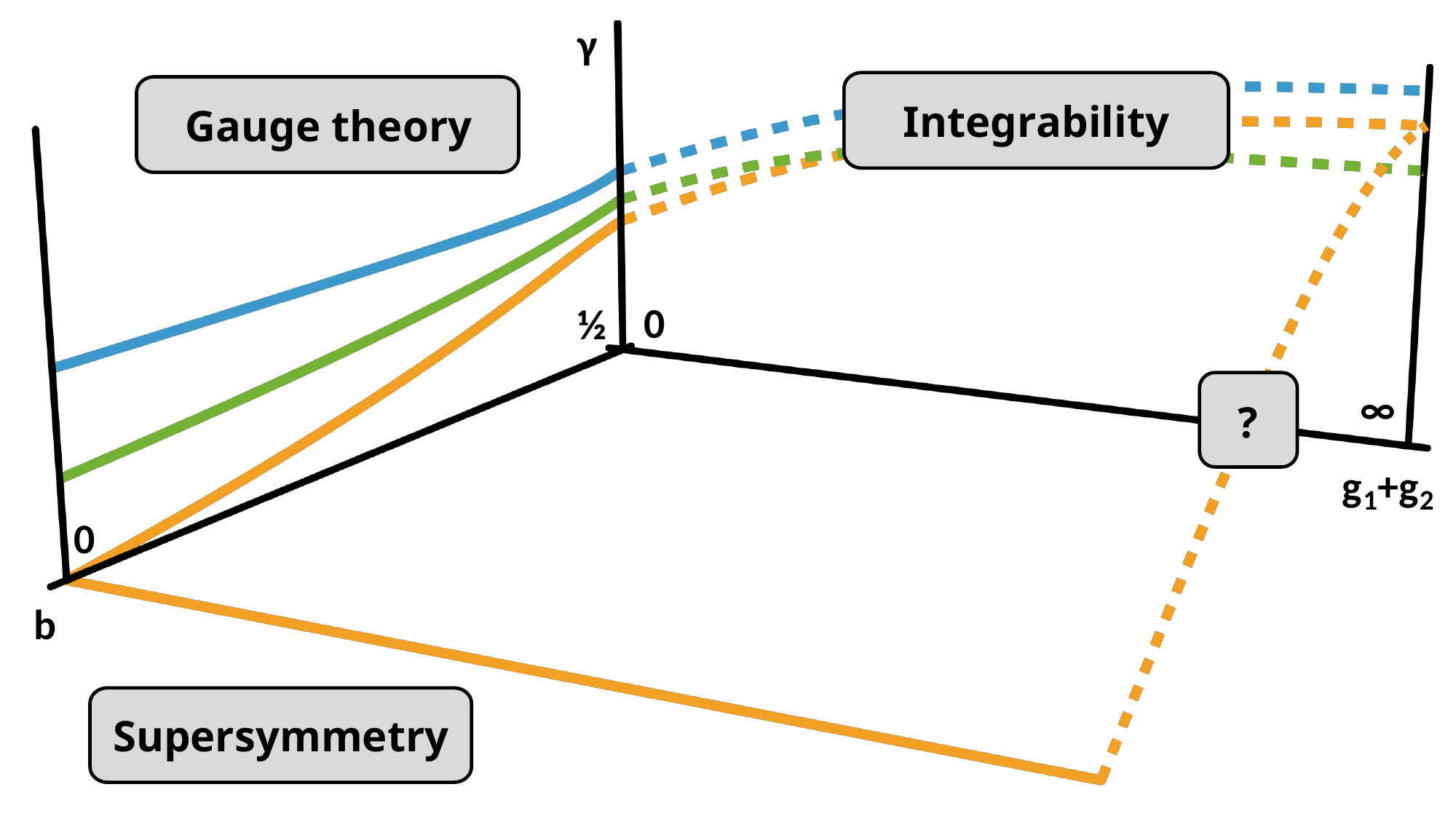}
\caption{\label{fig:idea} Behaviour of anomalous dimensions of a few operators over the conformal manifold spanned by $b$ and the overall coupling $g_1+g_2$. At small coupling we can explicitly diagonalise the one-loop Hamiltonian (see Figure \ref{fig:anomalousdimensions}). At the orbifold point $b=\tfrac{1}{2}$ the spectrum is accessible at any coupling via integrability techniques such as the quantum spectral curve, but since numerics are not available yet, we can only draw some schematic curves. In the decoupling limit $b=0$ some previously unprotected long multiplets saturate the BPS bound and become protected by supersymmetry, which is the effect described in this paper. An important task is to now identify the corresponding dual string theory states, which are best assessed at sufficiently large overall coupling $g_1+g_2=g_1\gg 0$\,.}
\end{figure}
\item The orbifold theory and its deformation can be studied in different duality frames. A T-dual type IIA description is given in terms of a Hanany--Witten construction involving two parallel NS5 branes localised on a circle and connected by stacks of D4 branes (see e.g. \cite{Giveon:1998sr} for a review). In this setup the deformation $b$ is geometrically realised as the separation between the NS5 branes. So far, however, the construction of the explicit background along the lines of \cite{Gaiotto:2009gz,Aharony:2012tz} is still an open problem.
\item As we approach the decoupling limit $b\to0$ the two NS5 branes approach each other in the type IIA picture. In absence of the D4 brane stack, the limit is well known to be represented by a linear dilaton model on the background 
\begin{equation} \label{eq: cigar}
 \mathbb{R}^{1,5} \times \frac{\text{SL}(2,\mathbb{R})_2}{\text{U}(1)} \Big / \mathbb{Z}_2 \ ,
\end{equation}
where the second component in the direct product goes under the name of \emph{cigar} background. In the large dilaton limit $\rho \to \infty$, \ie, in the long throat region of the cigar, the background \eqref{eq: cigar} is well approximated by the CHS background \cite{Callan:1991at}
\begin{equation} \label{eq: braneless}
 \mathbb{R}^{1,5} \times \mathbb{R}_{\rho} \times \text{SU}(2)_2\ .
\end{equation}
In line with this observation, \cite{Gadde:2009dj} argued that the holographic string dual of SCQCD in the Veneziano limit should be given by a similar non-critical string model\footnote{Non-critical w.r.t. the target space dimension, but still a critical worldsheet CFT.} after backreaction of the D4 brane stack (D3/D5 in a -dual type IIB frame). Recent works exploring this direction are \cite{Dei:2024frl,Dei:2025ilx}, but the full model is yet to be pinned down completely.

\item As already highlighted in \cite{Gadde:2009dj}, in the brane-less background \eqref{eq: braneless} it is already possible to find states dual to the ones belonging to the BPS multiplet inherited from the orbifold point (noticeably, the graviton and its KK modes). In the spirit of this work, one would be interested in finding the cigar states dual to the extra states identified in section \ref{ssec: scqcdhamiltonian} (see table \ref{tab: results}). Although a worldsheet approach may currently be out of reach, due to the unknown backreaction of the stack of $D4$ branes on the background \eqref{eq: cigar} \cite{Dei:2024frl}, it would already be interesting to study the possibility of a matching purely from a representation theory perspective. 

\item Another technical open question, already anticipated in the previous section, concerns the construction of the complete action of the one-loop (often referred to in the literature as $1/2$-loop) corrected supercharges. Determining their explicit form is necessary to fully fix the structure of the multiplets and the recombination of long multiplets, without relying on spectral arguments. To extend their action beyond the closed subsector studied in \cite{Liendo:2011xb}, one should formulate a general Ansatz for the action of the supercharges on all fields in the theory and impose closure of the algebra. Additional checks can be performed by acting with the supercharges on known protected multiplets and requiring that the resulting states remain BPS. While conceptually straightforward, this strategy involves several technical subtleties. We nevertheless hope to address these issues in future work.

\item Another tantalising question concerns the possible extension of our analysis to theories with further reduced supersymmetry. Although the recombination rules analysed in this paper pertains to the $\mathcal{N}=2$ superconformal algebra, similar mechanisms also arise in theories with $\mathcal{N}=1$ supersymmetry (see for example \cite{Pomoni:2019oib, Bourton:2020rfo} for the precise definition of the different multiplets),
\begin{equation}
\begin{split}
&\mathcal{A}^{\Delta=2+2j_1-\frac{3}{2}r}_{r<\frac{2}{3}(j_1-j_2),(j_1,j_2)} =\mathcal{C}_{r,(j_1,j_2)}\oplus\mathcal{C}_{r-1,(j_1-\frac{1}{2},j_2)}\,, \\
& \mathcal{A}^{\Delta = 2+2j_2+\frac{3}{2}r}_{r>\frac{2}{3}(j_1-j_2),(j_1,j_2)}=\overline{\mathcal{C}}_{r,(j_1,j_2)}\oplus\overline{\mathcal{C}}_{r+1,(j_1,j_2-\frac{1}{2})}\,,\\
&\mathcal{A}^{\Delta=2+j_1+j_2}_{\frac{2}{3}(j_1-j_2),(j_1,j_2)}=\hat{\mathcal{C}}_{(j_1,j_2)}\oplus\mathcal{C}_{\frac{2}{3}(j_1-j_2)-1,(j_1-\frac{1}{2},j_2)}\oplus\overline{\mathcal{C}}_{\frac{2}{3}(j_1-j_2)+1,(j_1,j_2-\frac{1}{2})}\,,\label{eqn:recomb2}
\end{split}
\end{equation}
or even without supersymmetry
\begin{equation}
\begin{split}
&\mathcal{A}^{\Delta = j + \bar{j} +2 }_{(j,\bar{j})} 
= \mathcal{C}_{(j,\bar{j})} \oplus \mathcal{A}^{\Delta = j + \bar{j} +3 }_{(j - \frac{1}{2},\bar{j}- \frac{1}{2})}, \quad \mathcal{A}^{\Delta = j + 1 }_{(j,\bar{j})} 
= \mathcal{B}^L_j \oplus \mathcal{C}_{(j - \frac{1}{2} , \frac{1}{2})} ,\\
&\mathcal{A}^{\Delta = \bar{j} + 1 }_{(j,\bar{j})} 
= \mathcal{B}^R_{\bar{j} } \oplus \mathcal{C}_{( \frac{1}{2}, j - \frac{1}{2} )}, \quad \mathcal{A}^{\Delta = 1 }_{(0,0)} 
= \mathcal{B} \oplus \mathcal{A}^{\Delta = 3 }_{(0,0)} \, .
\end{split}
\end{equation}
Unfortunately, both the existence of a conformal manifold and the techniques employed to demonstrate the mechanism under consideration rely crucially on supersymmetry. Consequently, extending this analysis to non-supersymmetric conformal field theories currently appears to be out of reach. However, the $\mathcal{N}=1$ case appears particularly promising. In particular, we plan to investigate the decoupling limit of the well-known $\mathbb{Z}_2 \times \mathbb{Z}_2$ orbifold of $\mathcal{N}=4$ SYM. This theory features four gauge couplings parameterising the conformal manifold.\footnote{A fifth parameter is supplied by a deformation of the superpotential.} By sending three of them to zero, one obtains $\mathcal{N}=1$ SCQCD at the edge of the conformal window. The techniques developed for the $\mathcal{N}=2$ case can also be adapted to the $\mathcal{N}=1$ setting, potentially allowing us to explore the corresponding string dual, which for $\mathcal{N}=1$ is comparatively less understood.  We will investigate this mechanism in detail in a forthcoming paper.
\end{itemize}

\paragraph{Acknowledgments}
We are particularly grateful to Florent Baume, Pieter Bomans, José Calderón Infante, Andrea Dei, Pietro Ferrero, Craig Lawrie, Volker Schomerus, Arkady A. Tseytlin, and Max Wiesner for valuable discussions on this work. EP is supported by ERC-2021-CoG - BrokenSymmetries 101044226. 
 TW is supported in part by Deutsche Forschungsgemeinschaft through a German-Israeli Project Cooperation (DIP) grant “Holography and the Swampland”.
 This work is funded in part by Deutsche Forschungsgemeinschaft under Germany’s Excellence Strategy – EXC 2121 Quantum Universe – 390833306 and by Deutsche Forschungsgemeinschaft under SFB 1624 – “Higher structures, moduli spaces and integrability” –506632645. 

\pagebreak
\appendix

\section{$\mathcal{N}=2$ superconformal multiplets}
\label{App:Dolan-Osborn}

In this appendix we review some basic facts regarding the superconformal algebra in $4$d $\mathcal{N}=2$ theories, following the conventions of \cite{Dolan:2002zh}. A generic long multiplet is generated by the action of the 8 Poincar\'e supercharges $\Qm$ and $\tilde \Qm$ on a superconformal primary, a state annihilated by all the superconformal charges $\Sm$ and $\tilde \Sm$, and is denoted as $\Am^{\Delta}_{R,r,(j,\bar{\jmath})}$. It can happen that some additional supercharge $\Qm$ (or a linear combination of them) annihilates the primary as well, generating a shorter representation. If this happens the conformal dimension of each state in the multiplet is protected against quantum corrections. Multiple such shortenings can happen at the same time. All the possible shortening conditions have been investigated in \cite{Dolan:2002zh}, and we summarise their findings in Table \ref{Tab:shortening}. In the first two columns we list all the types of shortening conditions, denoting them with a letter and a label $\Im =1,2$, indicating the supercharge killing the primary state of the short multiplet. In the two middle columns we list the protected quantum numbers of the highest weight state and in the last column we write the multiplet to which this highest weight state belongs. A multiplet denoted as $\Xm_{R,r(j, \bar{\jmath})}$ will have a highest weight state obeying the shortening condition $\Xm$, with $SU(2)_R$-charge $R$, $U(1)_r$-charge $r$, and Lorentz spin $(j,\bar{\jmath})$.

\begin{table}
\begin{centering}
\setlength{\tabcolsep}{4pt} 
\renewcommand{\arraystretch}{1.2} 
\begin{tabular}{|c|l|l|l|l|}
\hline 
\multicolumn{4}{|c|}{Shortening Conditions} & Multiplet\tabularnewline
\hline
\hline 
$\Bm_{1}$ & $\Qm_{\alpha}^{1}|R,r\rangle^{\text{h.w.}}=0$ & $j=0$ & $\Delta=2R+r$ & $\Bm_{R,r(0,\bar{\jmath})}$\tabularnewline
\hline 
$\bar{\Bm}_{2}$ & $\bar{\Qm}_{2 \dot{\alpha}}|R,r\rangle^{\text{h.w.}}=0$ & $\bar{\jmath}=0$ & $\Delta=2R-r$ & $\bar{\Bm}_{R,r(j,0)}$\tabularnewline
\hline 
$\Em$ & $\Bm_{1}\cap\Bm_{2}$ & $R=0$ & $\Delta=r$ & $\Em_{r(0,\bar{\jmath})}$\tabularnewline
\hline 
$\bar \Em$ & $\bar \Bm_{1}\cap \bar \Bm_{2}$ & $R=0$ & $\Delta=-r$ & $\bar \Em_{r(j,0)}$\tabularnewline
\hline 
$\hat{\Bm}$ & $\Bm_{1}\cap\bar{B}_{2}$ & $r=0$, $j,\bar{\jmath}=0$ & $\Delta=2R$ & $\hat{\Bm}_{R}$\tabularnewline
\hline
\hline 
$\Cm_{1}$ & $\epsilon^{\alpha\beta}\Qm_{\beta}^{1}|R,r\rangle_{\alpha}^{\text{h.w.}}=0$ & & $\Delta=2+2j+2R+r$ & $\Cm_{R,r(j,\bar{\jmath})}$\tabularnewline
 & $(\Qm^{1})^{2}|R,r\rangle^{\text{h.w.}}=0$ for $j=0$ & & $\Delta=2+2R+r$ & $\Cm_{R,r(0,\bar{\jmath})}$\tabularnewline
\hline 
$\bar \Cm_{2}$ & $\epsilon^{\dot\alpha\dot\beta}\bar\Qm_{2\dot\beta}|R,r\rangle_{\dot\alpha}^{\text{h.w.}}=0$ & & $\Delta=2+2\bar{\jmath}+2R-r$ & $\bar\Cm_{R,r(j,\bar{\jmath})}$\tabularnewline
 & $(\bar\Qm_{2})^{2}|R,r\rangle^{\text{h.w.}}=0$ for $\bar{\jmath}=0$ & & $\Delta=2+2R-r$ & $\bar\Cm_{R,r(j,0)}$\tabularnewline
\hline 
$\mathcal{F}$ & $\Cm_{1}\cap\Cm_{2}$ & $R=0$ & $\Delta=2+2j+r$ & $\Cm_{0,r(j,\bar{\jmath})}$\tabularnewline
\hline 
$\bar{\mathcal{F}}$ & $\bar\Cm_{1}\cap\bar\Cm_{2}$ & $R=0$ & $\Delta=2+2\bar{\jmath}-r$ & $\bar\Cm_{0,r(j,\bar{\jmath})}$\tabularnewline
\hline 
$\hat{\Cm}$ & $\Cm_{1}\cap\bar{\Cm}_{2}$ & $r=\bar{\jmath}-j$ & $\Delta=2+2R+j+\bar{\jmath}$ & $\hat{\Cm}_{R(j,\bar{\jmath})}$\tabularnewline
\hline 
$\hat{\mathcal{F}}$ & $\Cm_{1}\cap\Cm_{2}\cap\bar{\Cm}_{1}\cap\bar{\Cm}_{2}$ & $R=0, r=\bar{\jmath}-j$ & $\Delta=2+j+\bar{\jmath}$ & $\hat{\Cm}_{0(j,\bar{\jmath})}$\tabularnewline
\hline
\hline 
$\mathcal D$ & $\Bm_{1}\cap\bar{\Cm_{2}}$ & $r=\bar{\jmath}+1$ & $\Delta=1+2R+\bar{\jmath}$ & $\mathcal{D}_{R(0,\bar{\jmath})}$\tabularnewline
\hline 
$\bar{\mathcal{D}}$ & $\bar\Bm_{2}\cap{\Cm_{1}}$ & $-r=j+1$ & $\Delta=1+2R+j$ & $\bar{\mathcal{D}}_{R(j,0)}$\tabularnewline
\hline 
$\mathcal{G}$ & $\Em\cap\bar{\Cm_{2}}$ & $r=\bar{\jmath}+1,R=0$ & $\Delta=r=1+\bar{\jmath}$ & $\mathcal{D}_{0(0,\bar{\jmath})}$\tabularnewline
\hline
$\bar{\mathcal{G}}$ & $\bar\Em\cap{\Cm_{1}}$ & $-r=j+1,R=0$ & $\Delta=-r=1+j$ & $\bar{\mathcal{D}}_{0(j,0)}$\tabularnewline
\hline
\end{tabular}
\par\end{centering}
\caption{\label{Tab:shortening}Shortening conditions
and short multiplets ($\Nm=2$ superconformal algebra) \cite{Dolan:2002zh}.}
\end{table}

\begin{figure}
 \centering
 \includegraphics[width=1\linewidth]{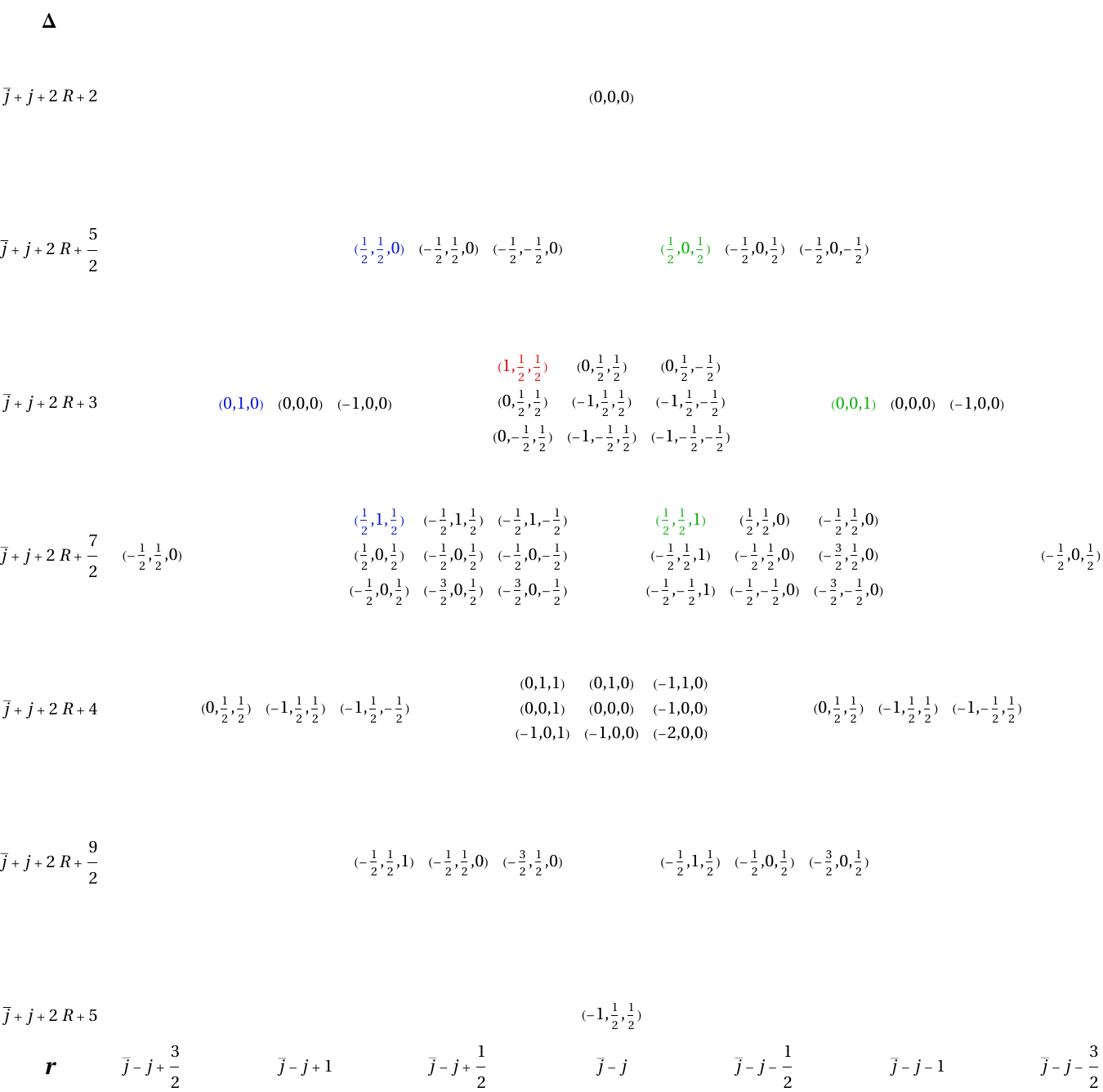}
 \definecolor{mygreen}{rgb}{0.05, 0.7, 0.05}
 \caption{States in a generic $\hat \Cm_{R(j,\bar{\jmath})}$ multiplet. We highlighted in \textcolor{blue}{blue} the contributions to the left index, in \textcolor{mygreen}{green} the ones to the right index and in \textcolor{red}{red} the ones contributing to both the indices. Only the shifts $(\delta R, \delta j, \delta \bar{\jmath})$ are reported.}
 \label{fig:CH_Index}
\end{figure}

\begin{figure}
 \centering
 \includegraphics[width=1\linewidth]{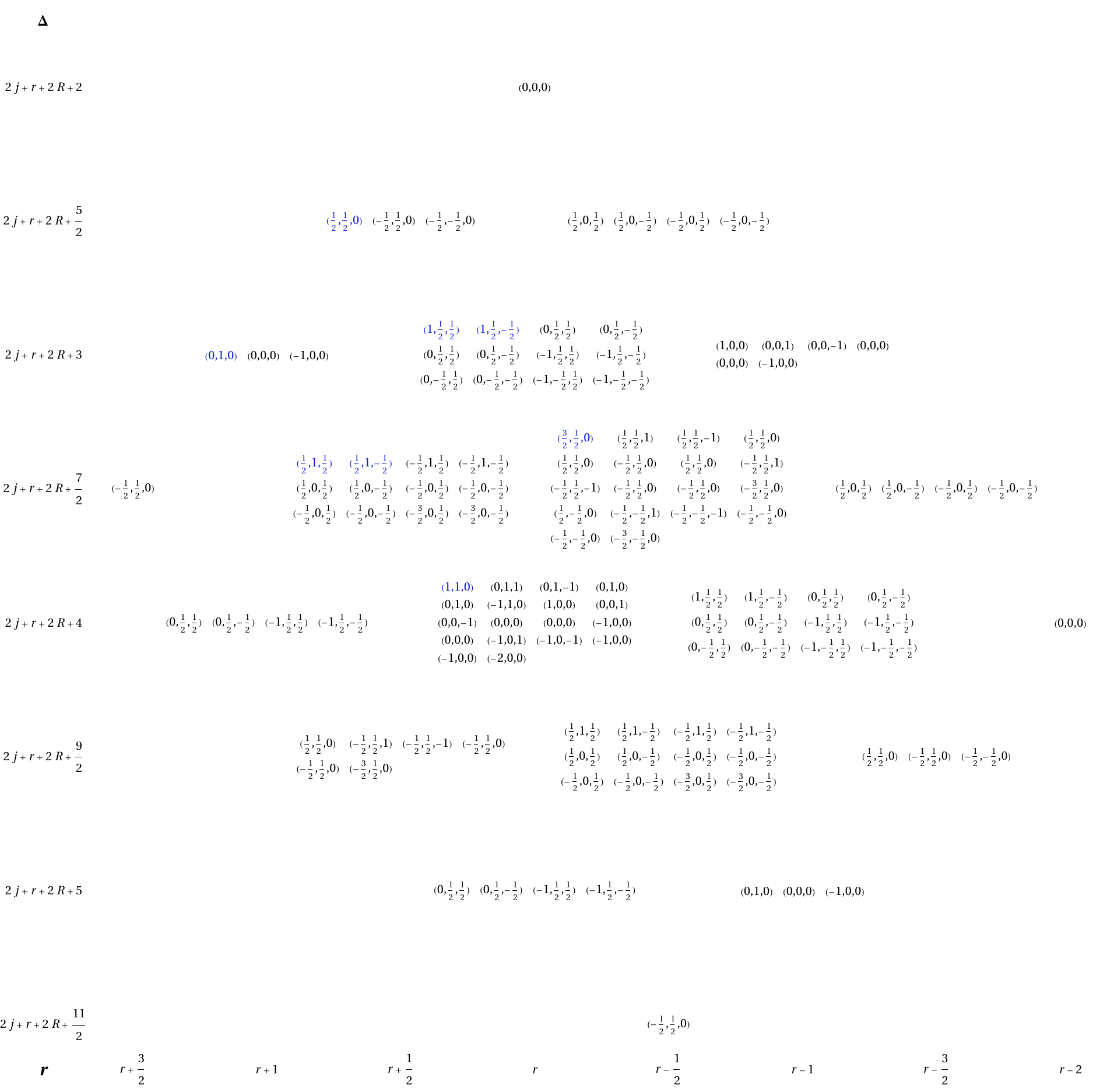}
 \caption{States in a generic $ \Cm_{R,r(j,\bar{\jmath})}$ multiplet. We highlighted in \textcolor{blue}{blue} the contributions to the left index. Only the shifts $(\delta R, \delta j, \delta \bar{\jmath})$ are reported.}
 \label{fig:C_Index}
\end{figure}

To be completely explicit, the content of generic $\hat{\mathcal C}_{R(j,\bar{\jmath})}$ and $\mathcal C_{R,r(j,\bar{\jmath})}$ multiplets is displayed in Figures \ref{fig:CH_Index} and \ref{fig:C_Index}. The notation used in the figures is as follows. A generic state in a multiplet $\mathcal X^{\Delta}_{R,r(j,\bar{\jmath})}$, following the conventions of~\cite{Dolan:2002zh}, is denoted by $(R+\delta R)_{(j+\delta j,\;\bar{\jmath}+\delta \bar{\jmath})}$ and placed on a grid, where the horizontal position denotes the $r$-charge and the vertical position the scaling dimension~$\Delta$. For brevity’s sake, in the figures we report only the three numbers $(\delta R,\delta j,\delta \bar{\jmath})$.
For completeness, we also collect here the recombination rules for long multiplet hitting the unitary bound (see Section \ref{recombination})
\begin{align}
 & \mathcal{A}^{2R + r +2j +2}_{R,r(j,\bar{\jmath})} \simeq \mathcal{C}_{R,r(j,\bar{\jmath})} \oplus \mathcal{C}_{R+\frac{1}{2},r+\frac{1}{2}(j-\frac{1}{2},\bar{\jmath})}\,,  \label{Eq:Recombination_Rules1} \\
 & \mathcal{A}^{2R - r +2\bar{\jmath} +2}_{R,r(j,\bar{\jmath})} \simeq \bar{\mathcal{C}}_{R,r(j,\bar{\jmath})} \oplus \bar{\mathcal{C}}_{R+\frac{1}{2},r-\frac{1}{2}(j,\bar{\jmath}-\frac{1}{2})} \,,  \label{Eq:Recombination_Rules2}\\
 & \mathcal{A}^{2R + j + \bar{\jmath} +2}_{R,j-\bar{\jmath}(j,\bar{\jmath})} \simeq \hat{\mathcal{C}}_{R(j,\bar{\jmath})} \oplus \hat{\mathcal{C}}_{R+\frac{1}{2}(j-\frac{1}{2},\bar{\jmath})} \oplus \hat{\mathcal{C}}_{R+\frac{1}{2}(j,\bar{\jmath}-\frac{1}{2})} \oplus \hat{\mathcal{C}}_{R+1(j-\frac{1}{2},\bar{\jmath}-\frac{1}{2})}\,.
 \label{Eq:Recombination_Rules3}
\end{align}
Once we consider the following identifications, formally allowing for the $j,\bar{\jmath}$ quantum numbers in the $\Cm/\hat{\Cm}$ multiplet to take also the value $-\tfrac{1}{2}$,
\begin{equation}
\begin{split}
& \Cm_{R,r(-\frac{1}{2},\bar{\jmath})}\simeq\Bm_{R+\frac{1}{2},r+\frac{1}{2}(0,\bar{\jmath})}, \\
& \hat{\Cm}_{R(-\frac{1}{2},\bar{\jmath})}\simeq \mathcal{D}_{R+\frac{1}{2}(0,\bar{\jmath})},\qquad\hat{\Cm}_{R(j,-\frac{1}{2})}\simeq\bar{\mathcal{D}}_{R+\frac{1}{2}(j,0)}\, , \\
&\hat{\Cm}_{R(-\frac{1}{2},-\frac{1}{2})}\simeq \mathcal{D}_{R+\frac{1}{2}(0,-\frac{1}{2})} \simeq
\bar{\mathcal{D}}_{R+\frac{1}{2}(-\frac{1}{2},0)} \simeq \hat{\Bm}_{R+1},
\label{translation}
\end{split}
\end{equation}
\begin{sidewaysfigure}
 \centering
 \includegraphics[width=0.8\linewidth]{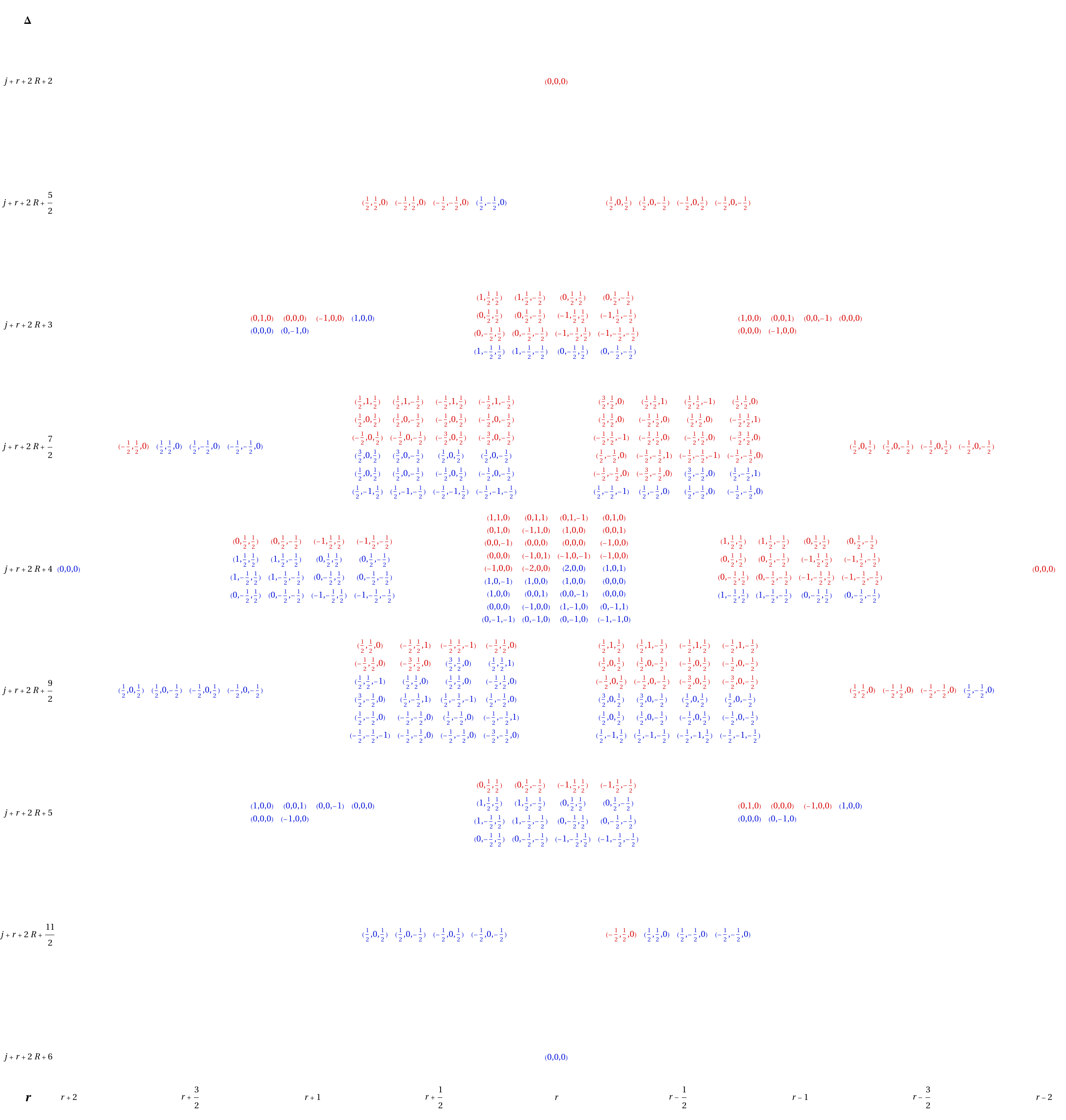}
 \definecolor{myred}{rgb}{0.85,0.05,0.05}
 \definecolor{myblue}{rgb}{0.05, 0.1, 0.85}
 \caption{Multiplet recombination following the rule $\mathcal{A}^{2R + r +2j +2}_{R,r(j,\bar{\jmath})} \simeq {\color{myred}\mathcal{C}_{R,r(j,\bar{\jmath})} }\oplus { \color{myblue} \mathcal{C}_{R+\frac{1}{2},r+\frac{1}{2}(j-\frac{1}{2},\bar{\jmath})}}$.}
 \label{fig:AC_Breaking}
\end{sidewaysfigure}

\begin{sidewaysfigure}
 \centering
 \includegraphics[width=0.8\linewidth]{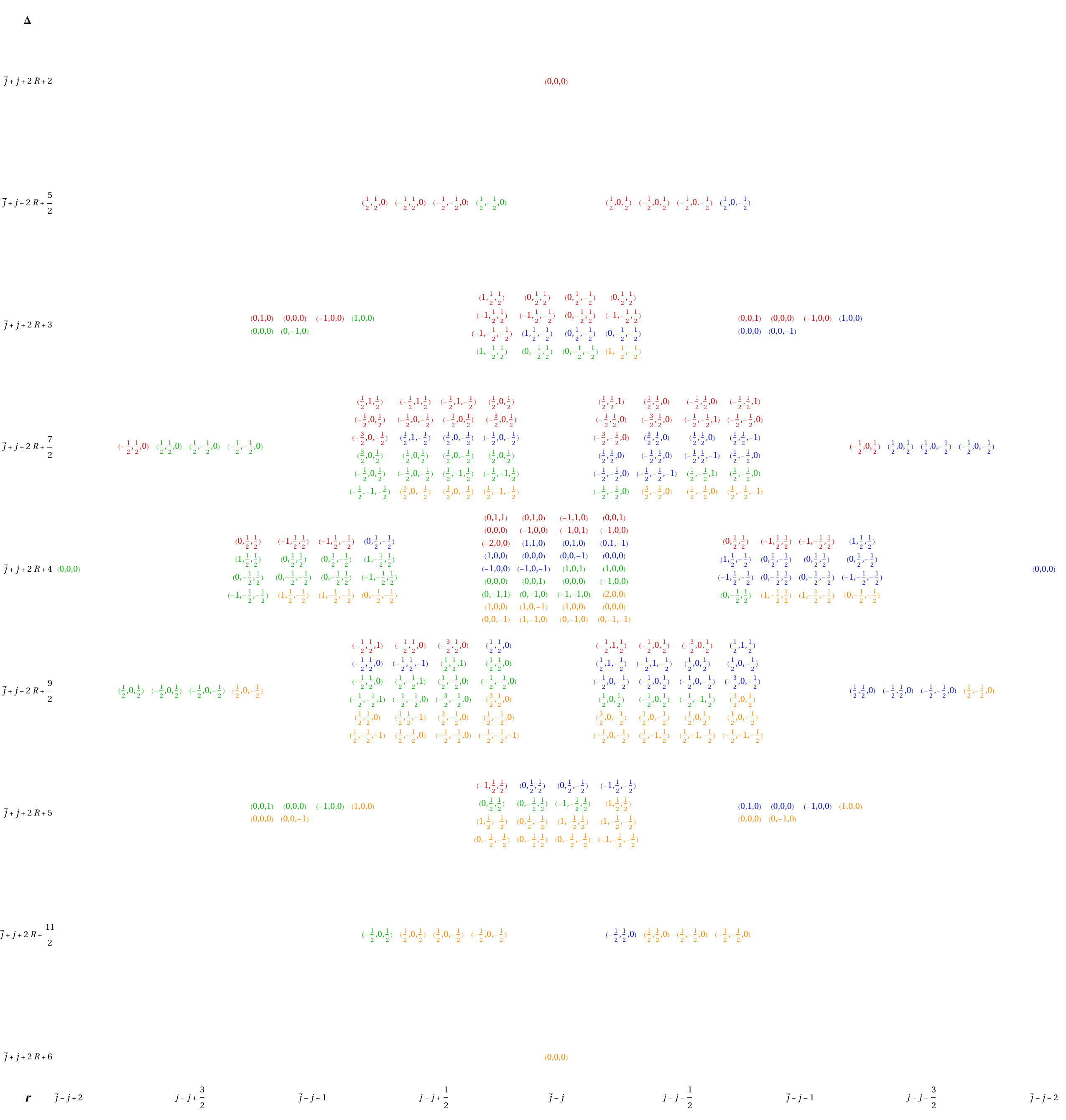}
 \definecolor{myred}{rgb}{0.85,0.05,0.05}
 \definecolor{myblue}{rgb}{0.05, 0.1, 0.85}
 \definecolor{mygreen}{rgb}{0.05, 0.7, 0.05}
 \definecolor{myorange}{rgb}{1.0, 0.55, 0.0}
 \caption{Multiplet recombination following the rule $\mathcal{A}^{2R + j + \bar{\jmath} +2}_{R,j-\bar{\jmath}(j,\bar{\jmath})} \simeq {\color{myred} \hat{\mathcal{C}}_{R(j,\bar{\jmath})}} \oplus {\color{mygreen}\hat{\mathcal{C}}_{R+\frac{1}{2}(j-\frac{1}{2},\bar{\jmath})}} \oplus {\color{myblue}\hat{\mathcal{C}}_{R+\frac{1}{2}(j,\bar{\jmath}-\frac{1}{2})}} \oplus {\color{myorange}\hat{\mathcal{C}}_{R+1(j-\frac{1}{2},\bar{\jmath}-\frac{1}{2})}}$.}
 \label{fig:ACH_Breaking}
\end{sidewaysfigure}
the formulas \eqref{Eq:Recombination_Rules1}-\eqref{Eq:Recombination_Rules3} describe the breaking for all the possible short multiplets. We give a pictorial representation of the breakings \eqref{Eq:Recombination_Rules1} and \eqref{Eq:Recombination_Rules3} in Figures \ref{fig:AC_Breaking} and \ref{fig:ACH_Breaking}, respectively. We employ the same conventions as before.

\section{Oscillator representation and harmonic action}\label{app: Harmonic}

We follow Appendix B in \cite{Liendo:2011xb} and introduce an oscillator representation of the $\Nm=2$ superconformal algebra $SU(2,2|2)$ in the free ($g_1=g_2=0$) theory. The letters composing the various multiplets of the free theory may be expressed in terms of two sets of bosonic oscillators $(\ab^\a,\ab^\dag_\a)$, $(\bb^{\ad},\bb^\dag_{\ad})$ and one
set of fermionic oscillators
$(\cb^{\Im},\cb^\dag_{\Im})$, where $(\a,\ad)$ are Lorentz indices and $\Im$ is an $SU(2)_R$ index. To distinguish the four multiplets at play, we need to introduce two auxiliary fermionic operators $(\db,\db^{\dag})$, $(\tilde{\db},\tilde{\db}^{\dag})$. These correspond to additional $(\cb,\cb^\dag)$ operators in $\mathcal{N}=4$ SYM which are broken by the orbifold projection. 
The non-zero (anti)commutation relations are
\begin{align} 
[\ab^\a,\ab^\dag_\b] = \delta^\a_\b\,,\qquad 
\lbr \bb^{\ad},\bb^\dag_{\bd} \rbr = \delta^{\ad}_{\bd}\, \qquad \{ \cb^{\Im},\cb^\dag_{\Jm} \} = \delta^{\Im}_{\Jm}\,,\qquad 
\{ \db,\db^\dag \} = \{ \tilde{\db},\tilde{\db}^\dag \} = 1\, .
\end{align}
We also need to add one bosonic twist operator $\gamma$ satisfying $\gamma^2=\mathbb{1}$, which distinguishes the gauge representations of fields. This operator was kept implicit in \cite{Liendo:2011xb}, but has been introduced in \cite{Bertle:2024djm} in the context of a twisted co-product. The generators of $SU(2,2|2)$ may be expressed in terms of these oscillators
\begin{equation}\begin{split}
\Qm^{\ph{k}\Im}_{\a} &= \ab^{\dag}_{\a}\cb^{\Im}\, ,
\qquad
 \Sm^{\ph{k}\a}_{\Im} = \cb^{\dag}_{\Im}\ab^{\a}\, ,
 \qquad
\tilde{\Qm}_{\ad\Im} = \bb^{\dag}_{\ad}\cb_{\Im}^{\dag}\, ,
\qquad 
\tilde{\Sm}^{\ad\Im} = \bb^{\ad}\cb^{\Im}\, ,
\\
\Pm_{\a \bd} &= \ab^{\dag}_{\a} \bb^{\dag}_{\bd}\, ,
\qquad
 \Km^{\a \bd} = \ab^\a \bb^{\bd}\, ,\qquad
\mathcal{D} = 1 + \frac{1}{2}\ab_{\g}^{\dag}\ab^\g + \frac{1}{2}\bb_{\gd}^{\dag}\bb^{\gd}\, ,
\\
\Lm^{\ph{\b}\a}_{\b} &= \ab^\dag_\b\ab^\a-\frac{1}{2}\delta^\a_\b \ab^\dag_\g\ab^\g\, ,
\qquad
\dot{\Lm}^{\ph{\bd}\ad}_{\bd} = \bb^\dag_{\bd}\bb^{\ad}-\frac{1}{2}\delta^{\ad}_{\bd} \bb^\dag_{\gd}\bb^{\gd}\, ,
\\
\Rm^{\ph{\Jm}\Im}_{\Jm} &= \cb^\dag_{\Jm}\cb^{\Im}-\frac{1}{2}\delta^{\Im}_{\Jm} \cb^\dag_{\Km}\cb^{\Km}\, ,
\qquad
r = -\frac{1}{2}\cb^\dag_{\Km}\cb^{\Km}+\frac{1}{2}\db^\dag\db+\frac{1}{2}\tilde{\db}^\dag \tilde{\db}\, ,
\\
C &= 1 - \frac{1}{2}\ab_{\g}^{\dag}\ab^\g + \frac{1}{2}\bb_{\gd}^{\dag}\bb^{\gd}-\frac{1}{2}\cb^\dag_{\Km}\cb^{\Km}
-\frac{1}{2}\db^\dag\db-\frac{1}{2}\tilde{\db}^\dag \tilde{\db}\,.
\end{split}\end{equation}
Here $C$ is a central charge that must annihilate any physical state. The global $SU(2)_L$ charge is captured by the operator
\begin{equation}
 Q_L=\left(\db^\dag\db-\tilde{\db}^\dag \tilde{\db}\right)\,.
\end{equation}
We define a vacuum state $|0\rangle$ annihilated by all the lowering operators and list the various letters that may be used to construct operators. Here we drop the Lorentz and $R$-symmetry indices for convenience, but indicate the $SU(2)_L$-charge on the hypermultiplets. 

\paragraph{Vector multiplets.}
The two chiral halves $\Vm$ and $\Vmb$ comprise the following letters and their oscillator identifications:
\begin{align} 
D^k \Fm &= (\ab^{\dag})^{k+2} (\bb^{\dag})^k (\cb^{\dag})^0 |0\rangle\, ,
\\
D^k\lambda &= (\ab^{\dag})^{k+1} (\bb^{\dag})^k (\cb^{\dag})^1 |0\rangle\, ,
\\
D^k \phi &= (\ab^{\dag})^{k\ph{+0}} (\bb^{\dag})^k (\cb^{\dag})^2 |0\rangle\, ,
\end{align}
and
\begin{align} 
D^k \bar{\Fm} &= (\ab^{\dag})^{k} (\bb^{\dag})^{k+2} (\cb^{\dag})^2 \db^{\dag} \tilde{\db}^{\dag} |0\rangle\, ,
\\
D^k\bar{\lambda} &= (\ab^{\dag})^{k} (\bb^{\dag})^{k+1} (\cb^{\dag})^1 \db^{\dag} \tilde{\db}^{\dag} |0\rangle\, ,
\\
D^k \bar{\phi} &= (\ab^{\dag})^{k} (\bb^{\dag})^{k\ph{+0}} (\cb^{\dag})^0\db^{\dag} \tilde{\db}^{\dag} |0\rangle\,.
\end{align}
The same identification applies to the second vector multiplet $\Vmc$ and $\Vmcb$, but we multiply by the twist operator $\gamma$ 
\begin{align} 
D^k \check{\Fm} &= (\ab^{\dag})^{k+2} (\bb^{\dag})^k (\cb^{\dag})^0 \gamma|0\rangle\, ,
\\
D^k\check{\lambda} &= (\ab^{\dag})^{k+1} (\bb^{\dag})^k (\cb^{\dag})^1 \gamma|0\rangle\, ,
\\
D^k \check{\phi} &= (\ab^{\dag})^{k\ph{+0}} (\bb^{\dag})^k (\cb^{\dag})^2 \gamma|0\rangle\, ,
\end{align}
and
\begin{align} 
D^k \check{\bar{\Fm}} &= (\ab^{\dag})^{k} (\bb^{\dag})^{k+2} (\cb^{\dag})^2 \db^{\dag} \tilde{\db}^{\dag}\gamma |0\rangle\, ,
\\
D^k\check{\bar{\lambda}} &= (\ab^{\dag})^{k} (\bb^{\dag})^{k+1} (\cb^{\dag})^1 \db^{\dag} \tilde{\db}^{\dag}\gamma |0\rangle\, ,
\\
D^k \check{\bar{\phi}}&= (\ab^{\dag})^{k} (\bb^{\dag})^{k\ph{+0}} (\cb^{\dag})^0\db^{\dag} \tilde{\db}^{\dag}\gamma |0\rangle\,.
\end{align}

\paragraph{Hypermultiplets.} We again split the multiplets into two halves, \eg~$\Hm_+$ and $\Hmb_+$ and identify the oscillator representations 
\begin{align} 
D^k Q_+ &= (\ab^{\dag})^{k} (\bb^{\dag})^k (\cb^{\dag})^1 \db^{\dag}|0\rangle\, ,
\\
D^k \psi_+ &= (\ab^{\dag})^{k+1} (\bb^{\dag})^k \db^{\dag} |0\rangle\, ,
\\D^k \bar{\tilde{\psi}}_+ &= (\ab^{\dag})^{k} (\bb^{\dag})^{k+1} (\cb^{\dag})^2 \db^{\dag} |0\rangle\,,
\end{align}
and 
\begin{align} 
D^k \bar{Q}_+ &= (\ab^{\dag})^{k} (\bb^{\dag})^k (\cb^{\dag})^1 \db^{\dag}\gamma|0\rangle\, ,
\\
D^k \tilde{\psi}_+ &= (\ab^{\dag})^{k+1} (\bb^{\dag})^k \db\gamma |0\rangle\, ,
\\
D^k \bar{\psi}_+ &= (\ab^{\dag})^{k} (\bb^{\dag})^{k+1} (\cb^{\dag})^2\db^{\dag} \gamma |0\rangle\, .
\end{align}
The $SU(2)_L$ conjugate multiplet ($\Hm_-$, $\Hmb_-$) is similarly identified as 
\begin{align} 
D^k Q_- &= (\ab^{\dag})^{k} (\bb^{\dag})^k (\cb^{\dag})^1 \tilde{\db}^{\dag}|0\rangle\, ,
\\
D^k \psi_- &= (\ab^{\dag})^{k+1} (\bb^{\dag})^k \tilde{\db}^{\dag} |0\rangle\, ,
\\D^k \bar{\tilde{\psi}}_- &= (\ab^{\dag})^{k} (\bb^{\dag})^{k+1} (\cb^{\dag})^2 \tilde{\db}^{\dag} |0\rangle\,,
\end{align}
and 
\begin{align} 
D^k \bar{Q}_- &= (\ab^{\dag})^{k} (\bb^{\dag})^k (\cb^{\dag})^1 \tilde{\db}^{\dag}\gamma|0\rangle\, ,
\\
D^k \tilde{\psi}_- &= (\ab^{\dag})^{k+1} (\bb^{\dag})^k \tilde{\db}^{\dag} \gamma|0\rangle\, ,
\\
D^k \bar{\psi}_- &= (\ab^{\dag})^{k} (\bb^{\dag})^{k+1} (\cb^{\dag})^2\tilde{\db}^{\dag} \gamma |0\rangle\, .
\end{align}
We do not split these multiplets according to their chirality but instead according to their representation with respect to the gauge groups to which they can be coupled. $\Hm_{\pm}$ has gauge indices $(a, \check{a})$ and $\Hmb_{\pm}$ has gauge indices $(\check{a},a)$, allowing coupling to both ($\Vm,\Vmb$) and ($\Vmc,\Vmcb$), but in opposite representations. In the oscillator representation this is captured by the twist operator $\gamma$. When building single-trace operators, we chain together these letters and ensure gauge invariance by hand. For example, we may never place two $\Hm_{\pm}$ multiplets next to each other when we want to create a gauge-invariant single trace operator (see \cite{Bozkurt:2024tpz,Bertle:2024djm,Bozkurt:2025exl} for recent discussions on these subtleties). The one-loop Hamiltonian respects these constraints and never creates such combinations when acting on a gauge-invariant operator.

\paragraph{The harmonic action.} With these identifications in place we may now create single-trace operators by chaining together the letters introduced above. The one-loop Hamiltonian acts on neighbouring letters by shuffling the oscillators as observed in \cite{Beisert:2003jj, Liendo:2011xb}. Schematically,
\begin{equation}\label{eq: shuffle}
 H \ket{A}\otimes \ket{B}=\sum_{\sigma}c_\sigma\,\delta_{C\ket{C(\sigma)},0}\,\delta_{C\ket{D(\sigma)},0}\ket{C(\sigma)}\otimes \ket{D(\sigma)}\,,
\end{equation}
where $\sigma$ denotes all reshufflings of the oscillators and $\ket{C(\sigma)}$ and $ \ket{D(\sigma)}$ are the resulting words. The Kronecker-deltas ensure that the resulting state is physical, \ie, annihilated by the central charge $C$. The coefficients $c_\sigma$ depend on the specific content of $\ket{A}$ and $\ket{B}$, but may be factorised when splitting the $SU(2,2|2)$ oscillators $(\ab^\dag,\bb^\dag,\cb^\dag)$ from the auxiliary ones $(\db^\dag,\tilde{\db}^\dag,\check{\db}^\dag)$. At this stage we resign ourselves to simply stating the Hamiltonian for every valid combination of multiplets $(\Vm,\Vmb,\Vmc,\Vmcb,\Hm_{\pm},\Hmb_{\pm} )$.

We introduce coefficients 
$c_{n,n_{12},n_{21}}$ which depend on the total number $n$ of oscillators $(\ab^\dag,\bb^\dag,\cb^\dag)$ present in a pair of letters, as well as the numbers $n_{12}$ and $n_{21}$ of such oscillators moved from position one to two and vice versa
\begin{equation}
 c_{n,n_{12},n_{21}}=(-1)^{1+n_{12}n_{21}}\frac{\Gamma(\tfrac{1}{2}(n_{12}+n_{21}))\Gamma(1+\tfrac{1}{2}(n-n_{12}-n_{21}))}{\Gamma(1+\tfrac{1}{2}n)}\,,\qquad c_{n,0,0}=h(\tfrac{n}{2})\,,
\end{equation}
where $h(x)$ is the $x$\textsuperscript{th} harmonic number. We also remind the reader of the definition $\kappa=\tfrac{g_2}{g_1}$.

With these definitions at hand, we may now list the action of the Hamiltonian on the various combinations of multiplets. We will be schematic and assume a summation over $(\ab^\dag,\bb^\dag,\cb^\dag)$ oscillators as in \eqref{eq: shuffle}:
\begin{align}
 H \ket{\Vm}\otimes \ket{\Vm}=&c_{n,n_{12},n_{21}}\ket{\Vm}\otimes \ket{\Vm}\,,\\
 H \ket{\Vmb}\otimes \ket{\Vmb}=& c_{n,n_{12},n_{21}}\ket{\Vmb}\otimes \ket{\Vmb}\,,\\
 H \ket{\Vm}\otimes \ket{\Vmb}=& c_{n+2,n_{12},n_{21}}\ket{\Vm}\otimes \ket{\Vmb}+ c_{n+2,n_{12}+2,n_{21}}\ket{\Vmb}\otimes \ket{\Vm}\nonumber\\
 &+c_{n+2,n_{12}+1,n_{21}}\left(\ket{\Hm_+}\otimes \ket{\Hmb_-}-\ket{\Hm_-}\otimes \ket{\Hmb_+}\right)\,,\\
 H \ket{\Vmb}\otimes \ket{\Vm}=& c_{n+2,n_{12},n_{21}}\ket{\Vmb}\otimes \ket{\Vm}+ c_{n+2,n_{12},n_{21}+2}\ket{\Vm}\otimes \bar{\ket{\Vm}}\nonumber\\
 &+ c_{n+2,n_{12},n_{21}+1}\left(\ket{\Hm_+}\otimes \ket{\Hmb_-}-\ket{\Hm_-}\otimes \ket{\Hmb_+}\right)\,,\\
 H \ket{\Vmc}\otimes \ket{\Vmc}=& \kappa^2 c_{n,n_{12},n_{21}}\ket{\Vmc}\otimes \ket{\Vmc}\,,\\
 H \ket{\Vmcb}\otimes \ket{\Vmcb}=&\kappa^2 c_{n,n_{12},n_{21}}\ket{\Vmcb}\otimes \ket{\Vmcb}\,,\\
 H \ket{\Vmc}\otimes \ket{\Vmcb}=& \kappa^2c_{n+2,n_{12},n_{21}}\ket{\Vmc}\otimes \ket{\Vmcb}+ \kappa^2c_{n+2,n_{12}+2,n_{21}}\ket{\Vmcb}\otimes \ket{\Vmc}\nonumber\\
 &+\kappa^2c_{n+2,n_{12}+1,n_{21}}\left(\ket{\Hmb_+}\otimes \ket{\Hm_-}-\ket{\Hmb_-}\otimes \ket{\Hm_+}\right)\,,\\
 H \ket{\Vmcb}\otimes \ket{\Vmc}=& \kappa^2c_{n+2,n_{12},n_{21}}\ket{\Vmcb}\otimes \ket{\Vmc}+ \kappa^2c_{n+2,n_{12},n_{21}+2}\ket{\Vmc}\otimes \ket{\Vmcb}\nonumber\\
 &+\kappa^2c_{n+2,n_{12},n_{21}+1}\left(\ket{\Hmb_+}\otimes \ket{\Hm_-}-\ket{\Hmb_-}\otimes \ket{\Hm_+}\right)\,, \\
 H \ket{\Vm}\otimes \ket{\Hm_\pm}=&c_{n+1,n_{12},n_{21}}\ket{\Vm}\otimes \ket{\Hm_\pm}+ \kappa c_{n+1,n_{12},n_{21}+1}\ket{\Hm_\pm}\otimes \ket{\Vmc} \,,\\
 H \ket{\Vmb}\otimes \ket{\Hm_\pm}=&c_{n+1,n_{12},n_{21}}\ket{\Vmb}\otimes \ket{\Hm_\pm}+ \kappa c_{n+1,n_{12},n_{21}+1}\ket{\Hm_\pm}\otimes \ket{\Vmcb} \,,\\
 H \ket{\Vmc}\otimes \ket{\Hmb_\pm}=&\kappa^2c_{n+1,n_{12},n_{21}}\ket{\Vmc}\otimes \ket{\Hmb_\pm}+ \kappa c_{n+1,n_{12},n_{21}+1}\ket{\Hmb_\pm}\otimes \ket{\Vm} \,,\\
 H \ket{\Vmcb}\otimes \ket{\Hmb_\pm}=&\kappa^2c_{n+1,n_{12},n_{21}}\ket{\Vmcb}\otimes \ket{\Hmb_\pm}+ \kappa c_{n+1,n_{12},n_{21}+1}\ket{\Hmb_\pm}\otimes \ket{\Vmb} \,,\\
 H \ket{\Hmb_\pm}\otimes\ket{\Vm} =&c_{n+1,n_{12},n_{21}}\ket{\Hmb_\pm}\otimes\ket{\Vm}- \kappa c_{n+1,n_{12},n_{21}+1}\ket{\Vmc}\otimes\ket{\Hmb_\pm}\,,\\
 H \ket{\Hmb_\pm}\otimes\ket{\Vmb}=&c_{n+1,n_{12},n_{21}}\ket{\Hmb_\pm}\otimes\ket{\Vmb}- \kappa c_{n+1,n_{12},n_{21}+1}\ket{\Vmcb}\otimes\ket{\Hmb_\pm}\,,\\
 H \ket{\Hm_\pm}\otimes\ket{\Vmc}=&\kappa^2c_{n+1,n_{12},n_{21}}\ket{\Hm_\pm}\otimes\ket{\Vmc}- \kappa c_{n+1,n_{12},n_{21}+1}\ket{\Vm}\otimes\ket{\Hm_\pm} \,,\\
 H \ket{\Hm_\pm}\otimes\ket{\Vmcb}=&\kappa^2 c_{n+1,n_{12},n_{21}}\ket{\Hm_\pm}\otimes\ket{\Vmcb}- \kappa c_{n+1,n_{12},n_{21}+1}\ket{\Vmb}\otimes\ket{\Hm_\pm} \,,\\
 H \ket{\Hm_\pm}\otimes \ket{\Hmb_\pm}=&\kappa^2 c_{n,n_{12},n_{21}}\ket{\Hm_\pm}\otimes \ket{\Hmb_\pm}\,,\\
 H \ket{\Hm_\pm}\otimes \ket{\Hmb_\mp}=&(\kappa^2 c_{n,n_{12},n_{21}}-c_{n+2,n_{12}+2,n_{21}})\ket{\Hm_\pm}\otimes \ket{\Hmb_\mp}\nonumber\\
 &+c_{n+2,n_{12}+2,n_{21}}\ket{\Hm_\mp}\otimes \ket{\Hmb_\pm}\nonumber\\
 &\pm c_{n+2,n_{12},n_{21}+1}\ket{\Vm}\otimes\ket{\Vmb} \pm c_{n+2,n_{12}+1,n_{21}}\ket{\Vmb}\otimes\ket{\Vm}\,,\\
 H \ket{\Hmb_\pm}\otimes \ket{\Hm_\pm}=& c_{n,n_{12},n_{21}}\ket{\Hmb_\pm}\otimes \ket{\Hm_\pm}\,,\\
 H \ket{\Hmb_\pm}\otimes \ket{\Hm_\mp}=&( c_{n,n_{12},n_{21}}-\kappa^2c_{n+2,n_{12}+2,n_{21}})\ket{\Hmb_\pm}\otimes \ket{\Hm_\mp}\nonumber\\
 &+\kappa^2c_{n+2,n_{12}+2,n_{21}}\ket{\Hmb_\mp}\otimes \ket{\Hm_\pm}\nonumber\\
 &\pm\kappa^2c_{n+2,n_{12},n_{21}+1}\ket{\Vmc}\otimes\ket{\Vmcb}\pm\kappa^2c_{n+2,n_{12}+1,n_{21}}\ket{\Vmcb}\otimes\ket{\Vmc}\,.
\end{align}
This action was derived in \cite{Liendo:2011xb} but not written out in full explicit detail, which was necessary for our code implementation. We provide a corresponding Wolfram Mathematica notebook as ancillary file to the ar$\chi$iv submission. It requires the freely available package grassmannOps.m by Jeremy Michelson
and Matthew Headrick, which was very useful for this task.

\bibliography{./auxi/biblio.bib}
\bibliographystyle{./auxi/JHEP}

\end{document}

%% file: auxi/packages.tex
\usepackage[ngerman, english]{babel}
\usepackage[T1]{fontenc}
\usepackage[titletoc]{appendix}
\usepackage{titletoc}
\usepackage{graphicx}
\usepackage{subcaption} 
\usepackage[dvipsnames]{xcolor}
\usepackage{amsmath,amssymb,amsthm}
\usepackage{bm}
\usepackage{float}
\usepackage[format=plain,labelfont={bf,it},textfont=it]{caption}
\usepackage{multirow}
\usepackage{slashed} 
\usepackage{mathtools} 
\usepackage{tabularx}
\usepackage{bbold}
\usepackage{enumitem}
\usepackage{tikz}
\usetikzlibrary{positioning,arrows,calc}
\usetikzlibrary{arrows.meta}
\usetikzlibrary{decorations.pathmorphing}
\usetikzlibrary{decorations.markings}
\usetikzlibrary{shapes.geometric}
\usetikzlibrary{patterns}
\tikzset{
  snake it/.style={
    decorate, 
    decoration=snake,
    segment length=3
  }
}
\usepackage{xparse}
\usepackage{pdfpages} 
\usepackage[makeroom]{cancel}

\usepackage{titlesec} 
\usepackage{fancyhdr} 
\usepackage{emptypage} 
\usepackage{chngcntr} 
\usepackage{etoolbox} 

\usepackage{dsfont} 

\usepackage{amsmath,amssymb,amsfonts,amsxtra, mathrsfs, makeidx,graphicx,amsthm,epsfig, bm,longtable,float, 
color,tikz,mathtools,xfrac,footnote,rotating, lscape, makecell, environ,mathtools, empheq, physics,cleveref,tensor,
slashed,subfiles,natbib,youngtab,multirow,slashed, longtable,graphicx, comment,verbatim,pdflscape}
\usepackage{colortbl}
\usepackage{braket}

%% file: auxi/commands.tex
\definecolor{DarkBlueGrey}{RGB}{76,94,107}
\definecolor{MediumBlueGrey}{RGB}{110,135,153}
\definecolor{LightBlueGrey}{RGB}{134,163,184}
\definecolor{WCOrange}{RGB}{242,146,29}
\definecolor{SCRed}{RGB}{179,48,48}
\definecolor{VertexColor}{RGB}{242,146,29}
\definecolor{GluonColor}{RGB}{255,172,172}
\definecolor{SEColor}{RGB}{134,163,184}

\definecolor{BGBox}{RGB}{255,254,230}
\definecolor{PlaneColor}{RGB}{230,230,230}
\definecolor{BlobColor}{RGB}{190,180,230}

\def\Am{{\mathcal{A}}}
\def\Bm{{\mathcal{B}}}
\def\Cm{{\mathcal{C}}}

\def\Em{{\mathcal{E}}}
\def\Fm{{\mathcal{F}}}

\def\Hm{{\mathcal{H}}}
\def\Im{{\mathcal{I}}}
\def\Jm{{\mathcal{J}}}
\def\Km{{\mathcal{K}}}
\def\Lm{{\mathcal{L}}}

\def\Nm{{\mathcal{N}}}

\def\Pm{{\mathcal{P}}}
\def\Qm{{\mathcal{Q}}}
\def\Rm{{\mathcal{R}}}
\def\Sm{{\mathcal{S}}}
\def\Tm{{\mathcal{T}}}

\def\Vm{{\mathcal{V}}}

\def\Xm{{\mathcal{X}}}

\def\Vmb{{\bar{\mathcal{V}}}}
\def\Vmc{{\check{\mathcal{V}}}}
\def\Vmcb{{\check{\bar{\mathcal{V}}}}}
\def\Hmb{{\bar{\mathcal{H}}}}

\def\a{{\alpha}}
\def\b{{\beta}}

\def\bd{{\dot{\beta}}}

\usepackage{bbm}

\def\ph{\phantom}

\newcommand\cb{{\bar{c}}}

\newcommand\g{{\gamma}}

\newif\ifstartcompletesineup
\newif\ifendcompletesineup
\pgfkeys{
    /pgf/decoration/.cd,
    start up/.is if=startcompletesineup,
    start up=true,
    start up/.default=true,
    start down/.style={/pgf/decoration/start up=false},
    end up/.is if=endcompletesineup,
    end up=true,
    end up/.default=true,
    end down/.style={/pgf/decoration/end up=false}
}
\pgfdeclaredecoration{complete sines}{initial}
{
    \state{initial}[
        width=+0pt,
        next state=upsine,
        persistent precomputation={
            \ifstartcompletesineup
                \pgfkeys{/pgf/decoration automaton/next state=upsine}
                \ifendcompletesineup
                    \pgfmathsetmacro\matchinglength{
                        0.5*\pgfdecoratedinputsegmentlength / (ceil(0.5* \pgfdecoratedinputsegmentlength / \pgfdecorationsegmentlength) )
                    }
                \else
                    \pgfmathsetmacro\matchinglength{
                        0.5 * \pgfdecoratedinputsegmentlength / (ceil(0.5 * \pgfdecoratedinputsegmentlength / \pgfdecorationsegmentlength ) - 0.499)
                    }
                \fi
            \else
                \pgfkeys{/pgf/decoration automaton/next state=downsine}
                \ifendcompletesineup
                    \pgfmathsetmacro\matchinglength{
                        0.5* \pgfdecoratedinputsegmentlength / (ceil(0.5 * \pgfdecoratedinputsegmentlength / \pgfdecorationsegmentlength ) - 0.4999)
                    }
                \else
                    \pgfmathsetmacro\matchinglength{
                        0.5 * \pgfdecoratedinputsegmentlength / (ceil(0.5 * \pgfdecoratedinputsegmentlength / \pgfdecorationsegmentlength ) )
                    }
                \fi
            \fi
            \setlength{\pgfdecorationsegmentlength}{\matchinglength pt}
        }] {}
    \state{downsine}[width=\pgfdecorationsegmentlength,next state=upsine]{
        \pgfpathsine{\pgfpoint{0.5\pgfdecorationsegmentlength}{0.5\pgfdecorationsegmentamplitude}}
        \pgfpathcosine{\pgfpoint{0.5\pgfdecorationsegmentlength}{-0.5\pgfdecorationsegmentamplitude}}
    }
    \state{upsine}[width=\pgfdecorationsegmentlength,next state=downsine]{
        \pgfpathsine{\pgfpoint{0.5\pgfdecorationsegmentlength}{-0.5\pgfdecorationsegmentamplitude}}
        \pgfpathcosine{\pgfpoint{0.5\pgfdecorationsegmentlength}{0.5\pgfdecorationsegmentamplitude}}
}
    \state{final}{}
}

\tikzset{
corner/.style={line width=1pt,dashed,draw=black,dash pattern=on 6pt off 4pt},
scalar/.style={line width=1pt,draw=black},
gluon/.style={line width=1pt,decorate, draw=GluonColor,
    decoration={complete sines,aspect=0,amplitude=1.25mm,segment length=1.5mm,start up,end up}},
gluontwo/.style={line width=1pt,decorate, draw=GluonColor,
    decoration={complete sines,aspect=0,amplitude=.7mm,segment length=1mm,start up,end up}},
ghost/.style={line width=1pt,loosely dotted,draw=black},
wilson/.style={line width=2pt,draw=black},
 }
\NewDocumentCommand\semiloop{O{black}mmmO{}O{above}}
{%
\draw[#1] let \p1 = ($(#3)-(#2)$) in (#3) arc (#4:({#4+180}):({0.5*veclen(\x1,\y1)})node[midway, #6] {#5};)
}

\global \long \def \ab{\textbf{a}}
\global \long \def \bb{\textbf{b}}
\global \long \def \cb{\textbf{c}}
\global \long \def \db{\textbf{d}}
\global \long \def \dag{\dagger}
\global \long \def \lbr{\lbrack}
\global \long \def \rbr{\rbrack}
\global \long \def \a{\alpha}
\global \long \def \b{\beta}
\global \long \def \g{\gamma}

\global \long \def \ad{\dot{\alpha}}
\global \long \def \bd{\dot{\beta}}
\global \long \def \gd{\dot{\gamma}}

%% file: auxi/topmatter.tex
\usepackage{dsfont}
\usepackage{float}
\usepackage{pgfplots}
\pgfplotsset{compat=1.14}

\let\oldbfseries=\bfseries
\let\oldmdseries=\mdseries
\let\oldnormalfont=\normalfont
\renewcommand{\bfseries}{\oldbfseries\boldmath}
\renewcommand{\mdseries}{\oldmdseries\unboldmath}
\renewcommand{\normalfont}{\oldnormalfont\unboldmath}

\makeatletter
\newlength{\apb@width}
\newcommand{\autoparbox}[2][c]{\settowidth{\apb@width}{#2}\parbox[#1]{\apb@width}{#2}}

\makeatother

\def \ph{\phantom}

\def\Am{{\mathcal{A}}}
\def\Bm{{\mathcal{B}}}
\def\Cm{{\mathcal{C}}}

\def\Em{{\mathcal{E}}}
\def\Fm{{\mathcal{F}}}

\def\Hm{{\mathcal{H}}}
\def\Im{{\mathcal{I}}}
\def\Km{{\mathcal{K}}}

\def\Nm{{\mathcal{N}}}

\def\Pm{{\mathcal{P}}}
\def\Qm{{\mathcal{Q}}}
\def\Rm{{\mathcal{R}}}
\def\Sm{{\mathcal{S}}}

\def\Xm{{\mathcal{X}}}

\def\a{{\alpha}}
\def\b{{\beta}}

\def\ad{{\dot{\alpha}}}
\def\bd{{\dot{\beta}}}

\def\ph{\phantom}

\newcommand{\beq}{\begin{equation}}
\newcommand{\eeq}{\end{equation}}

\definecolor{nicegreen}{rgb}{0.1,0.6,0.1}

\newmuskip\pFqskip
\mathchardef\pFcomma=\mathcode`,

\makeatletter
\renewcommand*\env@matrix[1][\arraystretch]{%
  \edef\arraystretch{#1}%
  \hskip -\arraycolsep
  \let\@ifnextchar\new@ifnextchar
  \array{*\c@MaxMatrixCols c}}
\makeatother